\documentclass[fleqn,usenatbib]{mnras}

\usepackage{mathptmx}

\usepackage[caption = false]{subfig}

\usepackage[T1]{fontenc}

\usepackage{comment}
\usepackage{soul}

\DeclareRobustCommand{\VAN}[3]{#2}
\let\VANthebibliography\thebibliography
\def\thebibliography{\DeclareRobustCommand{\VAN}[3]{##3}\VANthebibliography}

\usepackage{graphicx}	
\usepackage{amsmath}	
\usepackage{amssymb}	

\usepackage{pdflscape}	

\usepackage[dvipsnames]{xcolor}

\renewcommand{\deg}{^{\circ}} 

\defcitealias{Johnston2019}{JK19}

\title[TPA -- pulse widths]{The Thousand-Pulsar-Array programme on MeerKAT:  -- VI.\\ Pulse widths of a large and diverse sample of radio pulsars}

\author[B. Posselt et al.]{B. Posselt$^{1,2}$\thanks{E-mail: bettina.posselt@physics.ox.ac.uk (BP)},
A. Karastergiou$^{1,3}$,
S. Johnston$^{4}$, 
A.~Parthasarathy$^{5}$,
M.~J.~Keith$^{6}$,\newauthor
L. S. Oswald$^{1,7}$,
X. Song$^{6}$, 
P. Weltevrede$^{6}$, 
E.~D.~Barr$^{5}$,
S. Buchner$^{8}$,
M. Geyer$^{8}$,\newauthor
M. Kramer$^{5,6}$,
D.~J.~Reardon$^{9,10}$,
M.~Serylak$^{11,12}$
R.~M.~Shannon$^{9,10}$, 
R.~Spiewak$^{6,8}$,\newauthor
V.~Venkatraman~Krishnan$^{5}$\\
{$^{1}$ Department of Astrophysics, University of Oxford, Denys Wilkinson Building, Keble Road, Oxford OX1 3RH, UK}\\
{$^{2}$ Department of Astronomy \& Astrophysics, Pennsylvania State University, 525 Davey Lab, 16802 University Park, PA, USA}\\
{$^{3}$ Department of Physics and Electronics, Rhodes University, PO Box 94, Grahamstown 6140, South Africa}\\
{$^{4}$ CSIRO Astronomy and Space Science, Australia Telescope National Facility, PO~Box~76, Epping NSW~1710, Australia}\\
{$^{5}$ Max-Planck-Institut f\"{u}r Radioastronomie, Auf dem H\"{u}gel 69, D-53121 Bonn, Germany}\\
{$^{6}$ Jodrell Bank Centre for Astrophysics, Department of Physics and Astronomy, University of Manchester, Manchester M13 9PL, UK}\\
{$^{7}$ Magdalen College, University of Oxford, Oxford OX1 4AU, UK}\\
{$^{8}$  South African Radio Astronomy Observatory, Black River Park, 2Fir Street, Observatory, Cape Town, 7925, South Africa}\\
{$^{9}$ Centre for Astrophysics and Supercomputing, Swinburne University of Technology, Hawthorn, VIC, 3122, Australia}\\
{$^{10}$ ARC Centre of Excellence for Gravitational Wave Discovery (OzGrav)}\\
{$^{11}$  Square Kilometre Array Observatory, Jodrell Bank Observatory, Macclesfield, Cheshire SK11 9DL, United Kingdom}\\
{$^{12}$  Department of Physics and Astronomy, University of the Western Cape, Bellville, Cape Town, 7535, South Africa}\\
}

\date{Accepted XXX. Received YYY; in original form ZZZ}

\pubyear{2021}

\begin{document}
\label{firstpage}
\pagerange{\pageref{firstpage}--\pageref{lastpage}}

\maketitle

\begin{abstract}
We present pulse width measurements for a sample of radio pulsars observed with the MeerKAT telescope as part
of the Thousand-Pulsar-Array (TPA) programme in the MeerTime project.  
For a centre frequency of 1284~MHz, we obtain 762 $W_{10}$ measurements across the total bandwidth of 775\,MHz, where $W_{10}$ is the width at the 10\% level of the pulse peak. We also measure about 400 $W_{10}$ values in each of the four or eight frequency sub-bands.
Assuming, the width is a function of the rotation period $P$, this relationship can be described with a power law with power law index $\mu=-0.29\pm 0.03$. 
However, using orthogonal distance regression, we determine a steeper power law with $\mu=-0.63\pm 0.06$. A density plot of the period-width data reveals such a fit to align well with the contours of highest density. 
Building on a previous population synthesis model, we obtain population-based estimates of the obliquity of the magnetic axis with respect to the rotation axis for our pulsars.
Investigating the width changes over frequency, we unambiguously identify a group of pulsars that have width broadening at higher frequencies. The measured width changes show a monotonic behaviour with frequency for the whole TPA pulsar population, whether the pulses are becoming narrower or broader with increasing frequency. We exclude a sensitivity bias, scattering and noticeable differences in the pulse component numbers as explanations for these width changes, and attempt an explanation using a qualitative model of five contributing Gaussian pulse components with flux density spectra that depend on their rotational phase.
\end{abstract}



\section{Introduction}
Radio pulsars show a large variety of pulse profile widths, $W$. The width is one of the simplest descriptors of the pulse morphology.
The observed widths depend on the radio emission physics of the neutron star, the viewing geometry, the properties of the interstellar medium (ISM) through which the radio emission passes until detection, as well as the instrumental setup such as observing frequency and bandwidth.
Investigating some of these dependencies, many previous works in the literature have therefore studied the widths of pulsar profiles of various pulsar samples. Key works include  \citet{Lyne1988, Rankin1990, Rankin1993,Gil1993, Kramer1994, Gould1998, Tauris1998, Mitra2002}. The possible constraints on pulsar emission models and pulsar geometry provided by pulse width measurements remain a topic of intensive study with more recent works including \citet{Young2010,Maciesiak2011,Maciesiak2012,Skrzypczak2018,Johnston2019}, hereafter \citetalias{Johnston2019}. 
According to the textbook description, intrinsic factors that influence the measured pulse widths include the height of the radio emission above the surface of the star, $h_{em}$, the radio beam properties (e.g., its shape, discernable components, and whether it is filled or patchy), the number of radio emitting regions (i.e., multipolar or dipolar structures), but also the size of the region with open magnetic field lines, represented by the radius of the (assumed to be small) polar cap. For a dipole field, this is (e.g., \citealt{Ruderman1975}):
\begin{equation}
r_p = (R^3_{\rm NS}/R_{\rm LC})^{1/2}  = \left[2 \pi R^3_{\rm NS}/ (c P)\right]^{1/2}. 
\end{equation}
where $R_{\rm LC}= cP/ (2\pi)$ is the radius of the light cylinder, $c$ is the speed of light, $R_{\rm NS}$ is the radius of the neutron star, and $P$ the pulsar's 
rotation period.
For a constant emission height, the width of the assumed (circular and filled) radio emission beam cone, defined as the full opening angle $2 \rho$, can be expressed by $r_p$ and thus $P$, as illustrated by \citet{Rankin1990}:
\begin{equation}
w_{\rm beam} = 2 \rho=  3 h^{1/2}_{em} r_p R^{-3/2}_{\rm NS} = \left[(18 \pi h_{em}) / (c P)\right]^{1/2}. 
\end{equation}
The intrinsic beam width $w_{\rm beam}$ is modified by a geometric factor to the observable pulse profile width, $W$ as
\begin{equation}
\cos( w_{\rm beam}/2)  = \cos \alpha \cos \zeta + \sin \alpha \sin \zeta \cos( W/2), 
\end{equation}
where  $\alpha$ is the angle between the rotation axis and magnetic axis, $\zeta=\alpha +\beta$ with $\beta$ as the angle between the observer's line of sight and the magnetic axis \citep{Gil1984}. 
Even in the case of the simplest assumptions (a circular, filled beam,  and all emission originating from the same height), the geometric effects can modify the $W \propto P^{-1/2}$ relation that one would expect at $\alpha=90^{\circ}$ and $\beta=0^{\circ}$.
Despite this, the above mentioned studies 
found that the relation between $P$ and $W$ can be described by a power law (PL) with slightly varying slopes  
depending on samples, observation frequency and method of width measurement. 
These results indicated a more complex radio beam with  varying emission heights and potential dependency on the pulsar evolution as explored by, e.g., \citealt{Gupta2003}, \citetalias{Johnston2019}.

{\citet{Rankin1990} discovered the existence of a Lower Boundary Line (LBL) in the scatter plot of core-component pulse widths of interpulsars with a relation $W_{50} \propto P^{-1/2}$, where $W_{50}$ is the measured pulse width at 50\% of the pulse profile peak. The LBL width relation has been confirmed for core and conal emission of interpulsars in several follow-up studies, e.g., by  \citet{Maciesiak2011,Maciesiak2012,Skrzypczak2018}. \citet{Maciesiak2011} found that the lower boundary region is not only populated by interpulsars, a fact that was exploited by \citet{Skrzypczak2018} to confirm the $P^{-1/2}$ dependence of the LBL with 123 pulsars using quantile regression.}

A large, homogeneously measured $P-W$ sample over a wide range of periods has the potential to further illuminate the physics of the radio emission beam, and propel us towards a position of being able to disentangle intrinsic (beam structure, pulsar spin-down power, magnetic field strength, geometry $\alpha$) and exterior (viewing geometry $\beta$, ISM interaction) factors. 

Pulse profile widths are also interesting with respect to the multi-frequency view of neutron stars. 
For example, it was suggested that the so-called ``X-ray dim isolated neutron stars'' are radio-quiet because they are viewed well off axis from the radio beam (e.g., \citealt{Kaspi2010}). Such a suggestion implicitly makes use of the fact that these neutron stars have long periods ($>4$\,s), with expected small beam widths according to the period-width relation, hence likely narrow radio beams that can easily miss Earth. 
Considering radio and $\gamma$-ray emission detections,  \citet{Rookyard2017} and \citet{Johnston2020b} identified a demarcation line of radio pulse widths measured against the spin-down energy, above which $\gamma$-ray pulsars are rarely seen.

\subsection{Frequency-dependent pulse widths}
The availability of sensitive radio telescopes covering a broad range of radio frequencies has resulted in pulse width measurements at different frequencies. Recent works considering non-recycled pulsars include LOFAR-based measurements at 10--240\,MHz \citep{Pilia2016}, measurements with the Green Bank telescope at 350\,MHz \citep{McEwen2020} and, at the high frequency end, pulse width measurements at 5\,GHz and 8.6\,GHz by \citet{Zhao2017,Zhao2019}, and at 32\,GHz by \citet{Xilouris1996}.
Multifrequency investigations of integrated pulse profiles have been carried out, e.g., by \citet{Olszanski2019} for 46 pulsars at three frequencies (327\,MHz, 1.4\,GHz, and 4.5\,GHz), by \citet{Mitra2016} for 93 pulsars at two frequencies ($333$\,MHz, 618\,MHz),
by \citet{Johnston2008} for 34 pulsars at five frequencies ($243 - 1400$\,MHz), while \citet{Chen2014} studied the frequency dependence of the pulse widths of 150 normal pulsars, having at least four width measurements between 0.4\,GHz and 4.85\,GHz.
Based on earlier work by \citet{Thorsett1991}, \citet{Chen2014} used the measured pulse widths at the 10\% maximum of their pulse profiles, $W_{10}$, at different frequencies $\nu$ to carry out 
fitting of the relation 
\begin{equation}
\label{equ:thorsett}
W_{10}= A_T \nu^{\mu}+W_{10,0},
\end{equation}
where $A_T$ is their best-fitting coefficient A, $W_{10,0}$ is an asymptotic constant, and $\mu$ is the index reflecting (to some extent) the degree of broadening or narrowing of the pulse profile with frequency. As \citet{Chen2014} discussed, $\mu$ alone is not enough as a single parameter to classify different kinds of pulse width evolution. This is mostly due to measurement uncertainties and the existence of the third parameter $W_{10,0}$.
To comprehensively characterize their data, \citet{Chen2014} calculated fractional pulse width changes, defined as the $W_{10}$--difference between their highest and lowest frequency, normalized by the width at the lowest frequency.\\

The previous studies (e.g., \citealt{Johnston2008,Chen2014,Noutsos2015, Pilia2016,Zhao2019}) found narrowing of the pulse width (or component separation) with frequency for many pulsars as per the textbook scenario of radius-to-frequency mapping (e.g., \citealt{Komesaroff1970,Cordes1978}), where lower frequencies are thought to be emitted from higher in the pulsar magnetosphere than high frequencies. Assuming the emission comes from the same set of open field lines, the beam opening angle is greater at higher heights.
A complementary interpretation of this width change attributes the effect to propagation in the magnetosphere (e.g., \citealt{LyubarskiiPetrova1998,McKinnon1997,Noutsos2015}).  
However, for some pulsars the above studies also reported the opposite width change behaviour.
\citet{Johnston2008}, for example, found the pulse width increasing with frequency for one third of their sample.
\citet{Chen2014} reported about 20\% of their pulsars 
to show clearly increasing pulse width over frequency, and noted that these challenge the conventional picture where radio beam size is assumed to shrink with increasing frequency. 
Testing for a geometrical effect where emission beam shrinkage could lead to a steepening of the emission spectrum, \citet{Chen2014} reported negative results.
Instead, assuming a broadband nature for the radio emission as well as a fan beam model, \citet{Chen2014} suggested that the pulse
width change is a consequence of differences in the  spectrum across the emission region.\\

Here, we present measurements of a homogeneous large sample of pulse widths from the Thousand-Pulsar-Array (TPA) programme \citep{Johnston2020}
on the MeerKAT telescope. We concentrate on a population-wide interpretation of these measurements and 
also study their frequency dependence within the bandwidth of the MeerKAT L-Band receiver

\section{Observations}
\label{secObs}
Our pulsar observations were carried out as part of the TPA programme \citep{Johnston2020}
on the MeerKAT telescope, a 64-dish radio interferometer. MeerKAT is located in the Karoo region of South Africa and is operated by the South African Radio Astronomy Observatory (SARAO).
\citet{Bailes2020} presented in detail the instrumentation of MeerKAT for pulsar observations.

In this paper we use data obtained with the L-band receiver. It is centred at a frequency of 1284 MHz. We use a total bandwidth of 775\,MHz. 
The channelized time series were processed by the Pulsar Timing User Supplied Equipment (PTUSE) machines. There are 4 PTUSE machines that can each process one tied-array beam at a time. 
We use fold-mode observations obtained with $\sim$ all 64 antennas (full array) or about half the array (subarray) until 2020 October 30. 
Overall 1274 unique pulsars were observed between 2019 March 8 and 2020 October 30 (6277 individual observations), partly during the commissioning phase of the MeerKAT telescope. 
The population properties of the pulsar sample are discussed in an accompanying paper (Posselt et al., in prep.)

\section{Data Analysis}
\label{secAnalysis}
\subsection{Data reduction}
The software library
{\sc dspsr}\footnote{\href{http://dspsr.sourceforge.net}{http://dspsr.sourceforge.net}}\citep{vanStraten2011} provides the pipelines to process the data. The resulting data have 1024 frequency channels, sub-integration times of 8\,s, and all 4 Stokes parameters.\\ 

The data were folded using parameters obtained from {\sc psrcat} \citep{Manchester2005}\footnote{\href{http://www.atnf.csiro.au/research/pulsar/psrcat}{http://www.atnf.csiro.au/research/pulsar/psrcat}}, or a recent ephemeris from ongoing pulsar timing programmes at the Jodrell Bank Observatory or Parkes Observatory. Some of the catalogue ephemerides had errors in the pulse period large enough to create a phase drift across the longer observations. In these cases the pulsar frequency parameter was corrected by forming 8 time-of-arrival measurements over the longest single observation using {\sc psrchive} \citep{vanStraten2012}, and re-fitting the pulsar spin frequency parameter
with {\sc tempo2} \citep{Hobbs2006tempo2,Edwards2006tempo2}. 
After updating the pulsar parameters each observation was individually inspected to confirm that the phase drift was no longer present. 
We also computed and updated the dispersion measures, DM, of the pulsars. We are preparing details of these measurements in a catalogue (Posselt et al., in prep.) where the iterative process of profile template generation will be described as well. 
The pulsar data were de-dispersed, and cleaned of Radio Frequency Interference (RFI)
using {\sc CoastGuard} \citep{Lazarus2016}.\\ 

For a number of pulsars, data were taken on multiple epochs. Appropriate weighting of the data, e.g., by the number of antennas used per epoch and the observing time, was carried out. 
After the data were reprocessed with the updated ephemeris, we used {\sc tempo2} to obtain the phase shift of each individual observation in comparison to a reference epoch. Correcting for these phase shifts, data from individual observing epochs were then aligned and added using {\sc psrchive}.
The aligned individual and combined pulse profiles (averaged in frequency) were visually checked for consistency. 
There are 827 pulsars with a combined data set that we consider in addition to the individual observations. 
For the width measurements, the data were time-integrated to a single Stokes I profile per observation with our standard resolution of 1024 bins per pulse period. In order not to miss faint pulses, we additionally measured widths for a resolution of 256 bins per pulse period. In this paper we consider the frequency-averaged profiles, as well as data divided into four and eight frequency sub-bands. 
\subsection{Width measurements}
\label{sec:widthmeas}
In order to obtain smooth noiseless pulse profiles, we use the method presented by \citetalias{Johnston2019} and \citet{Brook2019} and derive a Gaussian Process (GP) for each pulse profile. 
The GP also allows us to determine the noise variances, $\sigma_{\rm GP}$, of the input (observed) pulse profiles. 
\citet{Roberts2012} and \citet{Rasmussen2006} describe the general features and applicability of the GP. Briefly, it is a Bayesian, non-parametric model that does not require any assumption about the functional form of the pulse profile. The method assumes that the data consists of a smooth signal and a (homoscedastic) white noise term. We employ the Python GP-package \texttt{George}\footnote{\href{https://george.readthedocs.io}{https://george.readthedocs.io}} by \citet{georgepython}.
Following \citetalias{Johnston2019}, we use a squared exponential kernel, resulting in a model with three hyper-parameters: magnitude and length-scale of the squared exponential and standard deviation of the white noise term $\sigma_{\rm GP}$. 
The model was found to perform very well in producing noiseless profiles of high fidelity. This GP method allows easy separation of signal and noise model components in the measurement without the requirement to pre-define on-pulse and off-pulse regions, and is therefore particularly useful for modelling profiles and obtaining a noise variance in pulsars with large duty cycles (e.g. millisecond pulsars). The profile model also allows for analytical computation of the derivatives of the signal, which we are experimenting with in counting the number of distinct components in a profile. \\  

\begin{figure}
\includegraphics[width=8.5cm]{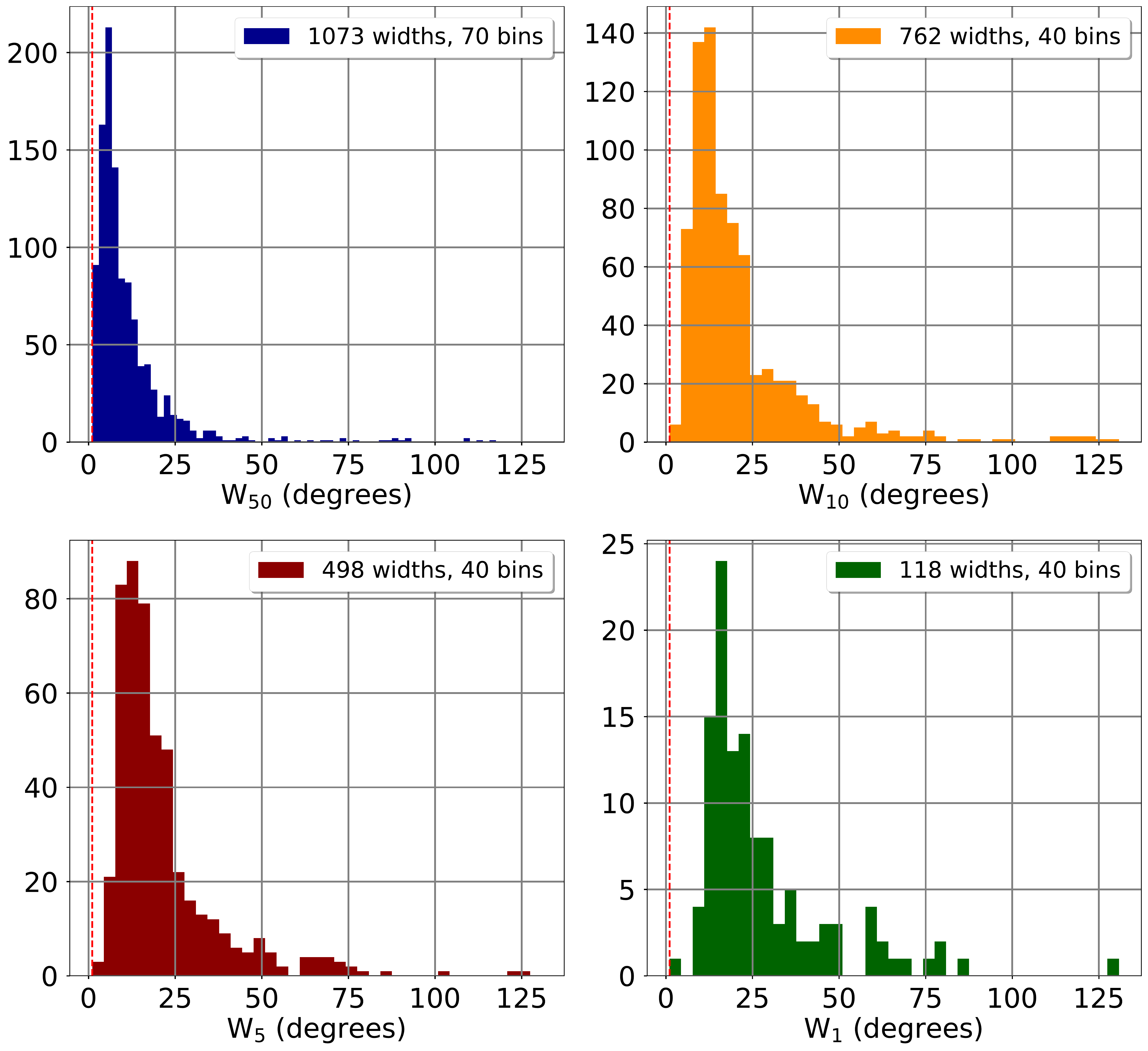}
\vspace{-0.5cm}
\caption{Width histograms for the frequency-averaged data after applying all filters described in Section~\ref{sec:widthmeas}. Only widths larger than the low limit (1\,degree), indicated by the dotted red line, are considered for further analysis.  \label{fig:widthsThisto}}
\end{figure}

In order to measure pulse widths, we define a contiguous analysis region (or on-pulse region) around the pulse. We use the pulse profile data (after baseline subtraction), the GP-derived noiseless profile, and $\sigma_{\rm GP}$, by requiring the edges to have a minimum signal-to-noise ratio of three in (at least) five subsequent bins. The peak of the noiseless profile and $\sigma_{\rm GP}$ define our \emph{peak signal-to-noise ratio}, $S/N$. The pulse widths $W_{50}$, $W_{10}$, $W_{5}$, $W_{1}$ correspond to measurements where the noiseless pulse profile is 50\%, 10\%, 5\%, and 1\% of its peak value respectively. 
Width uncertainties are obtained from taking measurements at $\pm 1 \sigma_{\rm GP}$, e.g., for $W_{50}$ the width measurement at 50\% of the peak value $\pm 1 \sigma_{\rm GP}$ defines the $\pm 1 \sigma$ uncertainty of $W_{50}$ respectively. This approach only returns whole phase bin values, and 
for very small uncertainties ($<0.5$\,bin) a value of 0 is returned. In such cases, we set the uncertainty to a conservative 0.5\,bin value for the following analysis steps.
The width uncertainties are often slightly asymmetric ($\sigma_{+},\sigma_{-}$) as can be seen from the listed values in the Table~\ref{tab:w50w10w5w1}. Where we applied symmetric errors in subsequent analysis (e.g. power-law fits), we chose the larger of the two uncertainties.
Based on the binning of our data and our error definitions, we require width measurements to be at least 3 bins (or approximately 1 degree in rotational phase) to be included in our further analysis. Figure~\ref{fig:widthsThisto} shows that this criterion together with our standard 1024 bin sampling for one full pulse period is sufficient to detect nearly all narrow instances of $W_{10}$, i.e., there is no significant ``bin bias'' for narrow profiles.\\

We have checked that our peak-based $W_{10}$ values are consistent with results from an alternative width measurement method described in \citet{Noutsos2015}. Their method excludes a two-tailed percentage (left and right bound) from the cumulative flux-density distribution and produces a smoother evolution of the pulse width with frequency. Considering a 90\% fraction of the total pulse energy for the method by \citet{Noutsos2015}, we evaluated the width differences (without errors) from the two methods, and conclude that, for this sample of profiles, both methods produce very similar results with negligible differences.\\

We carry out the pulse width measurements for all the data with an observing time covering at least 500 pulsar rotations.
The choice of this criterion and its implication for our analysis are discussed in more detail in Section~\ref{sec:widthvari}. We use observations from individual epochs as well as multi-epoch combined data for the respective pulsar.
For each pulsar, we obtain measurements for the total band (frequency-averaged) data as well as for each of the eight (or four) frequency sub-bands.
In some cases, bad data from one long epoch can influence a combined data set with a short clean observation from another epoch. The latter would provide more significant measurements than the combined data set. 
To avoid such cases, we select the observational data set with the highest $S/N$ in a chosen frequency band. In most cases, these are the combined multi-epoch data (if available), as expected. 
The minimum $S/N$s of pulsars with  $W_{50}$ and $W_{10}$ measurements in the frequency-averaged data are 8 and 40, respectively.
For pulsars with known interpulses or profiles with components that resemble interpulses\footnote{where there is no conclusive evidence we actually observe emission from both magnetic poles} (indicated in Table~\ref{tab:w50w10w5w1} with a flag),   
two width measurements (for main and interpulse) are carried out and reported if significant.
We exclude 113 pulsars from our ``scattered'' list \citep{Oswald2021} from further analysis.
Their width measurements are, however, included in the listed measurements.\\

A width measurement is selected as reliable enough for further investigation if the GP-defined on-pulse region includes $W_X + 1\sigma_{+}$. This ensures a reasonable uncertainty measurement. We further require the analysis region to be less than three quarters of the total rotational phase space (covering one full period), i.e. $<75$\,per cent duty cycle. This cut was introduced after visual cross checks of width measurements and corresponding profiles. This restriction in rotational phase lead to the automatic exclusion of the occasional contaminated observation where the baselines showed systematic ``wiggles'', and any missed strongly-scattered pulsars. The visual checks only found such scattered pulsars to have these extremely wide profiles. \\

In order to also obtain $W_{10}$ measurements for faint profiles in noisier data, we used wider bins, i.e., 256 phase bins per pulse period. These 256-bin measurements were only used when the 1024-bin data did not result in reliable $W_{10}$ widths. For the frequency-averaged data, for example, 67 out of the reliable 762 $W_{10}$ measurements are based on the 256-bin data.

\begin{figure}
\includegraphics[width=8.5cm]{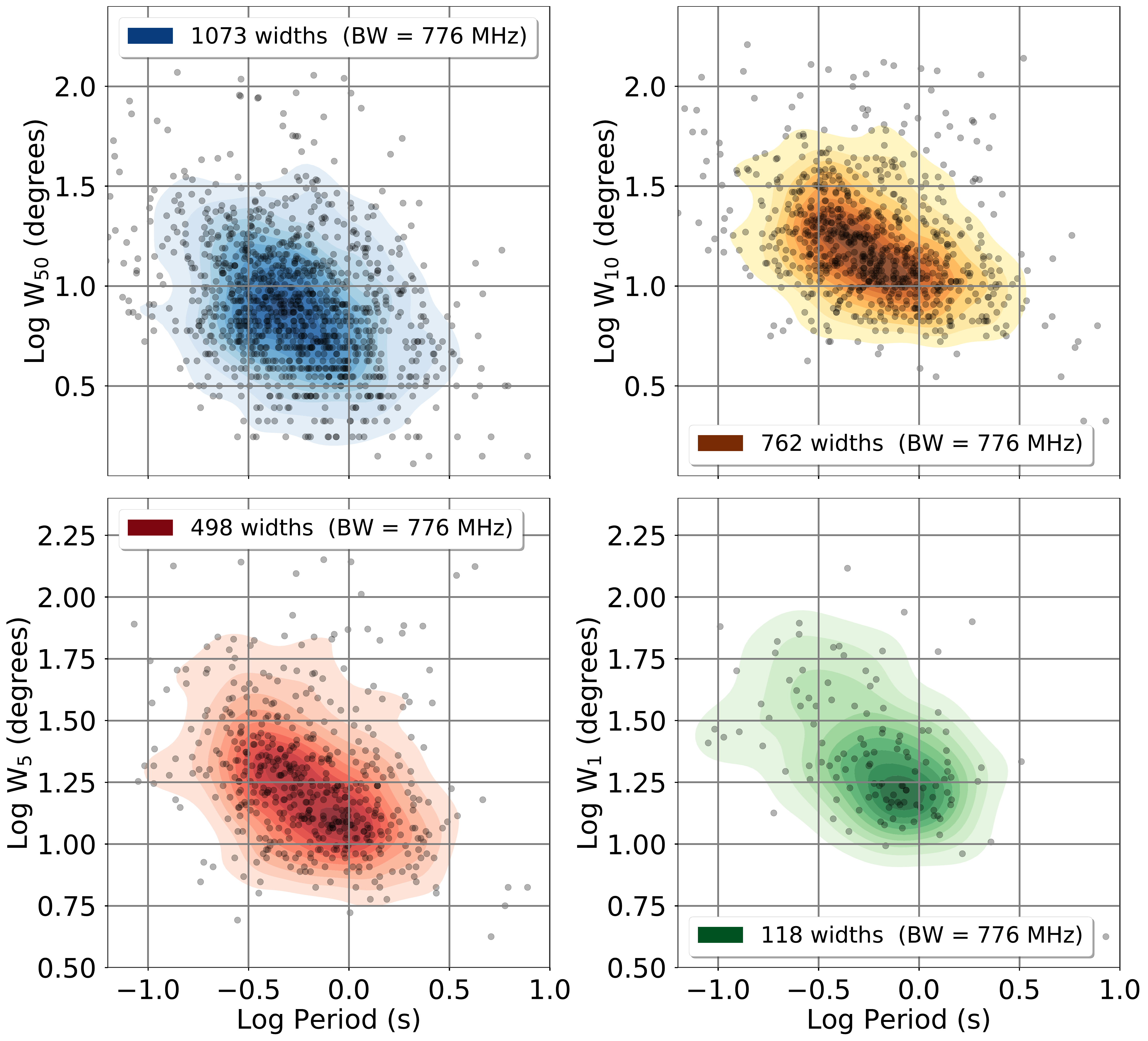}
\vspace{-0.5cm}
\caption{Widths over period plots for the frequency-averaged data. The underlying shaded areas represent the  kernel density estimate (KDE) plots (2D Gaussian, smoothing bandwidth chosen using Scott's rule, \citealt{Scott2015}), while the dots mark the individual width measurements.\label{fig:widthsTkde}}
\end{figure}

\subsection{Determining width relations and their uncertainties}
It is standard practice (e.g., \citealt{Rankin1990}) to assume power law relations of the width with period, $W_{10}=f(P)$, or  spin-down energy, $W_{10}=f(\dot{E})$.
We tested several objective functions for ordinary least-square minimization (OLS) for a power-law (PL) fit (e.g., $W_{10}= A \times (P/s)^{\mu} + c$) of the data with and without a constant offset, $c$, from zero. We considered a straight-line fit in logarithmic space, a PL-fit in linear space and its corresponding error-weighted variant, utilizing Python's \texttt{SciPy} and \texttt{lmfit} libraries \citep{2020SciPy,lmfit2016}.
The third parameter, the offset $c$, was only used in linear space. If an offset was considered, the fits were found to be highly dependent on the chosen initial conditions of the fit parameters. Since we were not able to obtain reliable fits, we restricted to a two-parameter fit with amplitude $A$ and PL-index $\mu$. 
Even in such a fit, the parameters are highly correlated and we found that slight changes in the sample size shifted the best-fitting results.
Typical uncertainty estimates, such as obtained using Markov chain Monte Carlo (MCMC), gave unreliably small values.
Therefore, we used bootstrapping (BT) to derive more reliable confidence intervals. Creating 10,000 random samples with replacements we studied the distribution of the resulting fit parameters, its general shape, and in particular the 16\%, 50\%, and 84\% quantiles. In the case of a Gaussian-shaped distribution of the fit parameters, we report in the following their standard deviation, otherwise the 50\%--16\%, 84\%--50\% quantile values as their confidence intervals. The BT-derived fit parameters are robust against slight changes in the size of the pulsar sample. 
The width uncertainties (only considered for the PL-fit in linear space) were found to have a negligible effect on the fit result since the actual spread of (well-constrained) width measurements in a chosen  period bin is much larger. Therefore, in the following we only discuss the results from the straight-line fits in logarithmic space.\\

As shown in Section~\ref{sec:Results}, the width-period sample can be represented by a contour or density plot instead of a scatter plot. For these contour plots, we employed a kernel density estimator within python's \texttt{seaborn}-package \citep{waskom2020seaborn}. In particular, we used a 2D-Gaussian kernel with bin widths according to Scott's rule \citep{Scott2015} after testing also for other bin width values. The resulting density plots (Fig.~\ref{fig:w10pfit}) clearly show a deviation of the OLS-fit 
from the line along the highest density region.
This indicates a problem with the OLS-fit due to the fact that the distribution of widths in a small period bin is typically not Gaussian and even difficult to describe with the same specific distribution over the whole period range.
To describe the line of highest source density, we used orthogonal distance regression (ODR) which minimizes the distances to the line for \emph{all} variables, in contrast to the OLS that only uses distances (residuals) of the dependent variable, i.e., the width.
For comparison of the minimisation statistics, our OLS objective function is:
\begin{equation}
OF_{OLS}(\mu, LA) =\sqrt{{\sum}_{i=1}^n ({\log W}_{calc,i} - {\log W}_i)^2 }
\end{equation}
where ${\log W}_{calc}$ correspond to the pulse widths that one would obtain with the current (linear) fit parameters (slope $\mu$ and intercept $LA$ which is related to the power law amplitude $A=10^{LA}$), and ${\log W}$ correspond to the width measurements for n pulsars.
The ODR objective function\footnote{This follows the ODRPACK documentation of the applied \texttt{scipy.odr} implementation.} is
\begin{equation}
OF_{ODR}(\mu, LA) = 
{\sum}_{i=1}^n ({\log W}_{calc,i} - {\log W}_i)^2 + ({\log P}_{calc,i} - {\log P_i})^2   
\end{equation}
, where $P_i$ is the $i$th pulsar's period, and 
\begin{equation}
{\log P}_{calc,i} = \frac{{\log W}_{calc,i} + \frac{1}{\mu}{\log P_i} - LA}{\mu+\frac{1}{\mu}}. 
\end{equation}
No uncertainties are considered for the ODR-fits of the width relations. The width scatter is much larger than their respective uncertainties and the periods have negligible uncertainties anyway. We use bootstrap to determine the confidence intervals in similar way as for the OLS-results.
These results and their implications are discussed in more detail in Section~\ref{sec:Discussion}.\\ 

{Following the approach by \citet{Skrzypczak2018}, we use quantile regression (QR) to estimate the LBL of our width-period sample. Similarly to these authors we use the QR-implementation in the python \texttt{statsmodel}-package \citep{seabold2010statsmodels} to carry out straight-line fits of the frequency-averaged data in logarithmic space.\footnote{{For the QR objective function, we refer to formula 14 in the Appendix of \citet{Skrzypczak2018}}} In this work, we do not differentiate between conal or core components.}\\

We attempted to fit width-frequency relations according to equation~\ref{equ:thorsett} \citep{Thorsett1991} using the measurements in the eight frequency sub-bands of the TPA pulsars. 
We found the fit results to be in general very uncertain and very dependent upon the choice of initial conditions. There are strong correlations between the fit parameters $\mu$, $A_T$, and $W_{10,0}$ which are difficult to break within the limited scope of our bandwidth. While some individual pulsars (such as those with high S/N and strong width-frequency dependency) may allow reasonable fits, the results for the overall population had uncertainties that were too large to be meaningful. Therefore, we investigated the frequency dependence of the pulse width indirectly by characterizing the properties of the width differences for two frequency sub-bands in Section~\ref{res:colours}.

\section{Results}
\label{sec:Results}
\begin{table*}
\centering
\caption{Pulse profile widths as measured for the frequency-averaged data of 1208 TPA pulsars (excerpt). Listed are period, $P$, and the dispersion measure, DM, used to obtain our folded pulse profiles, $W_{50}$, $W_{10}$, ($W_{5}$, $W_{1}$ online only), their respective positive and negative $1 \sigma$, in case of $W_{10}$, the larger of both, listed separately as $W_{10}$Err, was used when symmetric uncertainties were considered (see text). The Gflag (True if it is 1) indicates whether a width measurement fulfills our criterion for a ``valid'' measurement  (see text) to be included in our further analysis. The IPflag indicates pulsars that are thought to have interpulses, the component lists whether the measurement is for the major or minor peak in the pulse profile of the interpulse pulsar. The Sflag (True if it is 1) indicates scattered pulsars from the list of \citet{Oswald2021}. These pulsars are excluded in our analysis of $W_{10}$-relations. The full table is available as supplementary material.}
\label{tab:w50w10w5w1}
\begin{tabular}{lrrcccrrrcrrrrc}
        PSRJ &  $P$&  DM& IPflag &  comp. &  Sflag &    $W_{50}$ &   $\sigma^{+}_{W50}$ &   $\sigma^{-}_{W50}$ &  $W_{50}$Gflag &    
	$W_{10}$ &   $\sigma^{+}_{W10}$ &   $\sigma^{-}_{W10}$ &  $W_{10}$Err &  $W_{10}$Gflag \\

            &  s &  cm$^{-3}$  &   & 1/2 &   &  $\degr$   &   $\degr$ &   $\degr$ &   &    
	$\degr$ &   $\degr$ &   $\degr$ &  $\degr$ & \\   
  J0034-0721 & 0.94 &     10.92 &    $\cdots$ &         1 &      $\cdots$ &     20.39 &      0.18 &     -0.35 &            1 &     45.70 &      0.18 &     -0.70 &       0.70 &            1 \\ 
  J0038-2501 & 0.26 &      5.71 &    $\cdots$ &         1 &      $\cdots$ &     19.34 &     20.74 &     -2.81 &            1 &   $\cdots$ &   $\cdots$ &   $\cdots$ &    $\cdots$ & $\cdots$  \\
  J0045-7042 & 0.63 &     70.00 &    $\cdots$ &         1 &      $\cdots$ &     11.25 &      1.05 &     -1.05 &            1 &   $\cdots$ &   $\cdots$ &   $\cdots$ &    $\cdots$ & $\cdots$ \\ 
  J0108-1431 & 0.81 &      2.38 &    $\cdots$ &         1 &      $\cdots$ &     13.36 &      0.18 &     -0.35 &            1 &     27.07 &      4.22 &     -0.18 &       4.22 &            1 \\ 
  J0111-7131 & 0.69 &     76.00 &    $\cdots$ &         1 &      $\cdots$ &      7.73 &      1.05 &     -1.05 &            1 &   $\cdots$ &   $\cdots$ &   $\cdots$ &    $\cdots$ & $\cdots$ \\ 
  J0113-7220 & 0.33 &    125.49 &    $\cdots$ &         1 &      $\cdots$ &      4.92 &      0.35 &     -0.35 &            1 &     12.66 &      0.70 &     -0.70 &       0.70 &            1 \\ 
  J0131-7310 & 0.35 &    205.20 &    $\cdots$ &         1 &      $\cdots$ &      4.57 &      0.18 &     -0.70 &            1 &   $\cdots$ &   $\cdots$ &   $\cdots$ &    $\cdots$ & $\cdots$ \\ 
  J0133-6957 & 0.46 &     22.95 &    $\cdots$ &         1 &      $\cdots$ &      3.52 &      0.18 &     -0.35 &            1 &   $\cdots$ &   $\cdots$ &   $\cdots$ &    $\cdots$ & $\cdots$ \\ 
  J0134-2937 & 0.14 &     21.81 &    $\cdots$ &         1 &      $\cdots$ &      6.68 &      0.18 &     -0.70 &            1 &     18.63 &      0.35 &     -0.18 &       0.35 &            1 \\ 
  J0151-0635 & 1.46 &     25.66 &    $\cdots$ &         1 &      $\cdots$ &     29.88 &      0.18 &     -0.35 &            1 &     36.56 &      0.35 &     -0.18 &       0.35 &            1 \\ 
(continued online)\\
\hline
\end{tabular}
\end{table*}

Table~\ref{tab:w50w10w5w1} lists the $W_{50}$, $W_{10}$, $W_{5}$, $W_{1}$ measurements for the frequency-averaged data, while Table~\ref{tab:ovcolcon} reports $W_{10}$ in each frequency sub-band. {The Appendix~\ref{sec:profileplots} shows a few examples of the pulse profiles and the respective $W_{10}$ measurements.}
A summary histogram of the measured widths of the frequency-averaged data is shown in Figure~\ref{fig:widthsThisto}.
For $W_{10}$, this histogram confirms that there is no significant measurement bias at narrow widths.
Figure~\ref{fig:widthsTkde} shows the $W_{50}$, $W_{10}$, $W_{5}$, $W_{1}$ distributions with respect to the pulse period.

\begin{table*}
\centering
\caption{Overview table for pulse width measurements for the data in the individual frequency subbands, width differences, and width contrasts. The full tables are available as supplementary material (also at ViZieR at the CDS).}
\label{tab:ovcolcon}
\begin{tabular}{lrrr}
Label & Description & 4-band Table &  8-band Table \\
\hline
PSRJ & 		pulsar as in Table ~\ref{tab:w50w10w5w1} &	\\
IPflag &   	as in Table ~\ref{tab:w50w10w5w1}\\
component &	as in Table ~\ref{tab:w50w10w5w1}\\
fX &		center freq. in band X 		& X from 1 to 4 & X from 1 to 8\\ 
W10\_X & 	$W_{10}$ in subband X 			& X from 1 to 4  & X from 1 to 8 \\  	
psigW10\_X &	$\sigma^{+}_{W10}$ in subband X  & X from 1 to 4  & X from 1 to 8 \\    
nsigW10\_X &	$\sigma^{-}_{W10}$ in subband X & X from 1 to 4  & X from 1 to 8  \\   
uncW10\_X  &     $W_{10}Err$ as in Table ~\ref{tab:w50w10w5w1}, but in subband X & X from 1 to 4 & X from 1 to 8 \\   
W10Gflag\_X &	$W_{10}$Gflag in subband X & X from 1 to 4 & X from 1 to 8\\ 
colW10\_XY &    $C_{XY}$ (Sec.~\ref{res:colours}) & XY= 43,42,41,32,31,21 & XY= 87,86,85,84,83,82,81,\\ 
              &                                                            &                       & 76,75,74,73,72,71,65,64,63,\\ 
              &                                                            &                       & 62,61,54,53,52,51,43,42,41,\\
              &                                                            &                       & 32,31,21\\ 
uncColW10\_XY & conservative $C_{XY}$ uncertainty (Sec.~\ref{res:colours}) & XY= 43,42,41,32,31,21 & XY= 87,86,85,84,83,82,81,\\ 
              &                                                            &                       & 76,75,74,73,72,71,65,64,63,\\ 
              &                                                            &                       & 62,61,54,53,52,51,43,42,41,\\
              &                                                            &                       & 32,31,21\\ 
psigColW10\_XY & $\sigma^{+}_{C_{XY}}$ (Sec.~\ref{res:colours})  & XY= 43,42,41,32,31,21 & XY= 87,86,85,84,83,82,81,\\ 
              &                                                            &                       & 76,75,74,73,72,71,65,64,63,\\ 
              &                                                            &                       & 62,61,54,53,52,51,43,42,41,\\
              &                                                            &                       & 32,31,21\\ 
msigColW10\_XY & $\sigma^{-}_{C_{XY}}$ (Sec.~\ref{res:colours})  & XY= 43,42,41,32,31,21 & XY= 87,86,85,84,83,82,81,\\ 
              &                                                            &                       & 76,75,74,73,72,71,65,64,63,\\ 
              &                                                            &                       & 62,61,54,53,52,51,43,42,41,\\
              &                                                            &                       & 32,31,21\\ 
KonW10\_XY &    $K_{XY,t}$ (Sec.~\ref{res:colours})   & XY= 43,42,41,32,31,21 & XY= 87,86,85,84,83,82,81,\\ 
              &                                                            &                       & 76,75,74,73,72,71,65,64,63,\\ 
              &                                                            &                       & 62,61,54,53,52,51,43,42,41,\\
              &                                                            &                       & 32,31,21\\ 
uncKonW10\_XY & conservative $K_{XY,t}$ uncertainty (Sec.~\ref{res:colours})  & XY= 43,42,41,32,31,21 & XY= 87,86,85,84,83,82,81,\\ 
              &                                                            &                       & 76,75,74,73,72,71,65,64,63,\\ 
              &                                                            &                       & 62,61,54,53,52,51,43,42,41,\\
              &                                                            &                       & 32,31,21\\ 
\hline
\end{tabular}
\end{table*}

\subsection{Width variability and consistency of used $W_{10}$}
\label{sec:widthvari}
\begin{figure}
\includegraphics[width=8.5cm]{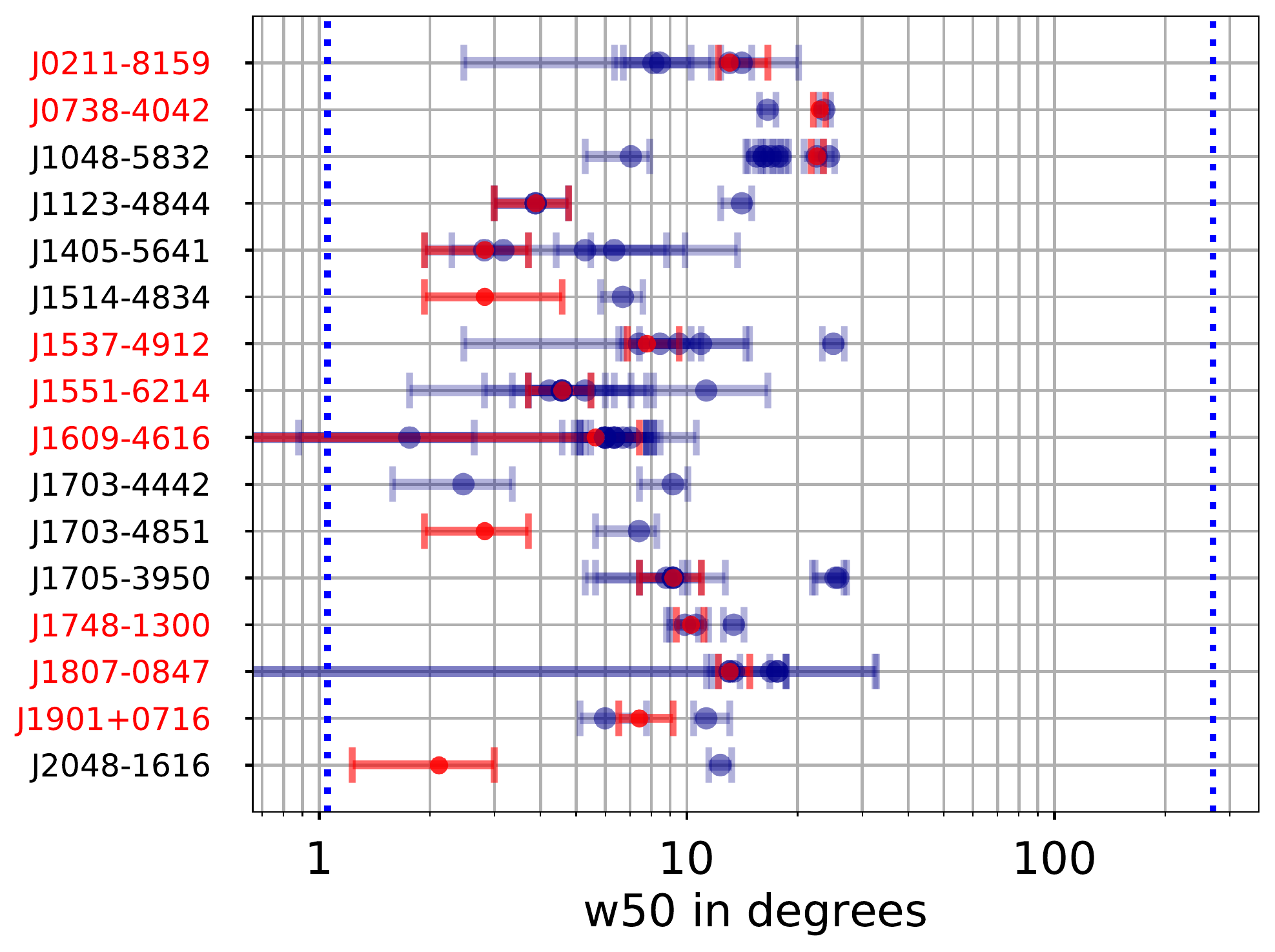}
\vspace{-0.5cm}
\caption{Variability of $W_{50}$ for those TPA pulsars where the $5\sigma$ uncertainties of $W_{50}$ do not overlap for at least two observations covering at least 500 rotation periods (all listed pulsar names), or 1000 rotation periods (black pulsar names only). Individual observations are plotted in blue, combined data sets in red. The dotted blue lines indicate the minimum and maximum considered widths according to Section~\ref{sec:widthmeas}). This plot shows $5\sigma$ uncertainties.\label{fig:w50vari}}
\end{figure}

\begin{figure}
\includegraphics[width=\columnwidth]{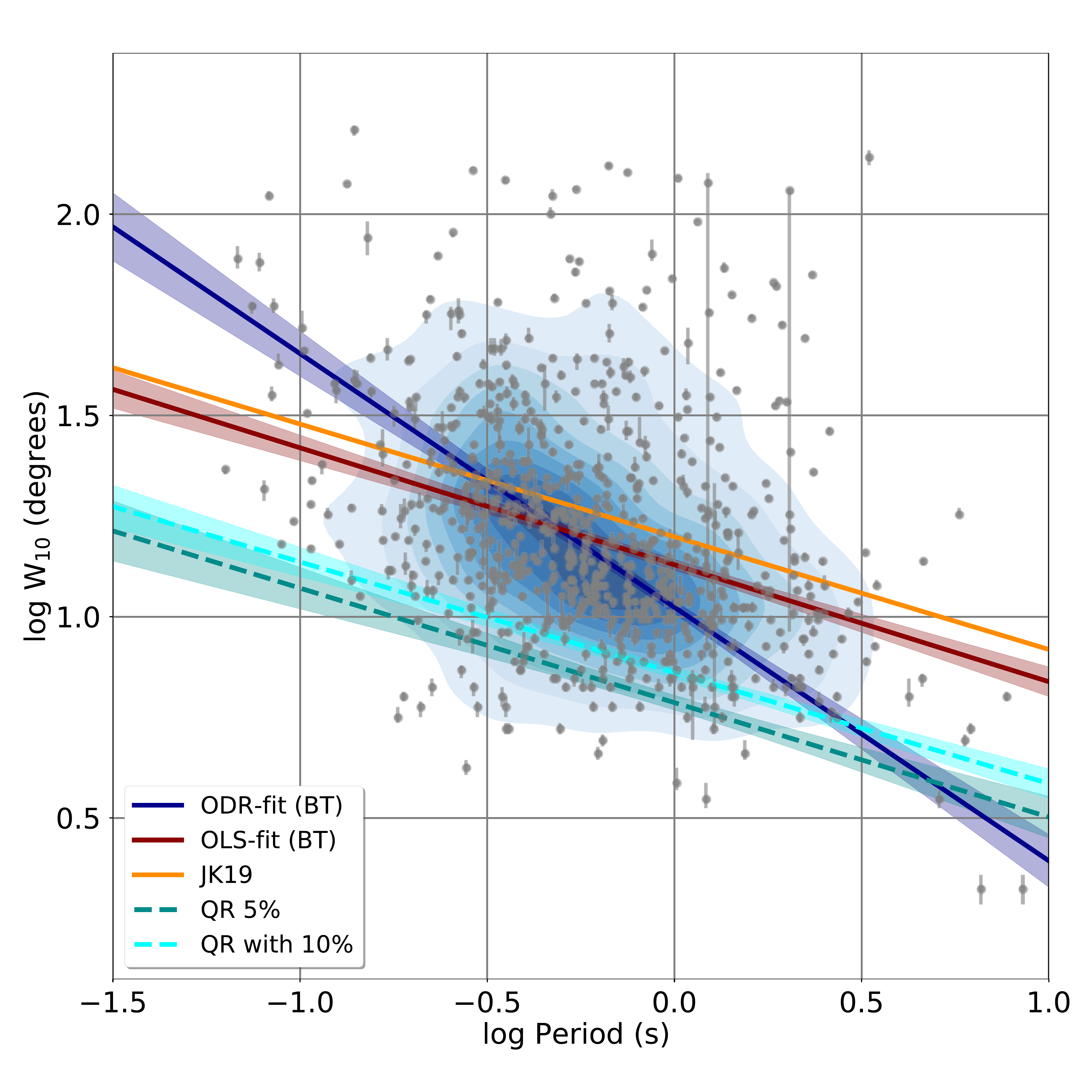}
\vspace{-0.5cm}
\caption{Profile widths $W_{10}$ vs. period for the total bandwidth data in logarithmic space. The 762 width measurements (including separate widths for main and interpulses) are shown as a scatter plot with uncertainties as well as a KDE-plot (2D Gaussian kernel). The red line and shaded area show the result of our OLS-fit and its uncertainties as obtained by using bootstrap.
The slope of the OLS-line (or the PL-index in linear space) agrees well with recent literature values. As an example, the slope derived by \citetalias{Johnston2019} is shown in orange. The blue line and shaded area represent the result of our ODR-fit and the respective BT-constrained uncertainty. {The two dashed cyan lines and their uncertainties show estimates of the LBL that were obtained via quantile regression as introduced by \citet{Skrzypczak2018}.} \label{fig:w10pfit}}
\end{figure}

Pulse shape variations and associated width variability can be due to different reasons. 
Short-term changes (within a few rotation periods) are often seen in single-pulse studies, e.g., due to stochastic pulse-to-pulse variability, nulling events, mode-changing or pulse drifting (e.g., \citealt{Lyne2010}).
For rotation-powered pulsars, a stable pulse profile is typically established after folding over a few hundred rotation periods, $N_{\rm fold}$.
The ideal minimum number of considered rotation periods depends on the chosen limits for the pulse profile fidelity. 
Using some simplifying assumptions, 
\citet{Song2020arXiv} discussed that the optimal observation lengths for the TPA pulsars should cover at least 200\,rotation periods for a $\sigma_{\rm shape}=0.1$ uncertainty of the shape parameter which measures the flux differences of two pulse profile bins of interest.  
For our analysis of pulse widths, we tested different $N_{\rm fold}$ values for potential influence on the width measurements, and decided to use $N_{\rm fold}$ larger than at least 500 for all considered observations.
Long-term (weeks, months, years) changes of the short-term averaged pulse profile are also seen for some pulsars, and regularly in magnetars. This can indicate, for example, interesting changes in individual pulse profile components due to large-scale changes in the magnetosphere, which may be correlated with observed changes in the spindown rate. 

For a consistent analysis of the width relations of the general pulsar population we want to avoid any potential strongly variable pulse widths. 
For this reason, we analysed all observations of an individual pulsar for width variability  (using the complete frequency bandwidth of 775\,MHz).
We considered $5\sigma$ uncertainties of the widths and checked for overlap of the uncertainty regions.
For $W_{50}$, we find slightly different numbers of pulsars with variable width if we consider $N_{\rm fold} > 500$ (16 pulsars) or $>1000$ (8 pulsars). These are shown for reference in Figure~\ref{fig:w50vari}.
However, in our analysis below we actually only consider $W_{10}$. We identify only two pulsars with long-term variable $W_{10}$, regardless of whether we consider $N_{\rm fold} > 500$ or $>1000$. 
PSR J1048-5832 has nine valid measurements, the largest deviation of $W_{10}$ is 43\%.
PSR J1057-5226 (main pulse component) has nine valid measurements, the largest deviation of  $W_{10}$ is 20\%.
We consider these differences small enough to not influence our statistical population width analysis. Therefore, for our studies below, we include these pulsars and their  $W_{10}$ measurements, using the data with the highest $S/N$ as outlined in ~\ref{sec:widthmeas}.

\subsection{Width relations}
\label{res:widthrelations}
\begin{table}
\centering
\caption{Results of the PL-fit, $W_{10}=A (P/s)^{\mu}$ with amplitude, $A$, and PL index, $\mu$, for the relation between period, $P$, and pulse width $W_{10}$. The listed results and their uncertainties correspond to the 50\%, 50\%-16\%, and 84\%-50\% quantiles of the respective BT simulations for the $N_{\rm W}$ valid $W_{10}$ measurements (only the highest S/N ones of each pulsar are used). Note that centre frequencies $\nu_{c}$ are median values since they slightly vary between observations in dependence of the number of good frequency channels in the nominal bandwidth, $BW$. Pulsars with and without interpulse components are considered.}
\label{tab:pw10fits}
\begin{tabular}{lrc|ll|r} 
channel & $\nu_{c}$ & $BW$ & $A$ & $\mu$ & $N_{\rm W}$\\
& MHz & MHz & deg & &\\
\hline
\multicolumn{6}{l}{OLS-fit for frequency-averaged data}\\
& 1284 & 775.75 & $14.2 \pm 0.4 $ & $ -0.29 \pm 0.03 $ & 762 \\
\hline
\multicolumn{6}{l}{OLS-fits for four channels}\\
ch1$_4$ & 996  & $\sim 194$  & $14.1 \pm 0.4 $       & $ -0.29 \pm 0.03 $ &  622\\
ch2$_4$ & 1179 & $\sim 194$  & $13.8^{+0.5}_{-0.4} $ & $ -0.26 \pm 0.03 $ &  532\\
ch3$_4$ & 1381 & $\sim 194$  & $13.7^{+0.5}_{-0.4} $ & $ -0.28 \pm 0.04 $ &  562\\
ch4$_4$ & 1576 & $\sim 194$  & $13.5 \pm 0.5 $       & $ -0.25 \pm 0.04 $ &  447 \\
\hline
\multicolumn{6}{l}{OLS-fits for eight channels}\\
ch1$_8$ & 944  & $\sim 97$  & $14.5 \pm 0.5       $ & $ -0.26 \pm 0.04	     $ & 531\\
ch2$_8$ & 1040 & $\sim 97$  & $13.9^{+0.5}_{-0.4} $ & $ -0.28 \pm 0.03	     $ & 531\\
ch3$_8$ & 1135 & $\sim 97$  & $14.0 \pm 0.5       $ & $ -0.23 \pm 0.04	     $ & 466\\
ch4$_8$ & 1232 & $\sim 97$  & $14.0 \pm 0.6       $ & $ -0.25 \pm 0.04	     $ & 399\\
ch5$_8$ & 1333 & $\sim 97$  & $13.8 \pm 0.5       $ & $ -0.25 \pm 0.04	     $ & 460\\
ch6$_8$ & 1429 & $\sim 97$  & $13.7 \pm 0.5       $ & $ -0.24 \pm 0.04	     $ & 443\\
ch7$_8$ & 1520 & $\sim 97$  & $13.5  \pm 0.6      $ & $ -0.22 \pm 0.05	     $ & 361\\
ch8$_8$ & 1627 & $\sim 97$  & $13.2  \pm 0.6      $ & $ -0.24 \pm 0.05	     $ & 332\\
\hline
\hline
\multicolumn{6}{l}{ODR-fit for frequency-averaged data}\\
& 1284 & 775.75 & $11.9\pm 0.4 $ & $ -0.63 \pm 0.06$  & 762 \\ 
\hline
\multicolumn{6}{l}{ODR-fits for four channels}\\
ch1$_4$ & 996  & $\sim 194$  & $12.0 \pm 0.4 $	     & $ -0.65 \pm 0.07 $ &  622\\
ch2$_4$ & 1179 & $\sim 194$  & $11.7^{+0.5}_{-0.4}$  & $ -0.60 \pm 0.08 $ &  532\\
ch3$_4$ & 1381 & $\sim 194$  & $11.5 \pm 0.5 $       & $ -0.65 \pm 0.08 $ &  562\\
ch4$_4$ & 1576 & $\sim 194$  & $11.6 \pm 0.5 $       & $ -0.56 \pm 0.08 $ &  447 \\
\hline
ch1$_8$ & 944  & $\sim 97$  & $12.2 \pm 0.5	  $ & $  -0.64^{+0.08}_{-0.09} $ & 531 \\
ch2$_8$ & 1040 & $\sim 97$  & $12.0 \pm 0.4       $ & $  -0.61 \pm 0.7         $ & 531 \\
ch3$_8$ & 1135 & $\sim 97$  & $11.9^{+0.6}_{-0.5} $ & $  -0.58^{+0.08}_{-0.10} $ & 466 \\
ch4$_8$ & 1232 & $\sim 97$  & $11.9 \pm 0.6       $ & $  -0.57^{+0.08}_{-0.09} $ & 399 \\
ch5$_8$ & 1333 & $\sim 97$  & $11.8 \pm 0.6       $ & $  -0.57 \pm 0.08        $ & 460 \\
ch6$_8$ & 1429 & $\sim 97$  & $11.9 \pm 0.5	  $ & $  -0.54 \pm 0.08        $ & 443 \\
ch7$_8$ & 1521 & $\sim 97$  & $11.8 \pm 0.6	  $ & $  -0.48 \pm 0.09        $ & 361 \\
ch8$_8$ & 1626 & $\sim 97$  & $11.4^{+0.7}_{-0.6} $ & $  -0.52 \pm 0.09        $ & 332 \\
\hline
\hline

\multicolumn{6}{l}{{QR-fit for frequency-averaged data}}\\
$q=5$\%& 1284 & 775.75 & $6.12\pm 0.03 $ & $ -0.28 \pm 0.05$  & 762 \\ 
$q=10$\%& 1284 & 775.75 & $7.26\pm 0.03 $ & $ -0.28 \pm 0.03$  & 762 \\ 
\hline
\end{tabular}
\end{table}

\begin{figure*}
\captionsetup[subfigure]{labelformat=empty}
\subfloat[]{\label{fig:w10pdot}\includegraphics[width = 0.40\textwidth]{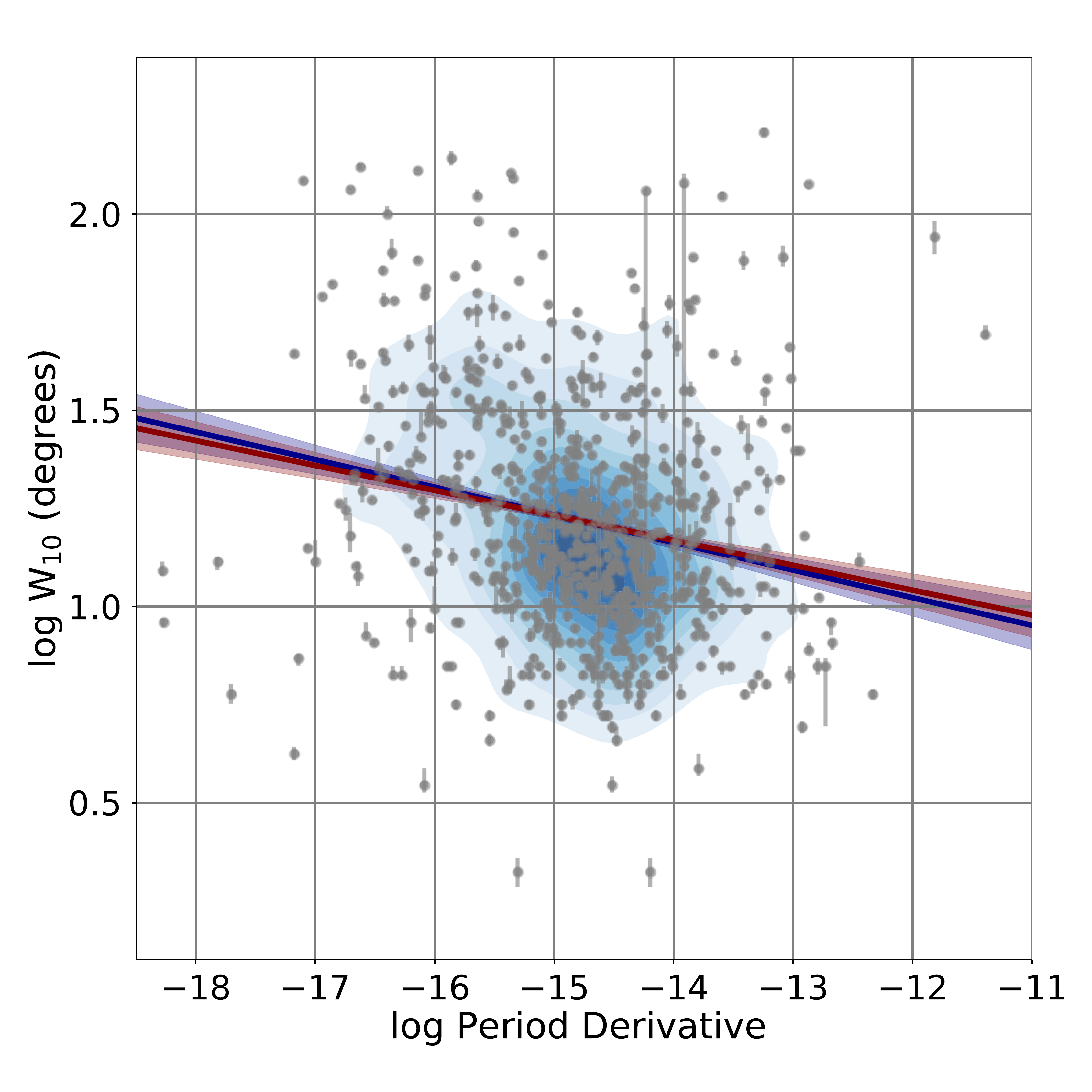}}
\subfloat[]{\label{fig:w10edot}\includegraphics[width = 0.4\textwidth]{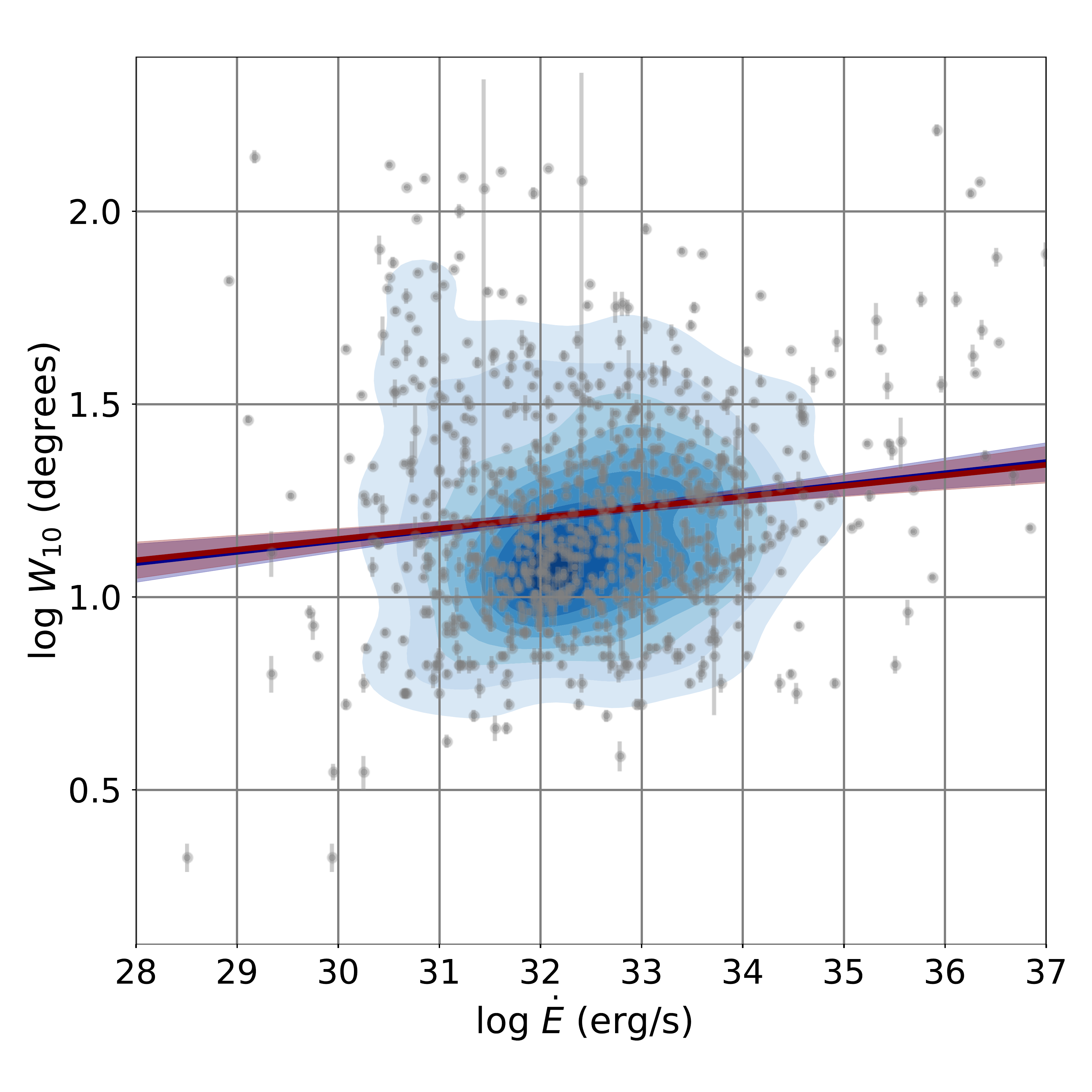}}
\vspace{-0.5cm}
\caption{Profile widths $W_{10}$ of the frequency-averaged data over $\dot{P}$ (left panel) and $\dot{E}$ (right panel). A KDE-plot (2D Gaussian kernel) and scatter plot in grey show all 762 pulse widths. The OLS fit results are in red, the ODR fit results are shown in blue. Bootstrapping was used to determine the uncertainties.}
\label{fig:w10PdotEdot}
\end{figure*}

$W_{10}$ was used to investigate correlations with other pulsar parameters. 
Figure~\ref{fig:w10pfit} shows the period-width relations derived from the measurements for the pulsars without obvious strong scattering for the frequency-averaged data. We obtained a slope (or PL-index) of 
$\mu =-0.29 \pm 0.03$ (amplitude of $14.2 \pm 0.4$\,degrees; see also Table~\ref{tab:pw10fits}) for our bootstrap of the OLS-fit. 

A density plot of our data in Figure~\ref{fig:w10pfit} indicates that the OLS-fit noticeably deviates from the region of highest densities, in particular for $P> 0.5$\,s. Mathematically, this is due to the minimization of residuals along the width axis only, whilst there is also a large spread of width values. This is shown in more detail in the multiple histogram plot in the the Appendix Figure~\ref{fig:pw10histos}.
Minimizing the orthogonal distances to the model curve, the ODR-fit (with bootstrap) results in a steeper slope of $\mu =-0.63 \pm 0.06$ (amplitude of $11.9 \pm 0.4$\,degrees) for our PL-fit. The ODR-fit is a better representation of the behaviour of the highest number densities in the KDE-plot.\\

Table~\ref{tab:pw10fits} and Figure~\ref{fig:w10pfit} also show the results of the QR for quantiles 5\% and 10\%. These results represent LBL estimates if we follow the approach by \citet{Skrzypczak2018}. We also employed the QR for lower quantiles (down to 1\%) and higher ones, and obtained similar slopes (agreement within $1\sigma$ of the 5\% fit).\\

For completeness, we also show  $W_{10}$ with respect to the period derivative $\dot{P}$ and spin-down energy $\dot{E}$ in Figure~\ref{fig:w10PdotEdot}. Assuming a power law relation between $W_{10}$ and $\dot{P}$, we obtain a PL-index 
of $\mu=-0.06 \pm 0.02$  with OLS ($\mu=-0.07 \pm 0.02$ with ODR). For the relation with $\dot{E}$, we obtained a slope of $\mu=0.03 \pm 0.01$ with OLS (the same for ODR).\\

\begin{figure*}
\captionsetup[subfigure]{labelformat=empty}
\subfloat[]{\label{fig:w10pfitch4ch8ta}\includegraphics[width = 0.5\textwidth]{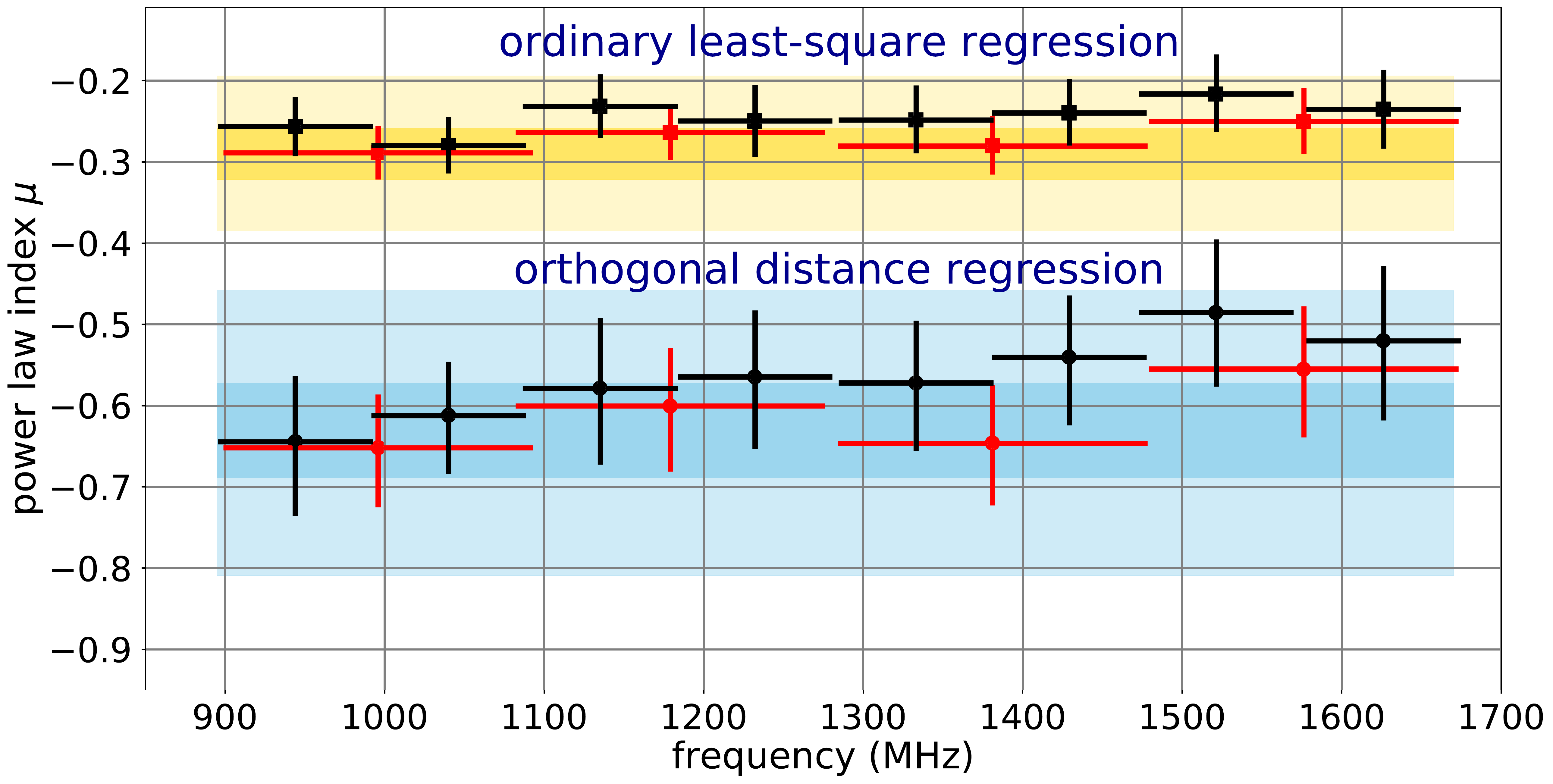}}
\subfloat[]{\label{fig:w10pfitch4ch8tb}\includegraphics[width = 0.5\textwidth]{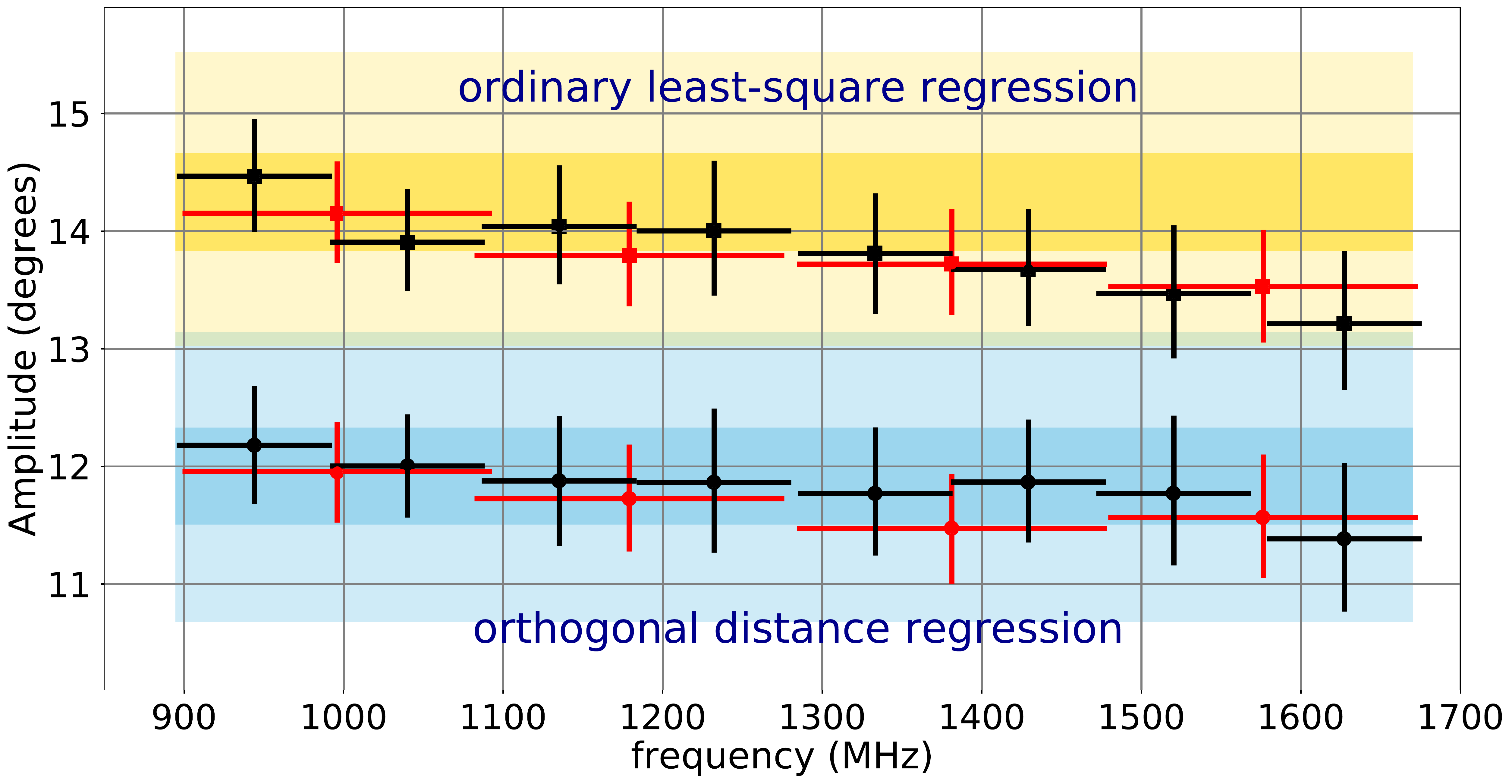}}
\vspace{-0.5cm}
\caption{The power law indices (left panel), amplitudes (right panel) and their respective $1\sigma$ uncertainties as derived from the period-width ($W_{10}$) relations in the four (red points) and eight (black points) frequency channels. The $1\sigma$ ($3\sigma$) uncertainty range of the frequency-averaged data is indicated with the dark (light)  areas. The values and their uncertainties (Table~\ref{tab:pw10fits}) were derived with the OLS (orange) and the ODR (blue) using bootstrap.}
\label{fig:w10pfitch4ch8t}
\end{figure*}

For the four and eight frequency sub-bands, we carried out a similar analysis for the period-width relations. Table~\ref{tab:pw10fits} lists all values, while Figure~\ref{fig:w10pfitch4ch8t} shows the power law indices over frequency. There is some fluctuation of the derived parameters over the sub-bands, but all values overlap considering the bootstrap-derived $3\sigma$ uncertainties. This is true for the OLS and the ODR. 
Since we chose for each width the file with the highest S/N in the respective frequency band, we checked whether the used files deviate from files used for the frequency-averaged data.
Typically, fewer than 7\% of the sub-band files are not the ones used for the fit over the total bandwidth, and the influence on the final result is  negligible.\\

\subsection{Geometry constraints based on the population statistics}
\label{sec:geometry}
The observed pulse width depends on the viewing geometry of the pulsar, in particular, the obliquity $\alpha$ which is the angle between magnetic and rotation axes, and the angle, $\beta$, between the magnetic axis and the observer's line of sight. 
In addition to the viewing geometry there are of course  other parameters, e.g., the properties of the radio beam,     influencing the observed pulse widths of pulsars. Assuming that pulsars have similar emission properties for similar physical parameters, one can use a statistical approach to describe the expected pulse widths  of the pulsar population. In principle, once a model is established, it can be used to invert the model to derive probabilistic geometry constraints for a given pulsar based on its data. Independent geometric constraints such as analysing the polarized radio emission within the framework of the Rotating Vector Model, the knowledge of interpulse pulsars ($\alpha \approx 90 \deg$), or other measurements such as the orientation of jets and tori in X-ray detected pulsar wind nebulae can then be used to test the underlying model assumptions.\\
 
\citetalias{Johnston2019} developed a parametric model to simulate the period-width relationship they observed for 600\,pulsars with pulse width measurements from the Parkes radio telescope.  
This work built on the results of \citet{Johnston2017} where a population of pulsars matching the observed $P-\dot{P}$ diagram was synthesized assuming a particular decay of the obliquity $\alpha$, but no magnetic field decay.
Employing a model for the obliquity $\alpha$  (with possible dependency on age and period), and simplified descriptions for emission height (with possible dependence from age and spin-down energy), beam filling factor (for a circular beam), and the ratio of the emission longitude to the size of the polar cap (with possible dependence on $\alpha$), \citetalias{Johnston2019} concluded that profile width measurements also support the statement that $\alpha$ decays with time. Only then were they able to explain the observed period-width distribution of the Parkes-observed pulsars.  The period-$W_{10}$ distribution of the frequency-averaged data of the TPA-pulsars resulted in a OLS-derived slope that is within $1\sigma$ of the slope reported by \citetalias{Johnston2019} (see listed relations \ref{eq:OLSslope} and \ref{eq:jk19slope}), while the amplitude has a small offset due to different frequency range covered. Thus, it can be expected that the \citetalias{Johnston2019} simulation for the model with $\alpha$-decay is a reasonable description of our data too. We tested this by producing probability distributions for the difference between simulated and observed widths for three period ranges, similar to the fig.~9 from \citetalias{Johnston2019}. 
Overall, we found a good agreement of our width distribution with their model prediction that considers $\alpha$-decay. Only for the fastest pulsars do we see a small difference in the peak location of the width-differences (indicating slightly smaller observed widths than simulated).  
A similar comparison is shown in more detail (nine period ranges instead of three) in the Appendix, in Figure~\ref{fig:pw10histos}, where the \citetalias{Johnston2019} predictions are plotted as black probability curves.\\

\begin{figure}
\includegraphics[width=\columnwidth]{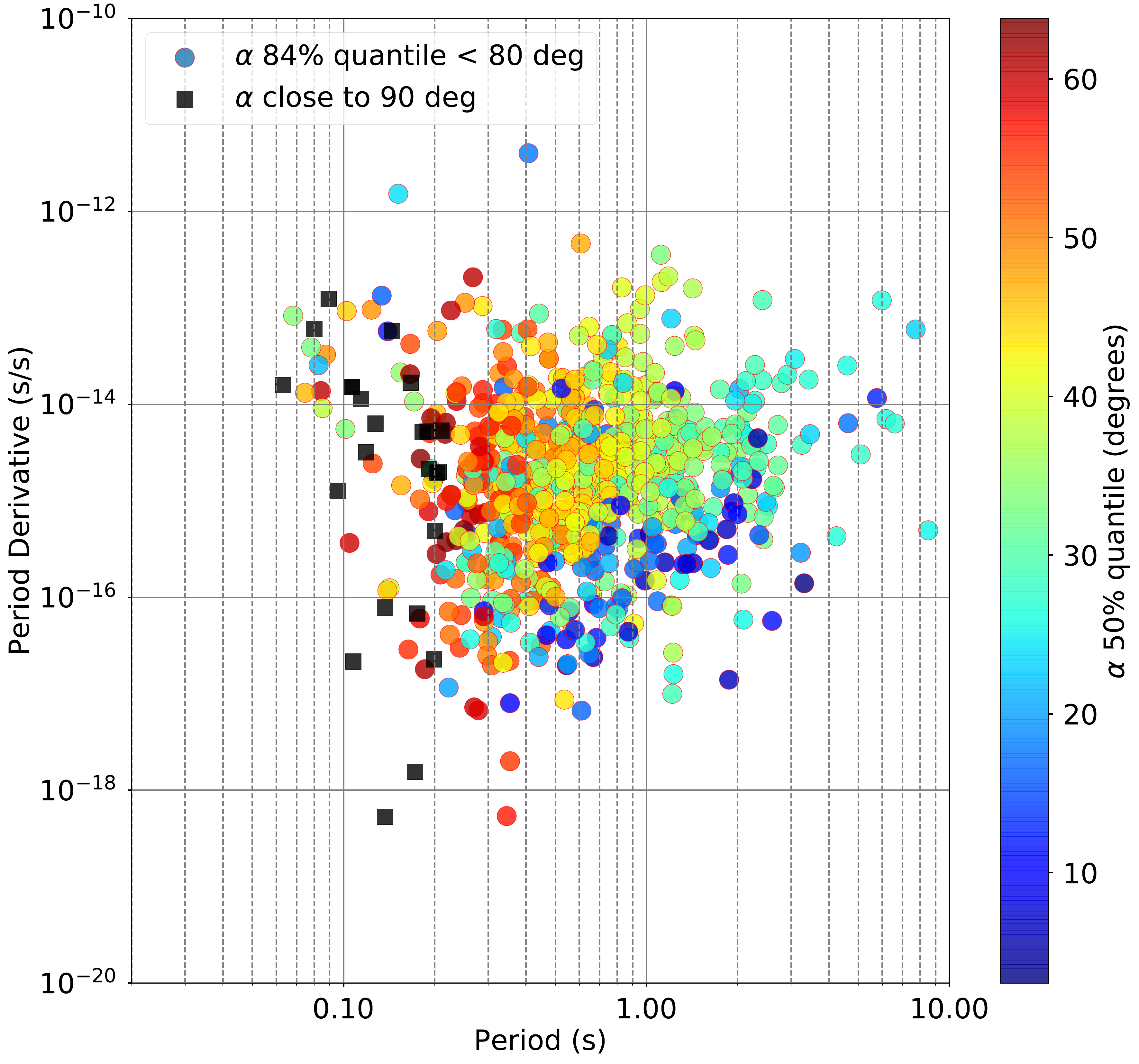}
\vspace{-0.5cm}
\caption{The obliquity ($\alpha$) distribution of the TPA pulsars in the $P-\dot{P}$ diagram if the $W_{10}$ measurements of the frequency-averaged data are combined with the simulation input of \citetalias{Johnston2019} (their figure\,8 which considers $\alpha$-decay with age).
\label{fig:ppdotJK19alpha}}
\end{figure}

We use the simulation data set by \citetalias{Johnston2019} to obtain estimates of the angles $\alpha$
for each pulsar based on its known period and measured $W_{10}$, by marginalizing over the parameters height and beam filling factor in the \citetalias{Johnston2019} model, and assuming $\geq 0$ for $\beta$\footnote{$\beta$ is typically around 0, or the positive and negative $\beta$ have very similar absolute values.} for simplicity.
Typically two values (i.e., $\alpha$ and $180\deg -\alpha$,
correspond to a specific $P$, $W_{10}$ pair. 
We obtain the 16\%, 50\%, and 84\% quantiles of the parameters using a range of 0--90$\deg$ for $\alpha$. 
Based on visual inspection of the 2D-density plots, we chose the following criteria for reporting:  If the 84\% quantile of $\alpha$ is smaller than $80\deg$, two peaks can be clearly separated, and we report the 16\%, 50\%, and 84\% quantiles for $<90\deg$ in Table~\ref{tab:alphas}. For the remaining cases, we report an $\alpha \approx 90\deg$. 
The distribution of $\alpha$ of the TPA pulsars as obtained from the input simulation by \citetalias{Johnston2019} is shown in Figure~\ref{fig:ppdotJK19alpha}.
We emphasize that these geometry estimates are based on a simplified model at the moment, and that this model uses a statistical population-based approach. For an individual pulsar, these estimates should be used with care.
Examples are shown in the Appendix~\ref{sec:obliquityappendix}.

\subsection{Width colours and contrasts}
\label{res:colours}
\begin{figure}
\includegraphics[width=\columnwidth]{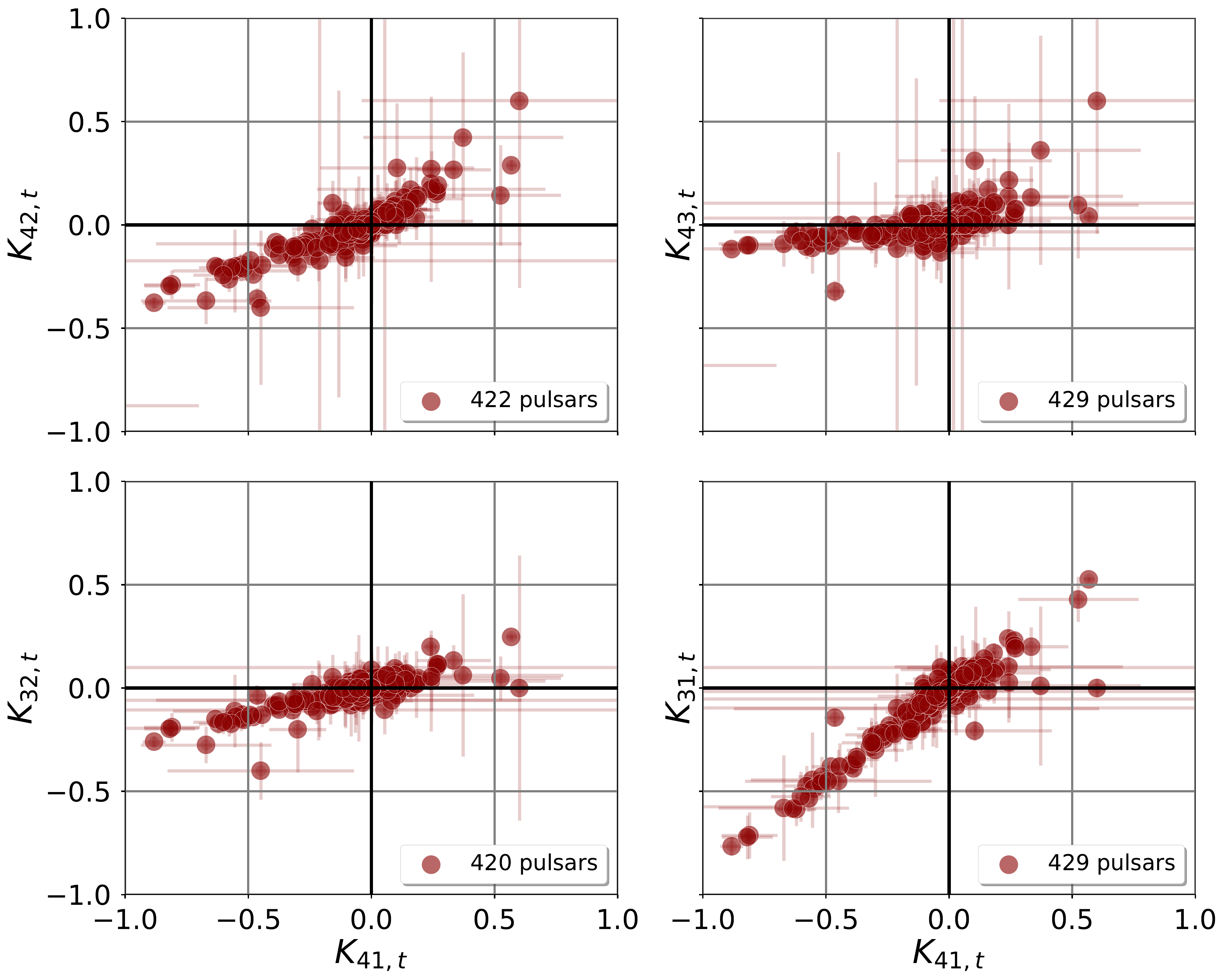}
\vspace{-0.5cm}
\caption{The $W_{10}$-contrast distribution for the 4-channel data. The x-axes represent the largest possible contrast $K_{41,t}$ (between frequency bands 1 and 4), while the y-axes show 4 out of the 5 remaining other contrasts (not shown: $K_{21,t}$). Only TPA pulsars with valid $W_{10}$ measurements in all the frequency channels of the respective contrasts are plotted.
\label{fig:contrast4ch}}
\end{figure}

\begin{table*}
\centering
\caption{Combination statistics for two colours ($W_{10}$ differences using two frequency sub-bands). A positive sign means the width gets broader with \emph{increasing} frequency, a negative sign indicates broadening of the width with \emph{decreasing} frequency. Here, ${\cal N}_{\rm comb}$ is the total number of combinations of two valid colours, $\sigma_{\rm Col}$ indicates the considered uncertainty level for the colour to distinguish significant positive and negative colours from those that are consistent with zero, ${\cal N}_{\rm comb}^{+,-}$ is the number of two-colour combination that are not zero for the chosen $\sigma_{\rm Col}$ level. The numbers ${\cal N}_{\rm comb}^{+ \Rightarrow +}$, ${\cal N}_{\rm comb}^{- \Rightarrow -}$, ${\cal N}_{\rm comb}^{+ \Rightarrow -}$, ${\cal N}_{\rm comb}^{- \Rightarrow +}$ indicate if the first colour and second colour are both positive, both negative, switch from positive to negative or from negative to positive. Each colour combination is only considered once. The (rounded) percentages are given with respect to  ${\cal N}_{\rm comb}^{+,-}$.}
\label{tab:colourcombi}
\begin{tabular}{ccrrr|rr} 
${\cal N}_{\rm comb}$ & $\sigma_{\rm Col}$ & ${\cal N}_{\rm comb}^{+,-}$ 
& ${\cal N}_{\rm comb}^{+ \Rightarrow +}$ & ${\cal N}_{\rm comb}^{- \Rightarrow -}$ 
& ${\cal N}_{\rm comb}^{+ \Rightarrow -}$ & ${\cal N}_{\rm comb}^{- \Rightarrow +}$ \\
\hline
\multicolumn{7}{l}{4-channel data}\\
6558 & 1   & 1758 & 463 (26\%) &  1278 (73\%) & 8 (0.5\%) & 9 (0.5\%) \\
6558 & 1.5 & 1095 & 249 (23\%) &  846 (77\%) & 0 & 0 \\
6558 & 3.0 &  477 &  75 (16\%) &  402 (84\%) & 0 & 0\\
\hline
\multicolumn{7}{l}{8-channel data}\\
126,174 & 1   & 26,395 & 6,657 (26\%) & 19,460  (74\%) & 181  (0.7\%) & 97 (0.4\%) \\
126,174 & 1.5 & 15,782 & 3,383 (21\%) & 12,352 (78\%) & 40 (0.3\%) & 7 (0.04\%) \\
126,174 & 3.0 & 6,631 & 997  (15\%) & 5634 (85\%) &0 & 0\\

\end{tabular}
\end{table*}

\begin{figure*}
\captionsetup[subfigure]{labelformat=empty}
\subfloat[]{\label{fig:J1559a}\includegraphics[width = 0.175\textwidth]{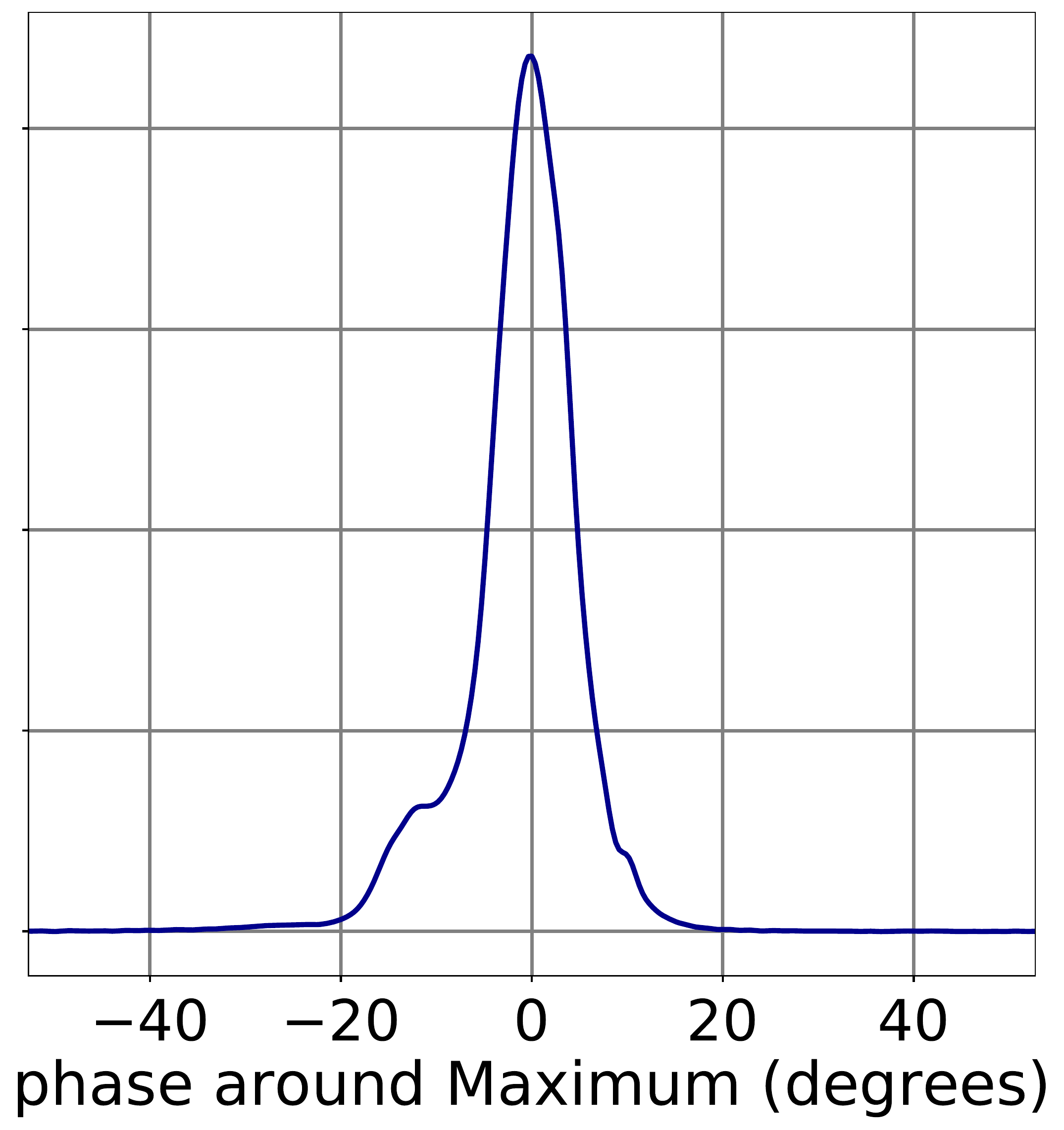}}
\subfloat[]{\label{fig:J1559b}\includegraphics[width = 0.325\textwidth]{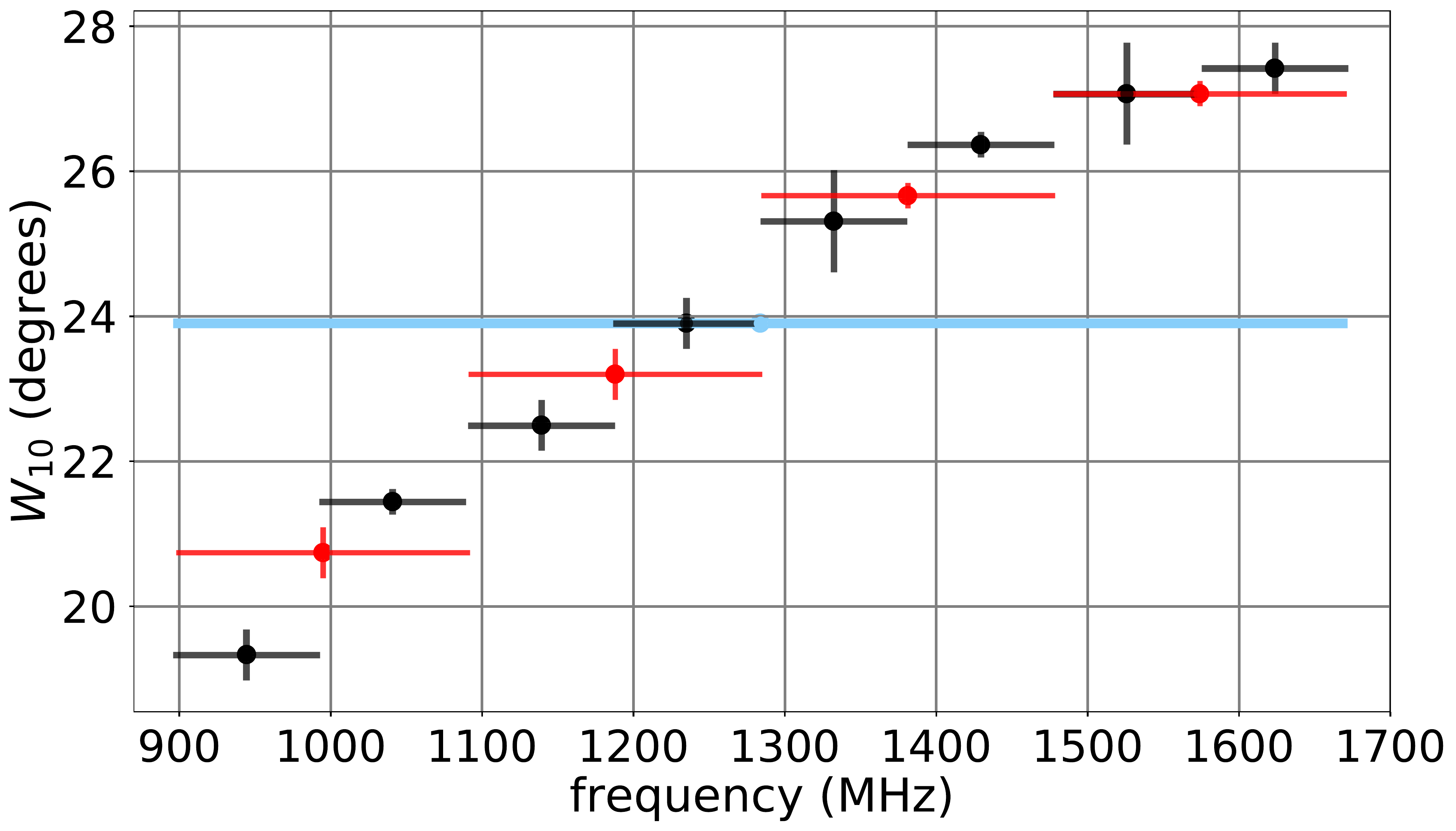}}
\subfloat[]{\label{fig:J0738a}\includegraphics[width = 0.175\textwidth]{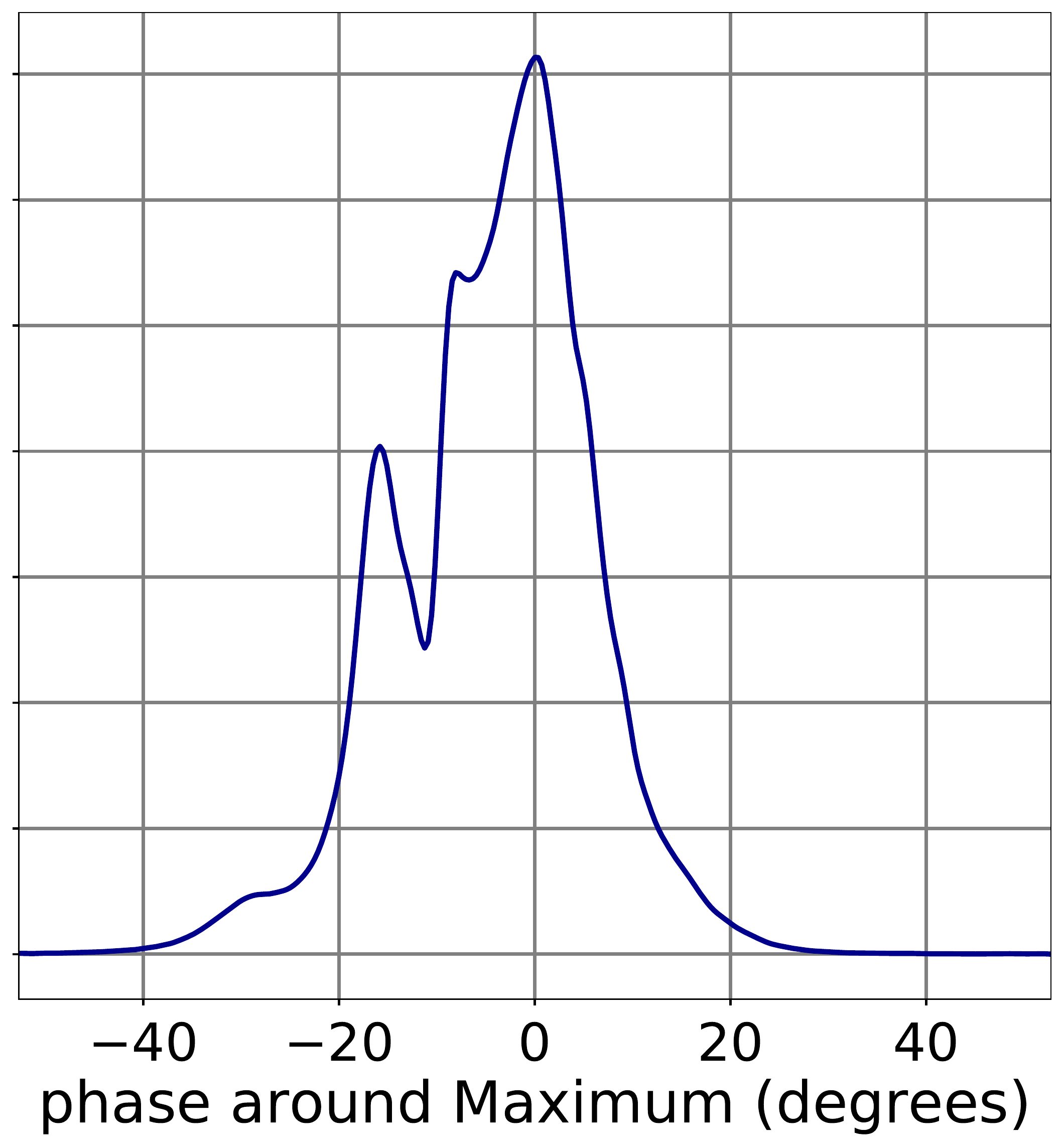}}
\subfloat[]{\label{fig:J0738b}\includegraphics[width = 0.325\textwidth]{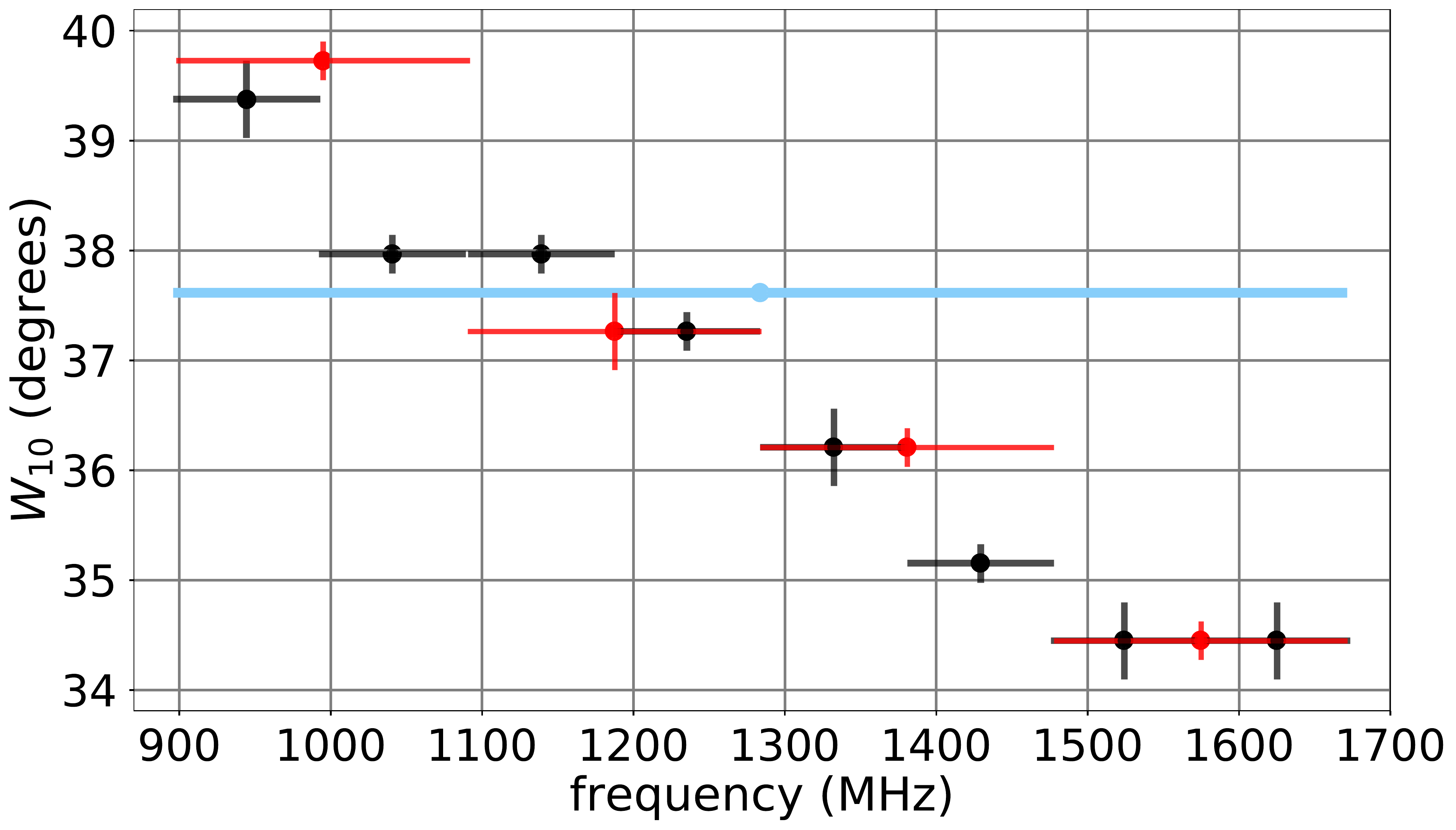}}
\vspace{-0.5cm}
\caption{The frequency-averaged noiseless profile and $W_{10}$ measurements for PSR\,J1559$-$4438 (left two panels) and PSR\,J0738-4042 (right two panels). The $W_{10}$ for four, eight, and the total frequency bands are plotted in red, black, light blue, respectively. {For plots of the pulse profiles in individual frequency channels see Figure~\ref{fig:pulseprofiles}.}}
\label{fig:twoexamples}
\end{figure*}

In order to describe frequency-dependent changes of the widths, we define \emph{width colours}, $C_{xy}$, and \emph{width contrasts}, $K_{xy}$. The width colour is the difference of two width measurements in two frequency channels $x$ and $y$, $C_{xy}=W_{10,x} - W_{10,y}$ where channels 1 to 4 (or 8) represent the frequency sub-bands with median values of 996\,MHz, 1179\,MHz, 1381\,MHz, and 1576\,MHz (or the respective values of the 8-channel data, see Table~\ref{tab:pw10fits}). 
The $W_{10}$ measurements typically have heteroscedastic uncertainties. We consider this heteroscedasticity in the colour uncertainty estimates in a conservative way. Thus, for the negative (positive) uncertainty of $C_{xy}$, we consider the negative (positive) error of $W_{10,x}$ and the positive (negative) error of $W_{10,y}$ for the error propagation.
For the width contrasts, we tested two definitions $K_{xy}= (W_{10,x} - W_{10,y})/(W_{10,x} + W_{10,y})$ and $K_{xy,t}= (W_{10,x} - W_{10,y})/W_{10,t}$ where $W_{10,t}$ is the width measurement in the total frequency band.  
Differences are minute, so we use $K_{xy,t}$. 
For the uncertainty estimate of the contrast, we use error propagation with the maximum error of each width measurement.
The contrast is useful to emphasize changes across the population since it normalizes the change for each individual pulsar. However, due to the additional term in the contrast formula, it also has a larger uncertainty.
Width colours and contrasts are available as supplementary material as outlined in Table~\ref{tab:ovcolcon}.\\ 

In order to look for a frequency-dependent effect on $W_{10}$, we plot contrasts in pairs of two in Figure~\ref{fig:contrast4ch}. This figure only shows contrasts of the 4-channel data, but the 8-channel data show a similar phenomenology. Of the $\sim 420$ pulsars with valid contrast values, almost all are located in the first (crudely about 1/3 of the population) and third quadrant (about 2/3 of the population). There is a strong clustering around contrast zero. For all contrast-pairs, there is a monotonic, seemingly linear rise towards positive values. This is easiest to see for the largest frequency differences, i.e., the  $K_{31,t}-K_{41,t}$ plot (lower-right in Figure~\ref{fig:contrast4ch}). 

The contrast figure shows in condensed (because normalized) form what is also true for the colours. Taking advantage of the smaller uncertainties of the colours, we analyse whether each pulsar indeed shows monotonic behaviour, i.e., whether a negative colour stays negative in the case of other frequency band combinations.  
Table~\ref{tab:colourcombi} summarizes the the number of two-colour combinations that stay positive, negative or switch their sign.
For the 4-channel data, for example, there are 6558 combinations of two valid $W_{10}$-colour measurements. Considering colour uncertainties of $1.5\sigma$, and excluding those with one colour being consistent with zero, there remain 1095 (17\%) ``interesting'' two-colour combinations. The sign of the colour stayed positive (negative) for each pulsar in 249 (846) of these two-colour combinations. 
This corresponds to 23\% positive and 77\% negative colours of the 1095 non-zero two-colour combinations. 
The statistical results in Table~\ref{tab:colourcombi} strongly support the notion that a pulsar's width within the observed band either increases or decreases with frequency, but not both. We show an example pulsar of each category in Figure~\ref{fig:twoexamples}.\\

\begin{figure}
\includegraphics[width=\columnwidth]{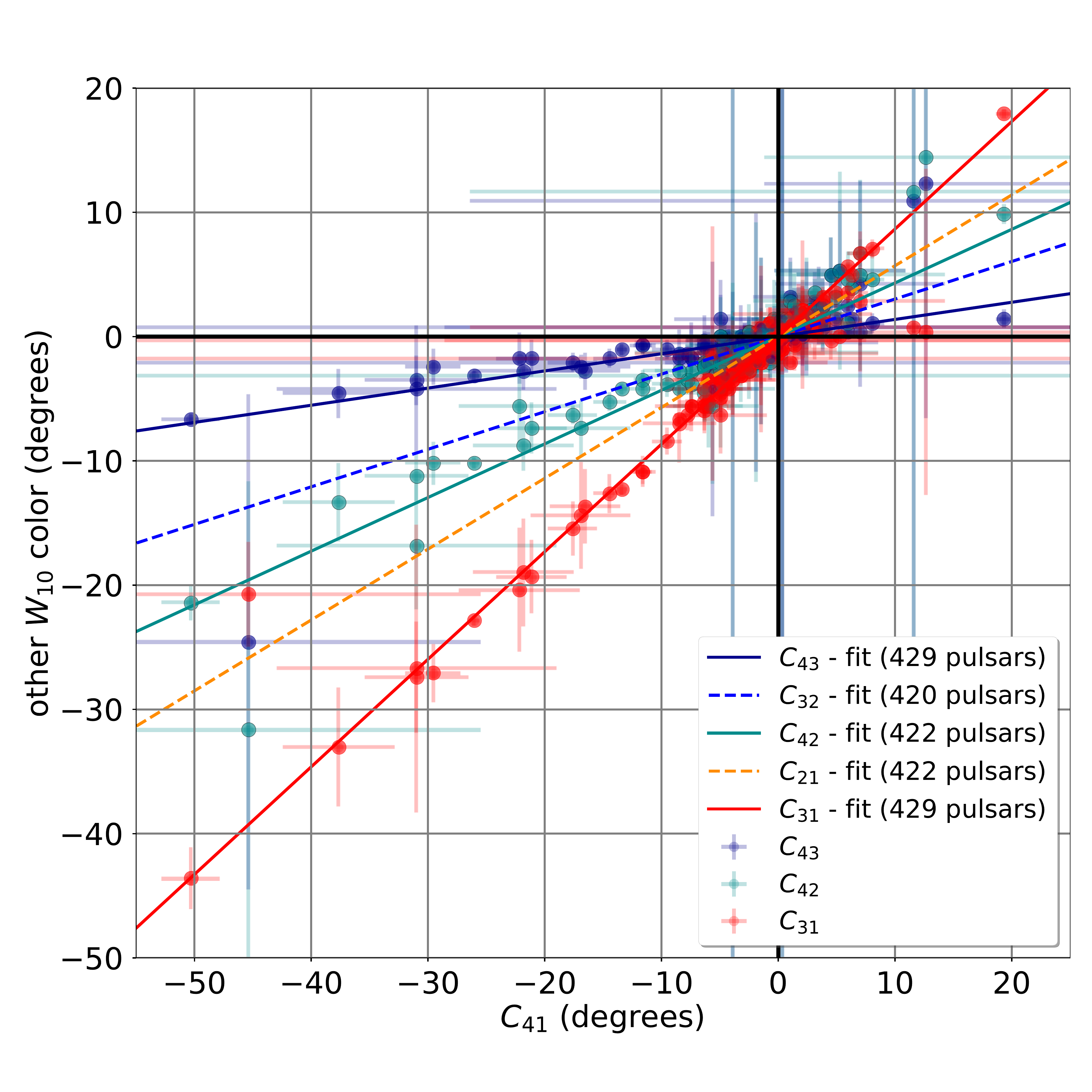}
\vspace{-0.5cm}
\caption{The $W_{10}$-colour distributions for the 4-channel data. The $x$-axis represents the largest possible colour $C_{41}$ (between frequency bands 1 and 4). Linear relations (ODR-fits, Table~\ref{tab:colourfits}) with the other colours are shown on the $y$ axis. For clarity, we only show the measured values and their uncertainties for three of the five relations. Only TPA pulsars with valid $W_{10}$ measurements in all the frequency channels of the respective colours are plotted.
\label{fig:colours4ch}}
\end{figure}

We employ two colours, their uncertainties, and ODR-fits to determine the slopes describing the monotonic behaviour.
We use the largest width difference as the reference ($x$-axis) for all relations, i.e.  $C_{41}$ for the 4-channel data.
We checked linear fits that involved a constant offset, but found the constants to be consistent with zero within their $3\sigma$ uncertainties. Thus, in Table~\ref{tab:colourfits}  we only report the results of linear fits through zero. They are also plotted in Figure~\ref{fig:colours4ch}. 
Colours between two adjacent frequency bands ($C_{43}$, $C_{32}$) have more shallow rises than colours with an additional frequency band in between ($C_{42}$ and  $C_{31}$). 
The $W_{10}$ difference $C_{21}$ has the second steepest rise, indicating that most of the profile width change happens at low frequencies.
The order from steepest to most shallow slope is:  $C_{31}$, $C_{21}$, $C_{42}$, $C_{32}$, and $C_{43}$.  
As we discuss in Section~\ref{sec:diskfreq}, we expect some scattered pulse profiles in our sample. Scattered profiles should appear in the third quadrant. They can influence the result of our colour fits. While there are pulsars in the third-quadrant (``negative-colour'') sample which are not obviously scattered (e.g., PSR\,J0738-4042), it cannot be entirely excluded that the negative-colour sample is dominated by scattered pulsars.
Restricting our fits to colours $\geq 0$, the general trend (larger slope for three sub-band coverage than for two sub-bands) remains, and power law indices are similar (Table~\ref{tab:colourfits}). 
However, $C_{21}$ is no longer an outlier in the general trend.
The order from steepest to most shallow slope is now:  $C_{31}$, $C_{42}$, $C_{21}$, $C_{32}$, and $C_{43}$. Width changes are stronger at lower frequencies, even considering only positive colours.\\  
\begin{table*}
\centering
\caption{Results of ODR-fits of two colours (or $W_{10}$ differences) assuming a linear relation through zero.
The third and fourth column list the results (slope and number of measurements) if all the valid colour measurements are used. 
For comparison, slope$^{-}_{\rm scat}$ lists the \emph{expected} value (for negative colours) assuming profile broadening due to scattering with $\alpha_{Scat}=-4$. 
The sixth and seventh column correspond to the results obtained by using colours equal or larger than zero, the last two columns show the results for the fits if only pulsars with DM $<170$\,cm$^{-3}$\,pc are considered.
For 4 frequency sub-bands, all colour combinations with $C_{41}$ are listed, for the 8 frequency bands we list an exemplary set of six out of 27 colour combinations with $C_{81}$.}
\label{tab:colourfits}
\begin{tabular}{cccc|c|cc|cc} 
colour 1 & colour 2 & slope & $N_{\rm PSR}$ & slope$^{-}_{\rm scat}$ & slope$^{0+}$ & $N^{0+}_{\rm PSR}$ & slope$^{DM}$ & $N^{DM}_{\rm PSR}$\\
\hline
\multicolumn{9}{c}{4-channel data}\\
$C_{41}$ & $C_{42}$  & $ 0.432 \pm 0.007$ & 422 & $ 0.416$  & $ 0.57 \pm 0.02$ & 183& $ 0.54 \pm 0.01$ & 280\\
$C_{41}$ & $C_{43}$  & $ 0.138 \pm 0.005$ & 429 & $ 0.132$  & $ 0.17 \pm 0.01$ & 181& $ 0.22 \pm 0.01$ & 285\\
$C_{41}$ & $C_{31}$  & $ 0.865 \pm 0.007$ & 429 & $ 0.868$  & $ 0.88 \pm 0.01$ & 174& $ 0.79 \pm 0.01$ & 285\\
$C_{41}$ & $C_{32}$  & $ 0.302 \pm 0.006$ & 420 & $ 0.284$  & $ 0.43 \pm 0.01$ & 170& $ 0.30 \pm 0.01$ & 279\\
$C_{41}$ & $C_{21}$  & $ 0.570 \pm 0.008$ & 422 & $ 0.584$  & $ 0.48 \pm 0.01$ & 171& $ 0.50 \pm 0.01$ & 280\\
\hline
\multicolumn{9}{c}{8-channel data}\\
$C_{81}$ & $C_{87}$  & $ 0.034 \pm 0.002 $ & 297 & 0.040 &  $ 0.05 \pm 0.01$ & 112 & $ 0.08  \pm  0.01  $ & 204\\
$C_{81}$ & $C_{86}$  & $ 0.080 \pm 0.004 $ & 313 & 0.087 &  $ 0.09 \pm 0.01$ & 121 & $ 0.15  \pm  0.01  $ & 214\\
$C_{81}$ & $C_{85}$  & $ 0.160 \pm 0.005 $ & 312 & 0.156 &  $ 0.22 \pm 0.02$ & 121 & $ 0.29  \pm  0.01  $ & 214\\
$C_{81}$ & $C_{84}$  & $ 0.266 \pm 0.008 $ & 304 & 0.261 &  $ 0.35 \pm 0.02$ & 116 & $ 0.37  \pm  0.02  $ & 206\\
$C_{81}$ & $C_{83}$  & $ 0.418 \pm 0.009 $ & 309 & 0.412 &  $ 0.47 \pm 0.02$ & 119 & $ 0.52  \pm  0.02  $ & 209\\
$C_{81}$ & $C_{82}$  & $ 0.661 \pm 0.008 $ & 314 & 0.638 &  $ 0.76 \pm 0.02$ & 124 & $ 0.84  \pm  0.01  $ & 214\\
\end{tabular}
\end{table*}
\label{conEdotKStest}

\begin{figure}
\includegraphics[width=\columnwidth]{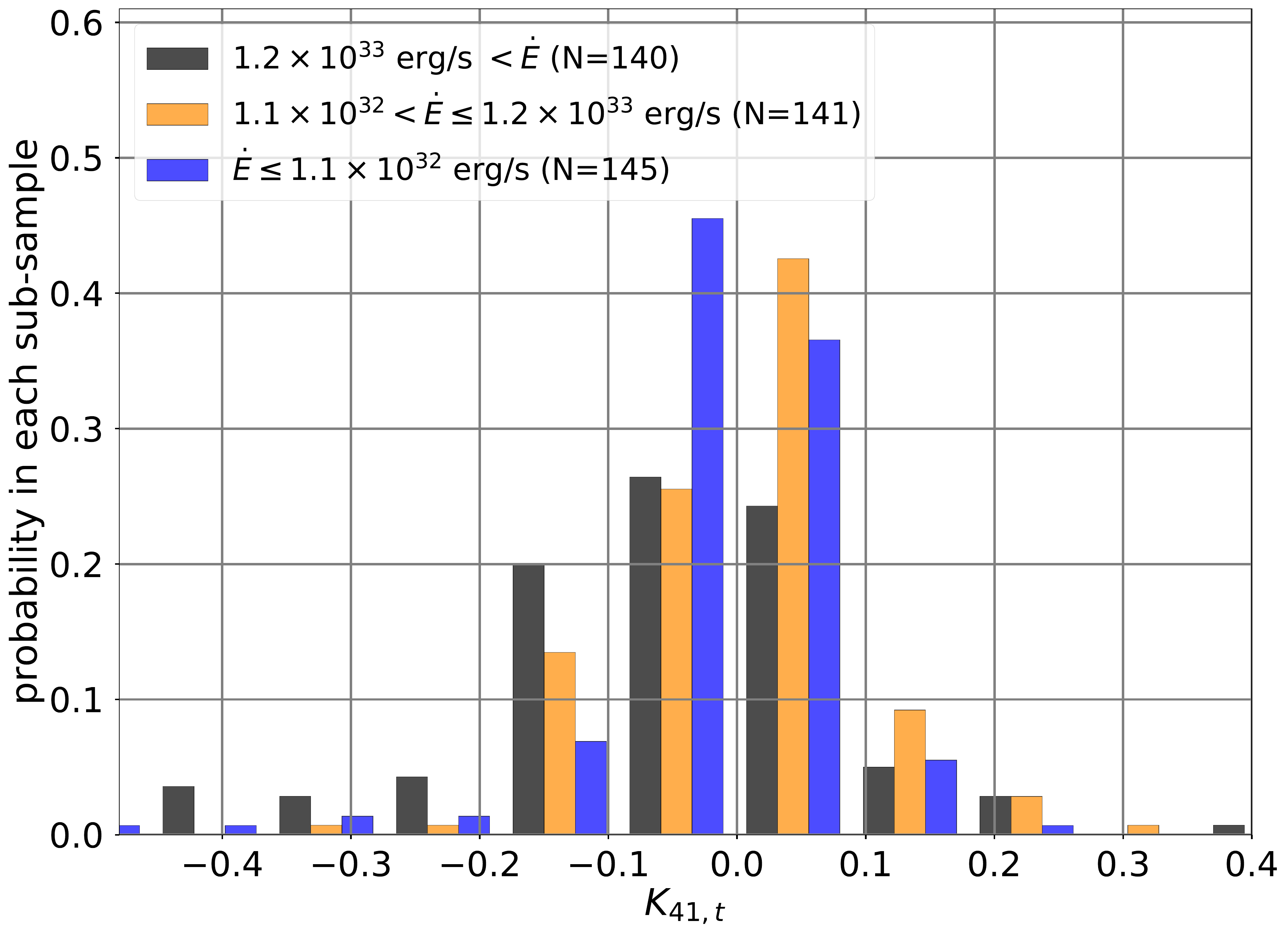}
\vspace{-0.5cm}
\caption{A zoom-in into the contrast $K_{41,t}$ distributions for three spin-down energy sub-samples based on the 4-channel data. Bin sizes of 1/30 have been used for the range of the 426 $K_{41,t}$ values ($-$1.09 to 1.54). \label{fig:conEdot}}
\end{figure}

We explored whether there is a dependence of the contrast on pulsar parameters. 
For this, we use the two-sample Kolmogorov-Smirnov (KS) test which gives a measure whether two samples are drawn from the same parent distribution. 
Using the width contrast between the two frequency bands furthest apart, we divide the sample of pulsars with valid $K_{41,t}$ values and $K_{41,t}$-uncertainties $<0.5$ into three ad-hoc subsamples:\\ 
\indent (a) $1.2 \times 10^{33}$\,erg\,s$^{-1}$\,$<\dot{E}$ (140 pulsars),\\ 
\indent (b) $1.1 \times 10^{32}$\,erg\,s$^{-1}$\,$<\dot{E} \leq 1.2 10^{33}$\,erg\,s$^{-1}$ (141 pulsars), and\\
\indent (c) $\dot{E} \leq 1.1\times 10^{32}$\,erg\,s$^{-1}$ (145 pulsars).\\ 
We obtain probabilities $p_{KS}(s1,s2)$ that two $(s1,s2)$ of them come from the same parent contrast distribution (Table~\ref{tab:ksadtest}).
The distribution of $K_{41,t}$ for the three $\dot{E}$-samples is shown in Figure~\ref{fig:conEdot}.
For $K_{41,t}$, the KS probabilities indicate a difference between the low-$\dot{E}$ and high-$\dot{E}$ pulsar populations.
The KS-test is insensitive to differences at the wings of the compared distributions\footnote{\href{https://asaip.psu.edu/articles/beware-the-kolmogorov-smirnov-test/}{https://asaip.psu.edu/articles/beware-the-kolmogorov-smirnov-test/}}.
Hence, we also employ the two-sample Anderson-Darling (AD) test to non-parametrically check whether our $\dot{E}$-subsamples come from the same distribution of width contrast values.
As Table~\ref{tab:ksadtest} shows, there is a hint of different distributions from the AD-test results as well. 
Since $\dot{E}$ is calculated from $P$ and $\dot{P}$, we explore the contrast in the $P$-$\dot{P}$ diagram for trends for the three $\dot{E}$-subsamples, but do not find anything significant (see Appendix~\ref{contrastEdot} for an example).
\begin{table}
\centering
\caption{Probabilities $p$ obtained from the Kolmogorov-Smirnov (KS) and Anderson-Darling (AD) tests that the $K_{41,t}$ values of three $\dot{E}$-subsamples come from the same parent distribution. The three $\dot{E}$-subsamples a (high $\dot{E}$), b, and c (low $\dot{E}$) are defined in the text. $^{(N)}$ Note that we used the \texttt{scipy.stats.anderson$\_$ksamp} implementation of the AD-test which is capped at a lowest value of 0.1\%.}
\label{tab:ksadtest}
\begin{tabular}{lrrr} 
$p$ & (a,c) & (a,b) & (b,c)\\
\hline
$p_{KS}$ ($K_{41,t}$) & $1 \times 10^{-7}$ & $4 \times 10^{-6}$ & $0.12$ \\
$p_{AD}$ ($K_{41,t}$) & $<0.001^{(N)}$ & $<0.001^{(N)}$ & $0.030$ \\
\end{tabular}
\end{table} 

\section{Discussion}
\label{sec:Discussion}
\subsection{Frequency-averaged data}
Our $W_{10}$ measurements from TPA observations have a good reproducibility and consistency over the many individual observing epochs. As described in Section~\ref{sec:widthvari}, there were only two pulsars for which changes at the $5\sigma$ significance level are seen, but even then the pulse width differences are $<45$\%, resulting only in a small relative change. 
The stability of $W_{10}$ is an important premise for the reliability of the inferred pulse-width relations for our TPA population.   
$W_{50}$ results are less stable with 16 pulsars showing significant changes (Figure~\ref{fig:w50vari}), potentially indicating interesting pulsar emission changes that warrant a closer look in future works.\\

We derived different slopes for the period-width relationship for the OLS and the ODR-fits, see Figure~\ref{fig:w10pfit}.
Our OLS-result, 
\begin{equation}
\label{eq:OLSslope}
W_{10} = (14.2 \pm 0.4) \deg (P/s)^{-0.29 \pm 0.03} 
\end{equation} 
is similar to previously reported results, e.g., these recent ones:
\begin{equation}
\label{eq:jk19slope}
W_{10} = (15.8 \pm 0.6) \deg (P/s)^{-0.28 \pm 0.03}
\end{equation} 
measured at 1.4\,GHz by \citetalias{Johnston2019},

\begin{equation}
\label{eq:McEwenslope}
W_{10} = (18.5 \pm 0.4) \deg (P/s)^{-0.270 \pm 0.001}
\end{equation} 
measured at 350\,MHz by \citet{McEwen2020},
\begin{equation}
\label{eq:Piliaslope}
W_{10} = (16 \pm 2) \deg (P/s)^{-0.3 \pm 0.4}
\end{equation} 
measured at 150\,MHz by \citet{Pilia2016}.
These PL-indices are also consistent with the $\propto P^{-1/3}$ relationship already reported by \citet{Lyne1988}.\\

It is, however, interesting to consider the striking difference to the ODR-fit result which appears visually to be a much better representation of the behaviour of the population density. Due to the non-Gaussian and changing distributions of widths in the period bins (e.g., Figure~\ref{sec:widthinpbin}), the OLS-derived pulse-width relationship is neither a good descriptor of the population trend, nor a good predictor of the expected width of an individual pulsar.
It is important to note that the non-Gaussian width distributions for small period slices are not indicative of ``bad'' data, rather they emphasize that the pulse width is not depending on the period alone. Higher dimensionality in dependent variables is projected on the period-width diagram. This is also true for the ODR-fit, whose residuals also do not follow a Gaussian distribution. However, the linear ODR-fit, mathematically a principal component analysis in 2D, points the way to identify other physical properties influencing the pulse width in a similar multidimensional analysis.\\

\noindent Our ODR-result, 
\begin{equation}
\label{eq:ODRslope}
W_{10} = (11.9 \pm 0.4) \deg (P/s)^{-0.63 \pm 0.06}, 
\end{equation} 
is however still subject to a strong positive correlation between amplitude and power law index, partly reflected in the relatively large $1\sigma$ uncertainties derived from bootstrapping.
Considering these uncertainties, the ODR-derived power law index lies within $3\sigma$ of the predicted relationship 
$W \propto P^{-1/2}$ whilst this is not the case for the OLS-result.\\

{As discussed by \citet{Isobe1990} for linear regression of astronomical data in general, there is no mathematical basis to prefer one regression method over another. Rather, it depends on the analysis goal. If the goal is the prediction of a dependent variable from one (assumed to be) independent variable and specific assumptions\footnote{The critical assumption in our case is whether the \emph{true} relation between $\log W_{10}$ and $\log P$ is indeed linear.} about the data are satisfied, then \citet{Isobe1990} recommend the use of the OLS result.
If, however, one wants to resolve underlying functional relations between the variables that describe, for example, an observed population, then a symmetric treatment of the variables such as the ODR\footnote{\citet{Isobe1990} prefer the OLS bisector regression over ODR because it produces smaller uncertainties than the ODR. This is however less of an issue for large samples such as ours.} is better suited.}\\

{Our QR-result for the LBL-description of the $W_{10}$ data shows a slope of ${-0.28 \pm 0.03}$ ($q=10$\%), thus it is very similar to the OLS-fit result (${-0.29 \pm 0.03}$). Since both, the QR and OLS, minimize the differences in $W_{10}$ versus the best-fit line, a similar slope is not that surprising. Our result is, however, quite different to the one by \citet{Skrzypczak2018} who found slopes of ${-0.51 \pm 0.07}$ and ${-0.51 \pm 0.05}$ at 333\,MHz and 618\,MHz, respectively (values for their $W_{\rm all}$ fits). This could be due to their individual measurements of conal and core components as opposed to our total pulse widths.}\\

Our OLS and ODR fit results agree within uncertainties for the relation of the pulse widths with $\dot{P}$ (or with $\dot{E}$). However, the best-fitting results are not aligned with the regions of highest number densities (blue areas in Fig.~\ref{fig:w10PdotEdot}). We suspect this is due to the rather asymmetric distribution of points. An example is the comparison of the two regions with $\dot{P} \approx 5\times 10^{-15}$ and $\dot{P} > 10^{-14}$ where the number of points below and above the best-fitting line are very different. The number density contours indicate that it may be still possible to identify a mathematical description of the correlation between these -- visually almost randomly distributed -- measurements. 
Since the $\dot{P}$ and $\dot{E}$ relations are not the main topic of this work, we defer a detailed investigation to a future paper, and caution against the use of our best-fitting results for $W_{10}-\dot{P}$ and $W_{10}-\dot{E}$.
  
\subsubsection{Pulsar evolution and the period-width diagram}
\label{sec:disJK19}
Previously discussed physical quantities influencing the pulse width include the shape and filled fraction of the emission beam, emission heights, and the geometry of the rotation and magnetic axes and the line of sight (e.g., \citealt{Mitra2002,Gupta2003,Weltevrede2009,Ravi2010,Wang2014}). There is observational evidence that the obliquity $\alpha$ decays with time, leading to an alignment of the spin and magnetic axes (\citealt{Tauris1998,Weltevrede2008,Young2010}).
Thus, neutron star evolution represented by the (ideally true) ages must influence the pulse-width relationships, too. 
Evidence for an effect of the $\alpha$-decay on the period-width diagram was presented by \citetalias{Johnston2019}.
In their population synthesis, \citetalias{Johnston2019} used $P$ as proxy for the poorly known true age and modified the random distribution $\alpha$ with a cut-off depending on $P^{2}$. 
By analysing slices of the simulated and observed distributions in the period-width diagram, \citetalias{Johnston2019} were able to corroborate the hypothesis of $\alpha$-decay, and thus a specific constraint on pulsar evolution. 
As we showed in similar plots in Figure~\ref{fig:pw10histos}, the TPA data already allows one to probe such simulations to finer detail, finding for example small deviations for the TPA pulsars from the \citetalias{Johnston2019}-simulated pulsar probability density function for small $P$.\\

The \citetalias{Johnston2019} simulation assumptions, in particular the simple $\alpha$-decay model, is also underlying our population-based obliquity estimates in Section~\ref{sec:geometry}.
As stated above, these estimates can be wrong for individual pulsars (see Appendix~\ref{sec:obliquityappendix} for examples). However, there is a great potential to refine parameters of pulsar evolution such as the functional shape of $\alpha$-decay if a statistically meaningful comparison with independent obliquity estimates (e.g., fits of observed polarization position angle swings) is connected with the population synthesis study of the period-width diagram.

\subsection{Frequency-dependent data}
\label{sec:diskfreq}
\subsubsection{Period-width relations over frequency}
Figure~\ref{fig:w10pfitch4ch8t} demonstrates that there is no significant ($>3\sigma$) dependence of the power-law spectral index $\mu$ or the amplitude on frequency for the period-width relation in the four or eight sub-bands of our total 775\,MHz bandwidth. 
There may be a hint that if the best-fitting value of one parameter stays constant ($\mu$  for the OLS, amplitude for ODR), then the best-fitting value of the other parameter shows an (insignificant) trend (decrease of amplitude for OLS, increase of $\mu$ for ODR). 
The scatter within the $>3\sigma$ uncertainty region can be explained with the strong correlation between parameters, the involved uncertainties, and  different numbers of pulsars in the respective frequency channels.

Equation~\ref{eq:OLSslope} lists our OLS-fit result for the frequency-averaged data in comparison to Equations~\ref{eq:jk19slope} to \ref{eq:Piliaslope} representing previous works at other frequencies.
A clear trend for the amplitude with frequency is difficult to spot in the listed relations 
due to the involved uncertainties and the difficulty in considering different bandwidths.
A frequency-dependent scaling, i.e., broadening with lower frequency, could in principle be expected from frequency-radius mapping and propagation effects in the magnetosphere.
However, as we show in Section~\ref{res:colours} and discuss in Section~\ref{dis:monchange}, a simple frequency dependence is not applicable to all pulsars. There is clearly a pulsar population where (unscattered) pulse widths get narrower with lower frequency.

\subsubsection{The monotonic change of width with frequency}
\label{dis:monchange}
The analysis of the width colours shows that pulse widths seem not to switch between an increase and a decrease over frequency in the considered 4 sub-bands. A small sample of pulsars shows an increase of pulse widths with frequency, a larger sample a decrease with frequency. 
Similar behaviours have been reported before in multi-frequency pulsar studies.
\citet{Johnston2008} found pulse width broadening with increasing frequency for one third of their 34 pulsars with high S/N.  
\citet{Chen2014} reported clear pulse width broadening with increasing frequency for about 20\% (and narrowing for 54\%) of their 150 pulsars. 
\citet{Noutsos2015} found pulse width broadening with increasing frequency for about 25\% (and narrowing for 56\%) of their 16 pulsars. \citet{Pilia2016} highlighted several cases of unexpected profile evolution where the widths broaden with increasing frequency, too.
In contrast to these studies, the much larger TPA data set (at least factor 4 more pulsars with information on width evolution) has been obtained at one telescope. This homogeneous data set has enabled us to look beyond individual pulsars and find a \emph{monotonic} behaviour of the width evolution over frequency across the whole population of non-recycled pulsars (Figure~\ref{fig:colours4ch}).
We can exclude a bias due to the observational setup as a cause of this monotonic change (see Appendix~\ref{sec:exccolbias}), and two questions arise (i) What can 
describe this monotonic change of width with frequency?, and (ii) What is its physical origin?\\ 

\begin{figure*}
\captionsetup[subfigure]{labelformat=empty}
\subfloat[]{\label{fig:colourDMa}\includegraphics[width = 0.5\textwidth]{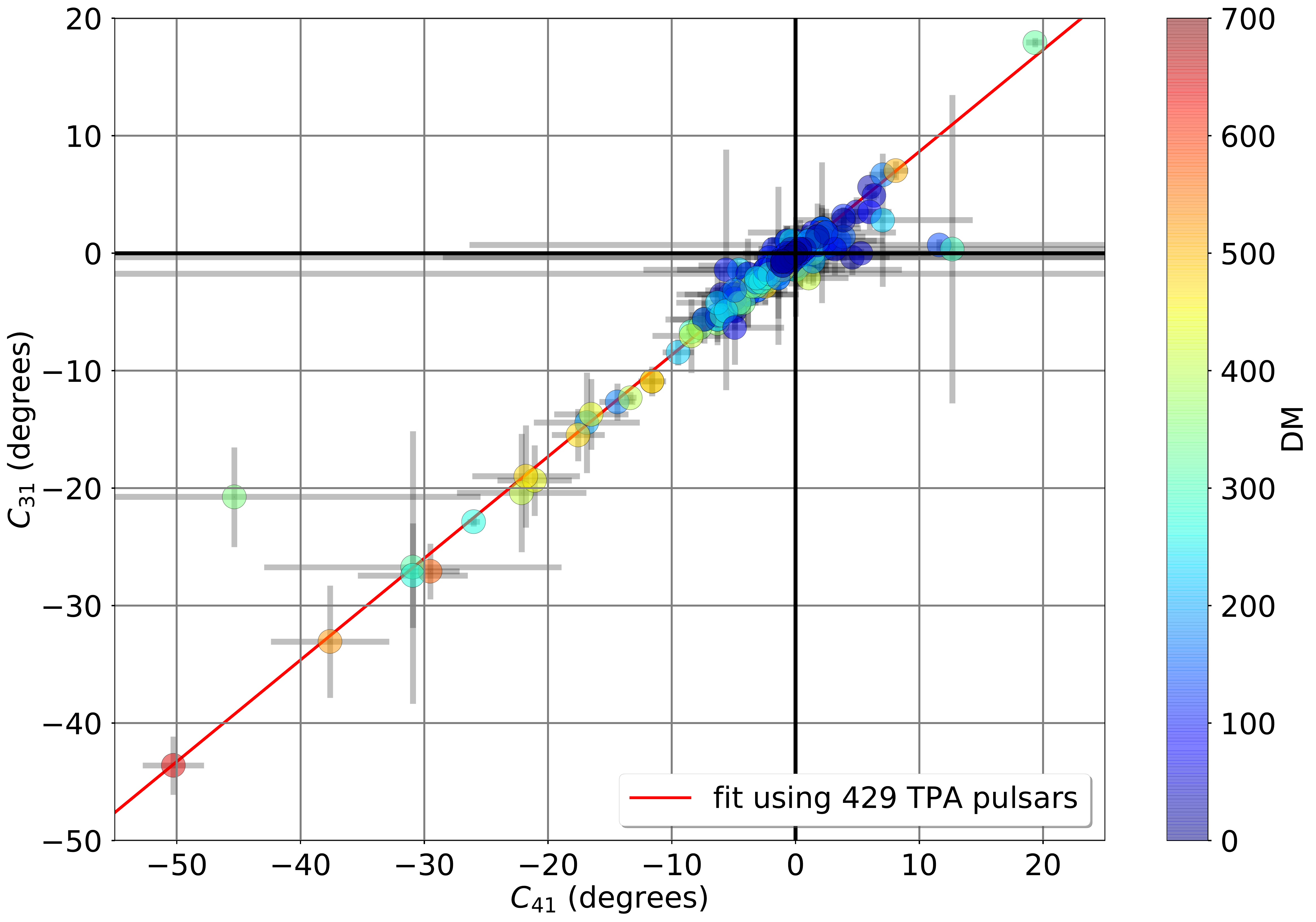}}
\subfloat[]{\label{fig:colourDMb}\includegraphics[width = 0.5\textwidth]{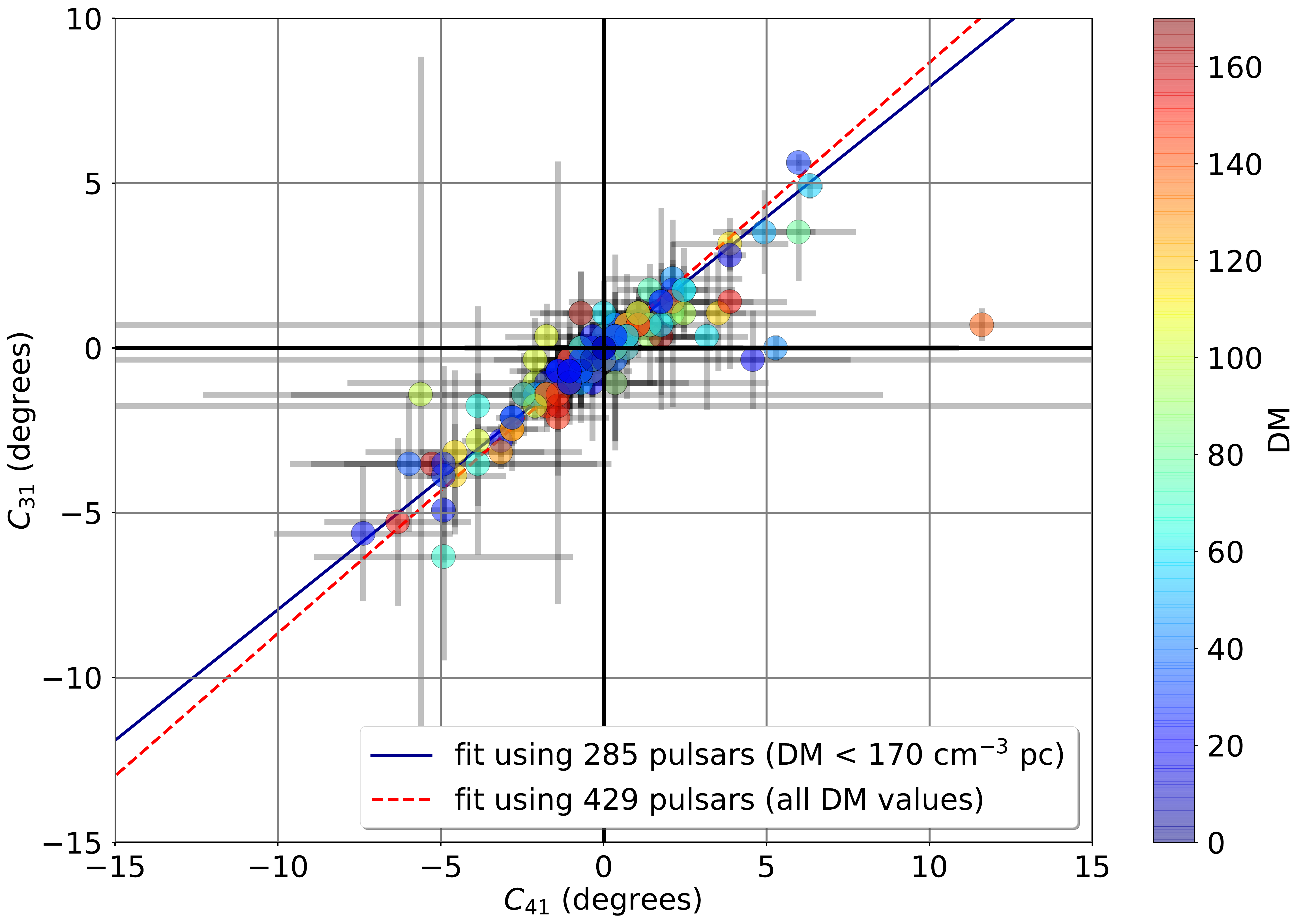}}
\vspace{-0.5cm}
\caption{The effect of DM on the fit results of the width differences for one exemplary colour combination. The left panel shows all the TPA pulsars with respective valid colour measurements. The DM value for each pulsar is indicated by the colour of the circles. The obtained slope for the relation between $C_{41}-C_{31}$ is shown with a red dashed line in the right panel too. This panel restricts to only those pulsars with low DM, and the obtained slope is indicated with the blue line.}
\label{fig:colourDM}
\end{figure*}

As illustrated in Section~\ref{res:colours}, the monotonic change of width with frequency can be described with a linear relation between colours (Figures~\ref{fig:colours4ch} and \ref{fig:colours8ch}, and Table~\ref{tab:colourfits}). A well established broadening of the widths with lower frequencies, i.e., negative colours in Figure~\ref{fig:colours4ch}, is expected from interstellar scattering.  
For the relation between two colours, one can derive the expected slope that only depends on the respective frequencies
 and the scattering spectral index $\alpha_{\rm Scat}$ if one uses a simplistic description of the scattered profile with frequency. We assumed $\alpha_{\rm Scat}=-4$ based on the results by \citet{Oswald2021}.  
The expected colour-colour slopes for scattered-broadened pulse profiles are listed in Table~\ref{tab:colourfits} together with our fit results. If all TPA pulsars with valid colours are considered, we obtain results which are surprisingly close to the ones expected for scattered pulsars. However, we emphasize here that for the positive side on the colour-colour diagram, scattering simply cannot be the underlying reason since these positive-side profiles get wider with \emph{higher} frequencies, the opposite behaviour of what is expected for scattering.\\ 

Nevertheless, scattered pulsars in the third quadrant may influence the exact values of the derived correlations.
While we excluded strongly scattered pulsars from our initial width analysis (Section~\ref{sec:widthmeas}), we may have missed a few scattered pulsars.
In order to exclude the potential effect of scattered pulsars, we do two further tests. Firstly, we only consider the positive colours (and those consistent with zero) for our fits, see Table~\ref{tab:colourfits} for the results. Secondly, we investigate the distribution of the Dispersion Measure ($DM$) and exclude all pulsars with $DM \ge 170$\,cm$^{-3}$\,pc.
The results from Figure~\ref{fig:colourDM} and Table~\ref{tab:colourfits} show only small changes in the derived slopes of the linear relations between colours. 
Our choice of $DM \ge 170$\,cm$^{-3}$\,pc was motivated by studies of the dependency between the scattering time scale $\tau_{\rm scat}$ and DM.  
Using the work by \citet{Krishnakumar2015}, for example, the majority of pulsars with such a DM-cut have  $\tau_{\rm scat}$(327\,MHz)$<100$\,ms.
Scaling with (1381\,MHz/327\,MHz)$^{\alpha_{\rm Scat}}$, appropriate to our third frequency channel (for the $C_{41}-C_{43}$ relation that has the flattest slope in Fig.~\ref{fig:colourDM}), gives typical $\tau_{\rm scat}$(1327\,MHz)$<0.3$\,ms.
Assuming a measurement sensitivity of 1\,deg (3 bins, here with respect to the width measured in the fourth frequency band because of the colour), implies that our width measurements in the third channel would typically \emph{not} be sensitive to scattering for pulsars with $P>0.107$\,sec. Such a period cut would only remove 3 (out of 285) pulsars and give nearly idential fit results for $C_{41}-C_{43}$ as listed in  Table~\ref{tab:colourfits}. Thus, the measured width evolution of the DM-selected pulsars should be insensitive to scattering. For the lower frequencies, similar estimates can be derived albeit with a lower number of pulsars. 
The results are similar if eight frequency sub-bands are employed (Figure~\ref{fig:colours8ch}, and Table~\ref{tab:colourfits}). Here, the change of the slope is more pronounced if one considers only positive colours or DM-selected pulsars. A factor of about two is seen, for example, for $C_{81}-C_{86}$. Visually, the deviation of the measurements from the original slope is also easily apparent for $C_{81}-C_{82}$ in Figure~\ref{fig:colours8ch}.
As an additional check, we plotted the difference of the measurements and the expected value from the scattering relation (using a range of $\alpha_{\rm Scat}$). These ``residuals''  showed a monotonic behaviour with negative slope which is close to zero if high-DM pulsars are included, but significantly different from zero if we employ our DM-cut. This also excludes scattering as the sole explanation of the observed trend.
While the exact mathematical description of the change of pulse width with frequency requires further investigation, the general monotonic trend is a robust result.
Our analysis shows that besides scattering, there must be at least one other underlying factor for the monotonic change of pulse width with frequency.\\

Figure~\ref{fig:conEdot}
and the respective KS and AD-tests showed reasonable hints of width-contrast sample deviations for different $\dot{E}$ sub-groups. 
Motivated by this, we employ the two-sample AD-test for the positive and negative $C_{41}$ samples to check whether the same distribution can describe these colour-subsamples with respect to the pulsar parameters $P$, $\dot{P}$, $\dot{E}$, $B$, age, DM, and $W_{10}$.
For this analysis, we consider the $1.5\sigma$ level of the colour uncertainties, and checked the result with and without the DM-cut $DM \ge 170$\,cm$^{-3}$\,pc. Sample numbers are rather small (on the order of $\sim 40-100$).
The derived $p_{AD}$-values are all above 0.04, and the null hypothesis cannot be excluded with these data. Thus, among the tested pulsar parameters our AD-tests do not identify an underlying factor for the monotonic width change.\\

Similar to scattering, propagation effects in the magnetosphere can only significantly contribute to the width broadening at \emph{lower} frequencies \citep{McKinnon1997}, i.e., pulsars in the third quadrant in the colour-colour diagram may show this effect. The monotonic broadening of the pulse width at higher frequencies (pulsars in first quadrant of Figure~\ref{fig:colours4ch}) is likely not due to this effect.

\citet{Pilia2016} pointed out that for a few of their pulsars with width broadening at higher frequency, new peaks appeared in the profile at higher frequencies. Thus, 
an obvious question is whether we see a different number of pulse components in the positive-colour and negative-colour samples. For example, more pulse profile components could appear at higher frequencies in the positive-colour sample. We visually checked pulse profiles of the different samples and did not notice any abnormalities in number of pulse components or clusters of multiple-component profiles. Thus, we exclude newly appearing or disappearing pulse components as the sole explanation for the observed monotonicity of the colours.\\

This leaves the change of the spectral index, in particular the change of the spectral indices of individual (not necessarily discernible) pulse components as an explanation. This possibility was already investigated by \citet{Chen2014} on the basis of the fan beam model. In their interpretation, the frequency dependence of the pulse width is not caused  by emission frequency dependence on altitude. Rather, the pulse width change is merely a byproduct
of an inhomogeneous broadband emission spectrum across the emission region.
{Evidence for this was, for example, presented recently by \citet{Basu2021}, who showed the difference in spectral index between central and outer components for 21 pulsars.} 
Here, we conducted a simple simulation to explore the hypothesis of the spectral index effect independent of the exact emission process. 
In particular, we want to see whether all the observed pulse width changes could be reproduced from two simple assumptions -- (i) the widths of the pulse \emph{components} get wider with lower frequency (as observed, e.g., by \citealt{Thorsett1991}), and (ii) the outer pulse components have flatter spectra than the inner pulse components (as is often observed, e.g. \citealt{Rankin83}).    
We generated fake profiles at the same 8 channel frequencies as the data used in this paper. For 
each simulated pulsar, we generate a profile at the highest frequency first.
Each profile is made up of 5 Gaussian components, whose centroids are positioned randomly within a fixed window of $35\degr$. The width of each component, given as the standard deviation of the Gaussian, is also chosen randomly from a uniform distribution from $1.75\degr$ to $12\degr$, which are representative values from pulsar data. The amplitudes are chosen randomly between 0 and some arbitrary maximum value. To generate the remaining 7 frequency channels, we scale the amplitudes and widths of these components as follows. For the amplitudes, we assign a spectral index depending on the distance of the component centroid from the fixed pulse phase that corresponds to the middle of the allowed window. We assign a spectral index of $-2.0$ to the central pulse phase, decreasing linearly to $-1.0$ at the edges of the window. {For comparison, \citep{Basu2021} show the average difference between spectral indices of outer and inner components to be $\sim-0.7$}. This qualitatively emulates the known observational property of central components having steeper spectra compared to leading and trailing edge components. For the component widths, we assign a spectral index of $-0.1$ (wider with lower frequency). As the data are noiseless, width measurements are trivial. The frequency dependence can easily be captured by the difference in width between highest and lowest frequency. We generated many sets of 1000 pulsars with multi-frequency profiles, and find that, as expected, depending on the relative amplitude 
and position of the components at the reference frequency, the majority of pulsar profiles become narrower at high frequencies compared to a minority that become broader. Crucially, the simulation results in a uni-modal distribution of the width difference parameter (i.e., colour), as we see in our data. The parameters we have used in our simulation are chosen to demonstrate this qualitative result, and should not be attributed higher significance.\\

With respect to the two questions at the beginning of this section, our simple simulation indicates a possibility to phenomenologically describe the found monotonic change of the pulse width with frequency. We emphasize that it is crucial to explain the behaviour of pulsars where we see broader widths at higher frequencies. While for the classically expected behaviour (broadening with lower frequencies), there can be several contributing factors (propagation effects in magnetosphere, scattering, different emission heights following the radius-frequency-mapping scenario), the former width changes are more challenging to explain. We excluded the influence of observation sensitivities on the finding of monotonicity, and excluded the description by scattering as the sole reason, as well as the explanation of visible emergence of new pulse components.
Our tests for dependence of the width changes on other pulsar properties were inconclusive.
For quantitative constraints regarding the emission model and thus the physical origin of the monotonicity, individual pulse components of our TPA-sample need to be investigated with full polarisation information which we defer to a future paper.

\section{Summary}
The MeerKAT telescope provided the TPA with an exquisite data set which enabled us to homogeneously measure pulse-widths for over thousand pulsars. For pulsars with multiple observations, we verified that the pulse width at 10\% does typically not significantly vary. The 50\% pulse widths indicated 16 pulsars with significant changes. About half of the $\sim 760$ pulsars with measured $W_{10}$ in the frequency-averaged data, also have such measurements in four and eight frequency sub-bands.\\ 

We revisited the period-width diagram and found that\\ 
(i) the general trend for the frequency-averaged data, $W_{10} \propto (P)^{-0.29 \pm 0.03}$ is consistent with previous estimates if $P$ is used as the only independent variable for a relation $W_{10}=f(P)$.\\
(ii) this period-width relationship is not a good description of the correlation seen between $W_{10}$ and $P$ if the population density of our large data set is considered. An ODR-fit resulted in a much steeper slope  $W_{10} \propto P^{-0.63 \pm 0.06}$. This, and the large spread of $W_{10}$ over any period-bin indicate that instead of a pure 2D-relation ($W_{10},P$) the {pulsar population in the} width-period diagram {is better described by} a projection of higher dimensionality of variables.\\
(iii) the population synthesis by \citetalias{Johnston2019} is an example of such multi-variable approach to explain the physical underpinnings of the period-width diagram. Deriving population-based obliquity estimates from their model, we make the argument that independent obliquity estimates together with the period-width relationship can be used to constrain pulsar evolution.\\

Utilizing the $W_{10}$-measurements in the frequency sub-bands, we showed that\\
(i) the period-width relationships in the sub-bands are not significantly different in comparison to the result from the frequency-averaged data.\\
(ii) by introducing colours (width differences) and contrasts (width differences normalized by the frequency-averaged measurement) 
it is possible to derive interesting conclusions about the pulsar population.
We confirm the previous reports, e.g. by \citet{Chen2014}, that there is a sizeable pulsar population for which the pulse widths broaden with higher frequencies. This is difficult to explain with radius-to-frequency mapping in the conal beam model.\\
(iii) the width change is monotonic if the whole pulsar population is considered. \\  
(iv) instrumental effects can be excluded as explanation of the observed monotonicity. Scattering or magnetospheric propagation alone cannot describe such a relationship, in particular not for the pulses that broaden with higher frequency. We did not see any noticeable pulse shape differences in the two pulsar groups that had either widths increasing or decreasing with frequency. 
A simple qualitative model considering a spectral index distribution over several contributing pulse components was able to produce an uni-modal distribution of the width difference. This indicates that, independent of the emission process, the spectral index warrants further studies in connection with the monotonically changing pulse width of the pulsar population.\\

The large homogeneous data set of TPA width measurements will be available as supplementary material and made accessible at the {\sl CDS}.

\section*{Acknowledgements}

We would like to thank Eric Feigelson
for the helpful discussions. 
The MeerKAT telescope is operated by the South African Radio Astronomy Observatory, which is a facility of the National Research
Foundation, an agency of the Department of Science and Innovation.
MeerTime data are housed and processed on the OzSTAR supercomputer at Swinburne University of Technology with the support of
ADACS and the gravitational wave data centre via AAL. 
BP and LO acknowledge funding from the UK Science and Technology Facilities
Council (STFC) Grant Code ST/R505006/1. AK also acknowledges
funding from the STFC consolidated grant to Oxford Astrophysics,
code ST/000488.

\section*{Data availability}
An upcoming TPA census paper will summarize the fully calibrated TPA data set, including the pulse profiles used here.   
The TPA data will be made pulic at that time. The code used to measure pulse widths is available on request to the corresponding author.



\bibliographystyle{mnras}
\bibliography{TPApops} %



\appendix
\section{Pulse profiles}
\label{sec:profileplots}
\begin{figure*}
\captionsetup[subfigure]{labelformat=empty}
\subfloat[]{\label{fig:profilesa}\includegraphics[width = 0.32\textwidth]{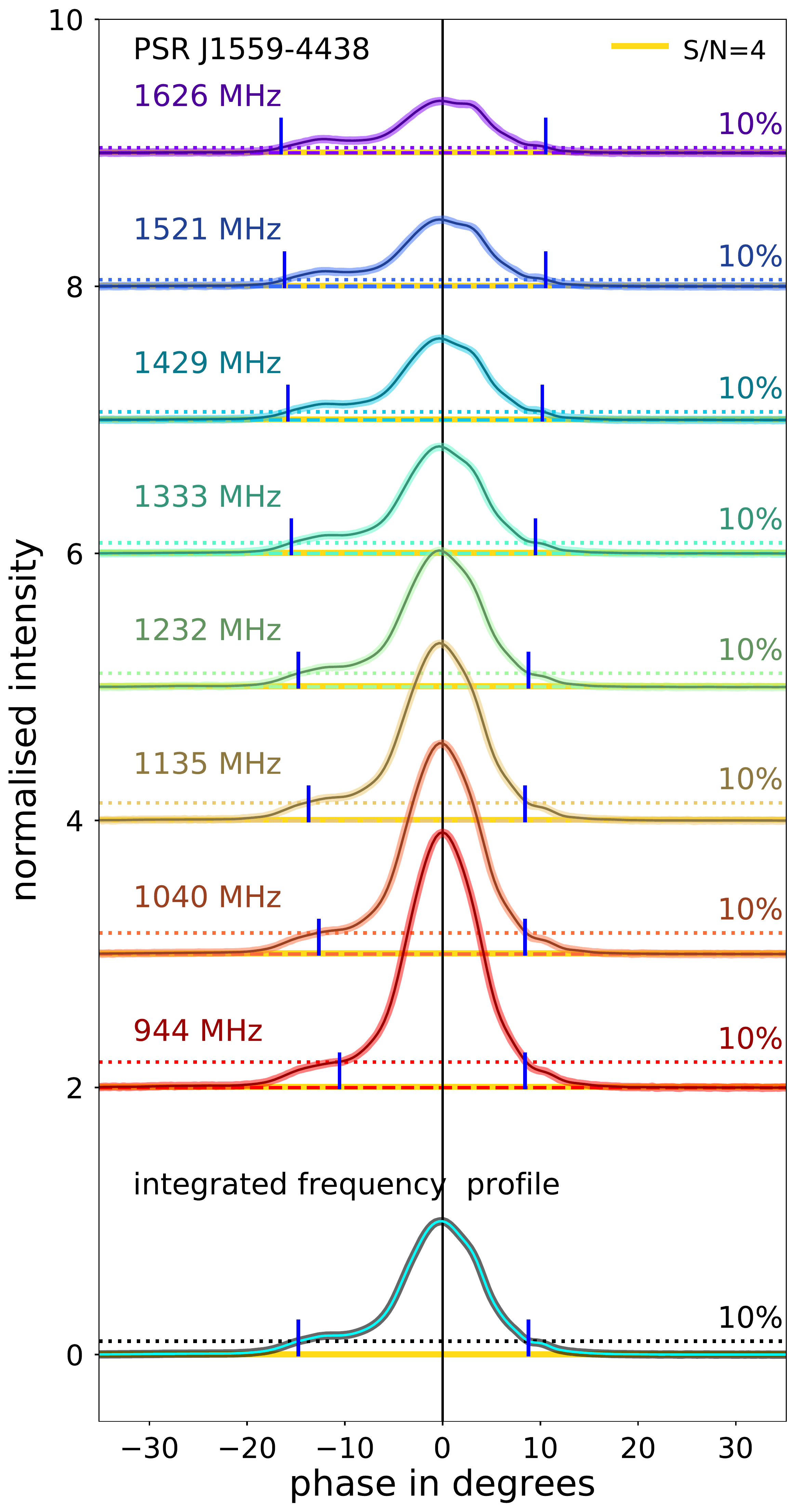}}
\subfloat[]{\label{fig:profilesb}\includegraphics[width = 0.32\textwidth]{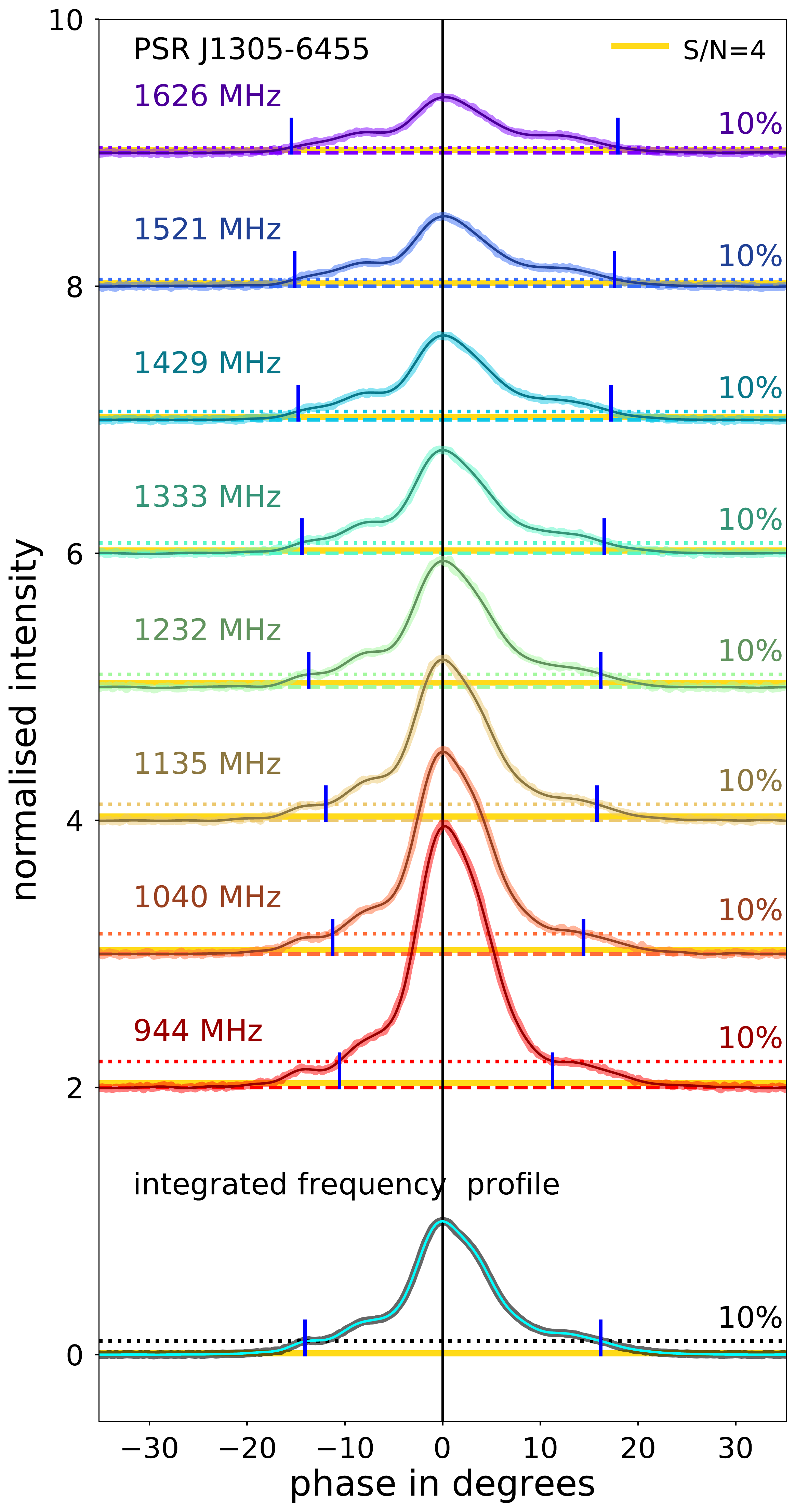}}
\subfloat[]{\label{fig:profilesc}\includegraphics[width = 0.32\textwidth]{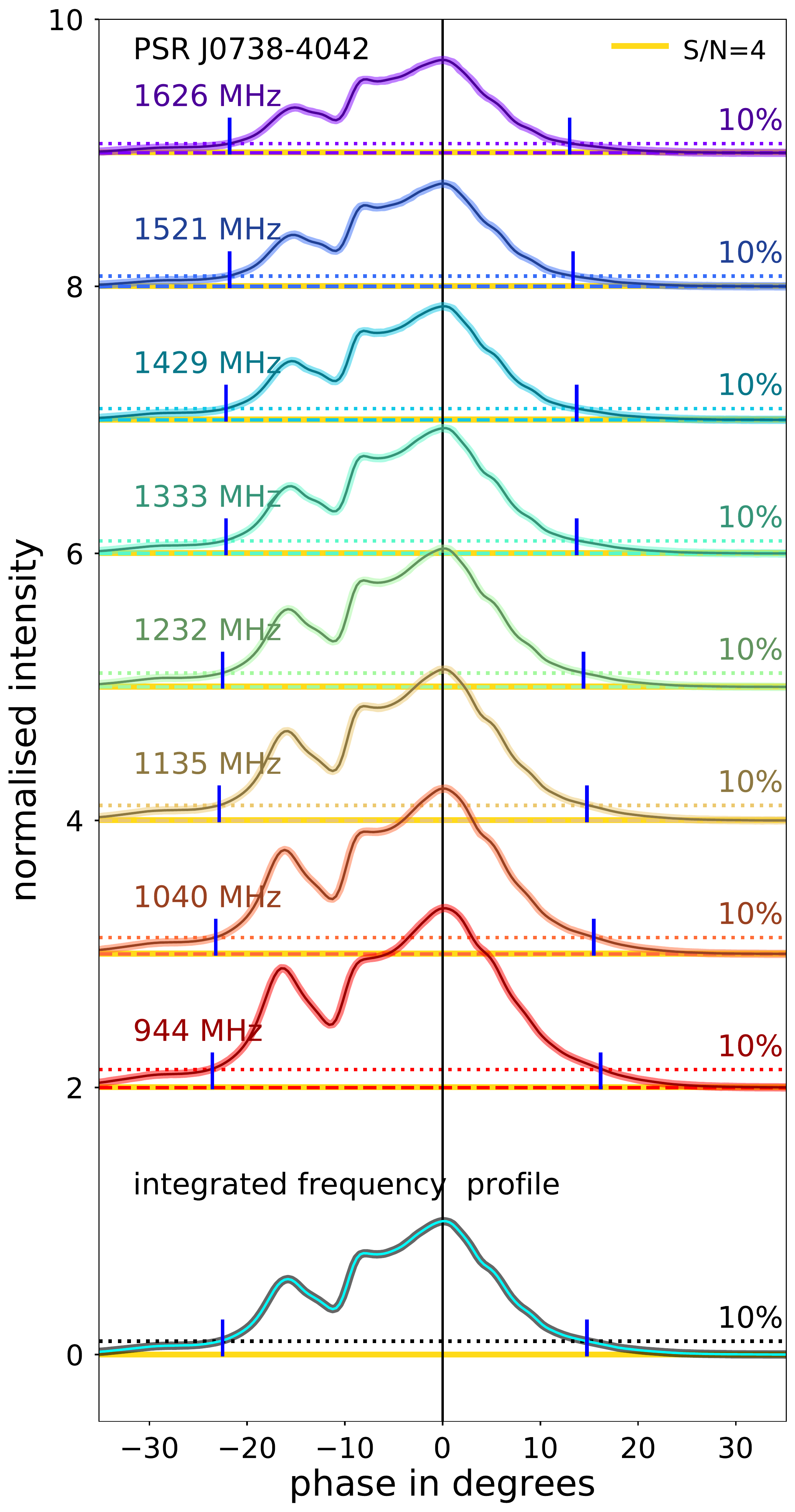}}\\
\vspace{-0.5cm}
\subfloat[]{\label{fig:profilesd}\includegraphics[width = 0.32\textwidth]{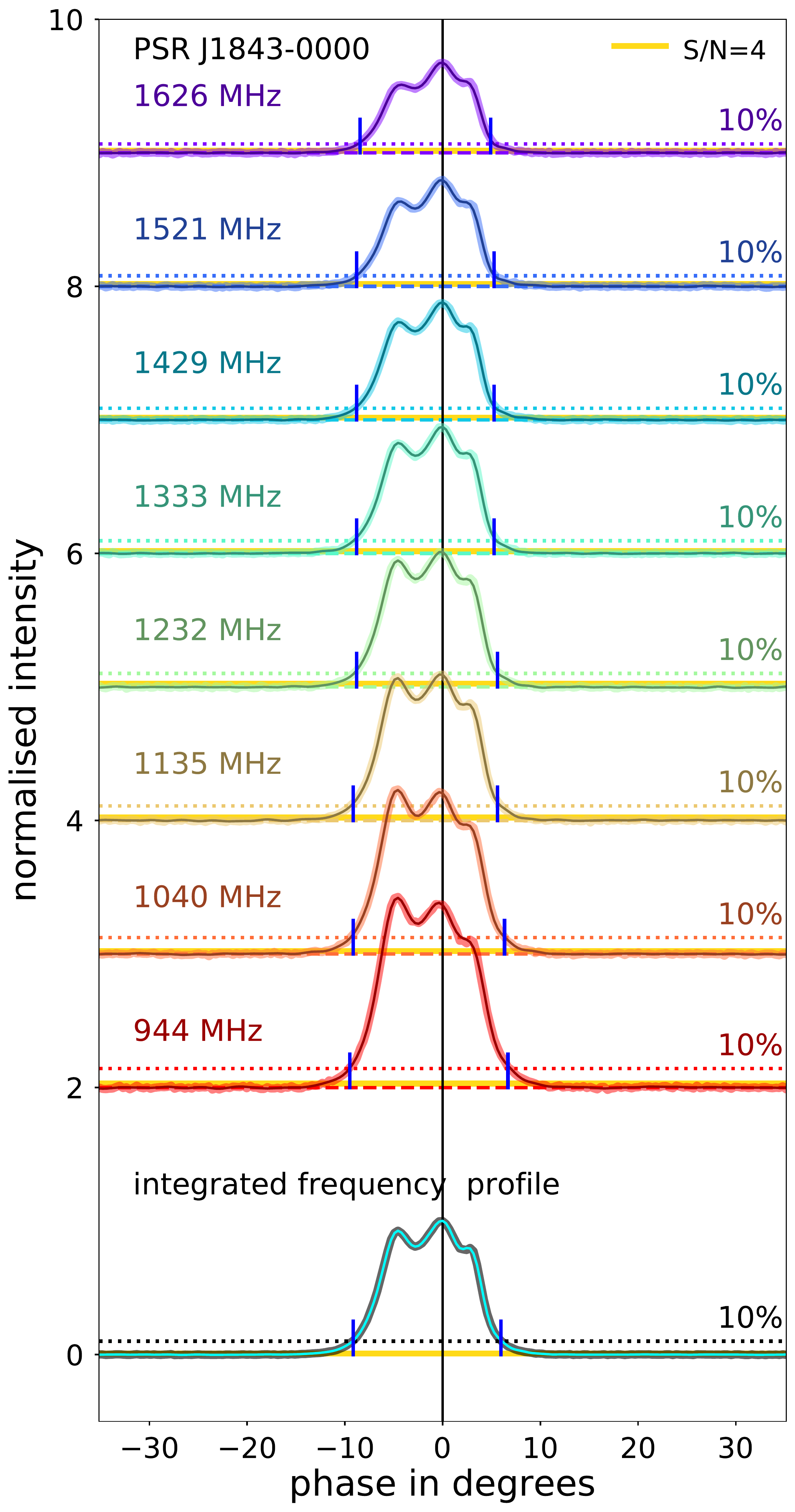}}
\subfloat[]{\label{fig:profilese}\includegraphics[width = 0.32\textwidth]{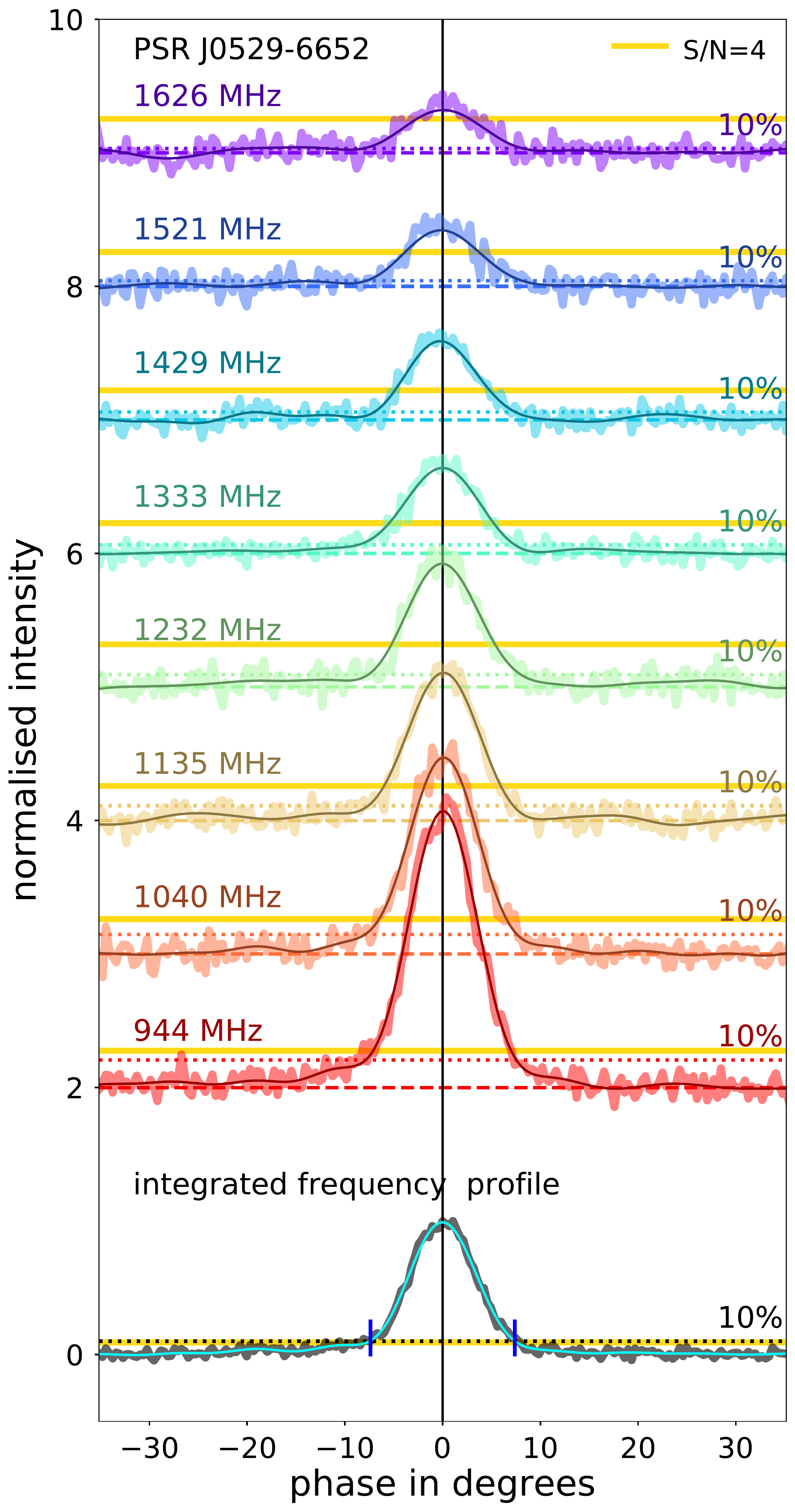}}
\subfloat[]{\label{fig:profilesf}\includegraphics[width = 0.32\textwidth]{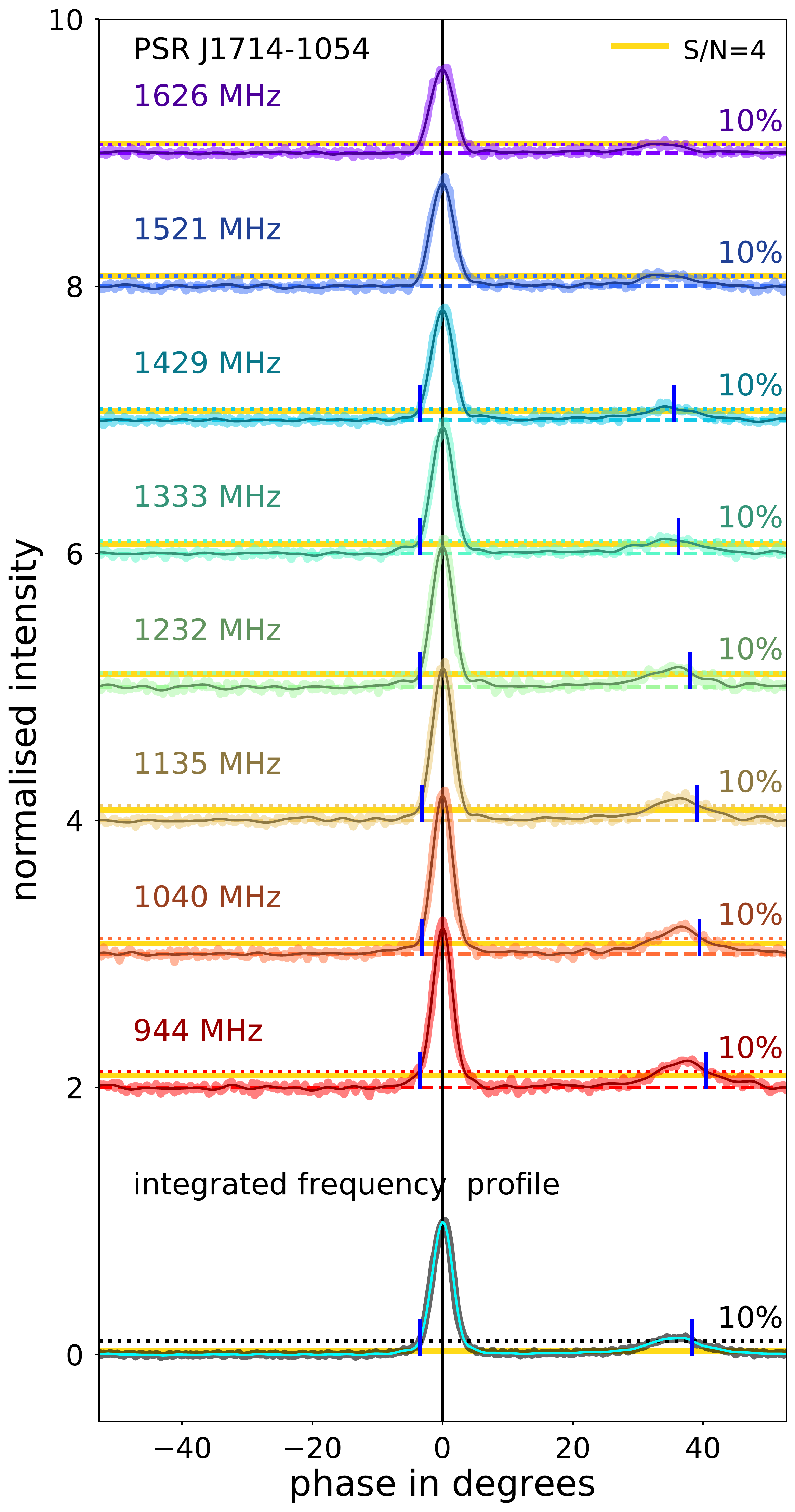}}
\vspace{-0.5cm}
\caption{{Data and overplotted GP-profiles of 6 representative TPA pulsars, centered on the maximum of the frequency-averaged profile. Each plot shows the frequency-averaged profile on the bottom, and eight frequency channels above (shifted by constants indicated with dashed lines). The profiles are normalised to the peak of the frequency-averaged profile. The dotted lines indicate the 10\% of the respective maxima, the yellow lines mark the $S/N=4$ level for each profile. If there are valid $W_{10}$ measurements (see Sec.~\ref{sec:widthmeas}), these are indicated by the two blue vertical lines. PSRs J1559$-$4438 and J1305$-$6455 show increasing $W_{10}$, PSRs J0738$-$4042 and J1843$-$0000 show decreasing $W_{10}$ trends for increasing frequency. PSRs J0529$-$6652 and J1714$-$1054 are weak TPA pulsars for which we were able to measure $W_{10}$ for the frequency-averaged profiles, but not for all eight frequency channels.}}
\label{fig:pulseprofiles}
\end{figure*}

{Here, we show in Figure~\ref{fig:pulseprofiles} a few representative pulsar examples to illustrate how their data and the noiseless profiles produced by the GP are related to our $W_{10}$ measurements.}

\section{Width distribution in period bins}
\label{sec:widthinpbin}
\begin{figure*}
\includegraphics[width=13cm]{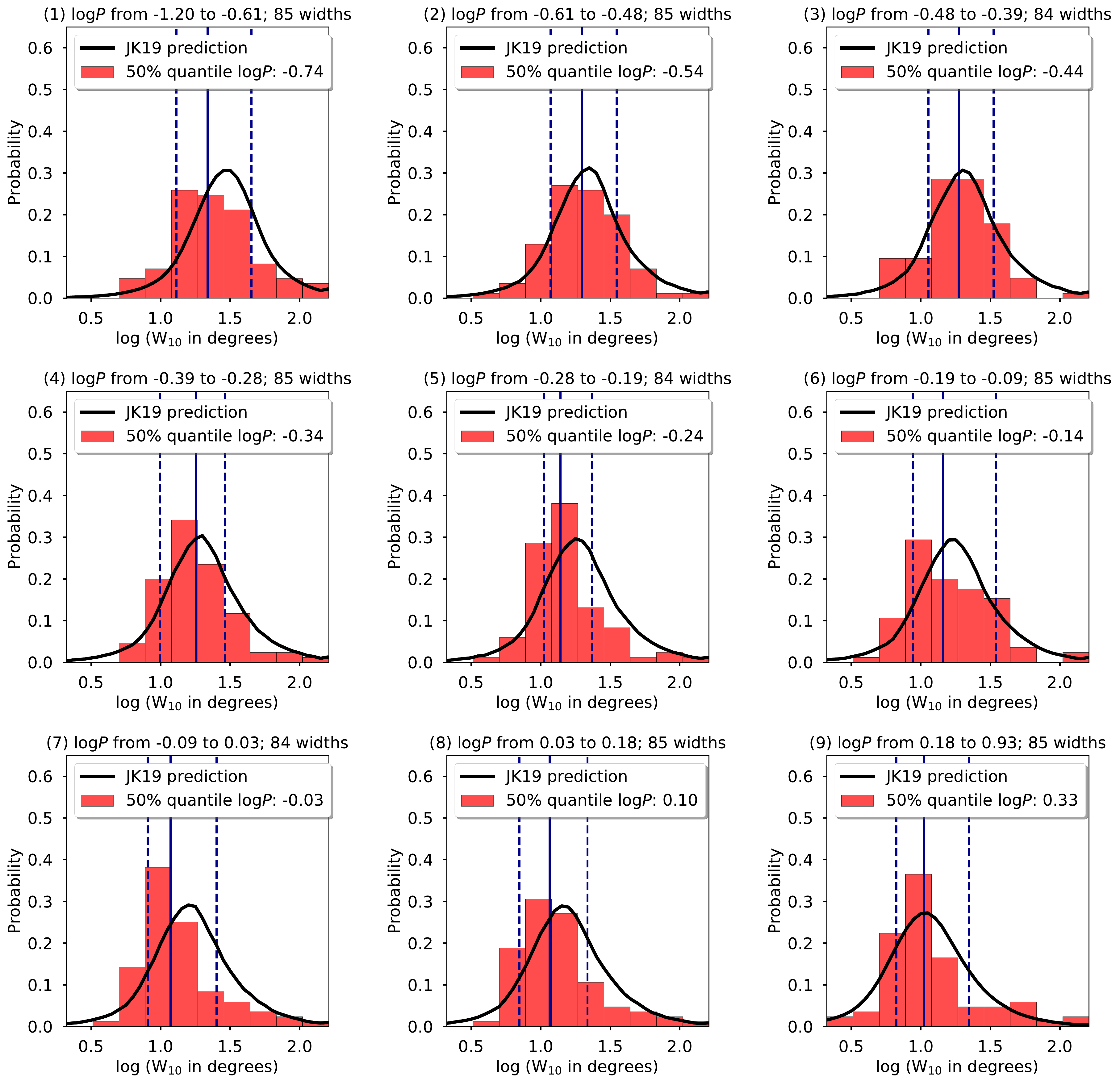}
\vspace{-0.2cm}
\caption{The distribution of $\log W_{10}$ for different $\log P$ bins for the frequency-averaged data. Ten quantiles are used to divide the period range in 9 samples of roughly equal size of 85.  The $\log P$ range is listed on top of each histogram, its 50\% quantile as label. The solid and dashed blue lines indicate the 50\%, 16\% and 84\%  quantiles of the $\log W_{10}$ distribution. The black line shows the expected distribution for the model by \citetalias{Johnston2019} (their figure\,8).
\label{fig:pw10histos}}
\end{figure*}

While the density and scatter plots of the period-width diagrams already indicate a non-Gaussian distribution of the width values, Figure~\ref{fig:pw10histos} shows this behaviour even more clearly. The histograms, representing width distributions of equal-sized sized samples, show very different shapes depending on the considered period bin. The large width spread and the different shapes of these histograms lead to a strong deviation of the OLS-derived best power law model from the line of highest density in the logaritmic period-width.

\section{Obliquity constraints}
\label{sec:obliquityappendix}

\begin{table}
\centering
\caption{Population-based obliquity estimates (the angle $\alpha$ between the magnetic and the rotation axes) for 762 TPA pulsars (excerpt). These estimates use the population synthesis model by \citetalias{Johnston2019} together with the measured $W_{10}$ of the frequency-averaged data. See text for details. The component indicates the major or minor component as in Table~\ref{tab:w50w10w5w1}, i.e., it can only be 2 if the pulsar is identified as interpulse pulsar. $\alpha (Q50)$ lists the 50\% quantile of $\alpha$, while  $\sigma^-_\alpha$ and $\sigma^+_\alpha$ indicate the uncertainties based on the 16\%, 50\% and 84\% quantiles. The full table is available as supplementary material.}
\label{tab:alphas}
\begin{tabular}{lcrrr}
        PSR &  component &  $\alpha (Q50)$ & $\sigma^-_\alpha$   &  $\sigma^+_\alpha$ \\
	& & $\degr$ & $\degr$ &$\degr$ \\
\hline
  J0034-0721 &         1 & 12.94 &  -3.93 &  3.78 \\
  J0108-1431 &         1 & 18.50 &  -5.89 &  7.48 \\
  J0113-7220 &         1 & 55.51 & -15.97 & 15.69 \\
  J0151-0635 &         1 & 12.84 &  -3.83 &  3.65 \\
  J0152-1637 &         1 & 41.16 & -11.89 & 10.71 \\
  J0206-4028 &         1 & 44.97 & -12.71 & 11.74 \\
  J0211-8159 &         1 & 19.82 &  -6.21 &  6.33 \\
  J0255-5304 &         1 & 51.91 & -15.54 & 12.84 \\
  ...\\
\end{tabular}
\end{table}

\begin{figure}
\includegraphics[width=\columnwidth]{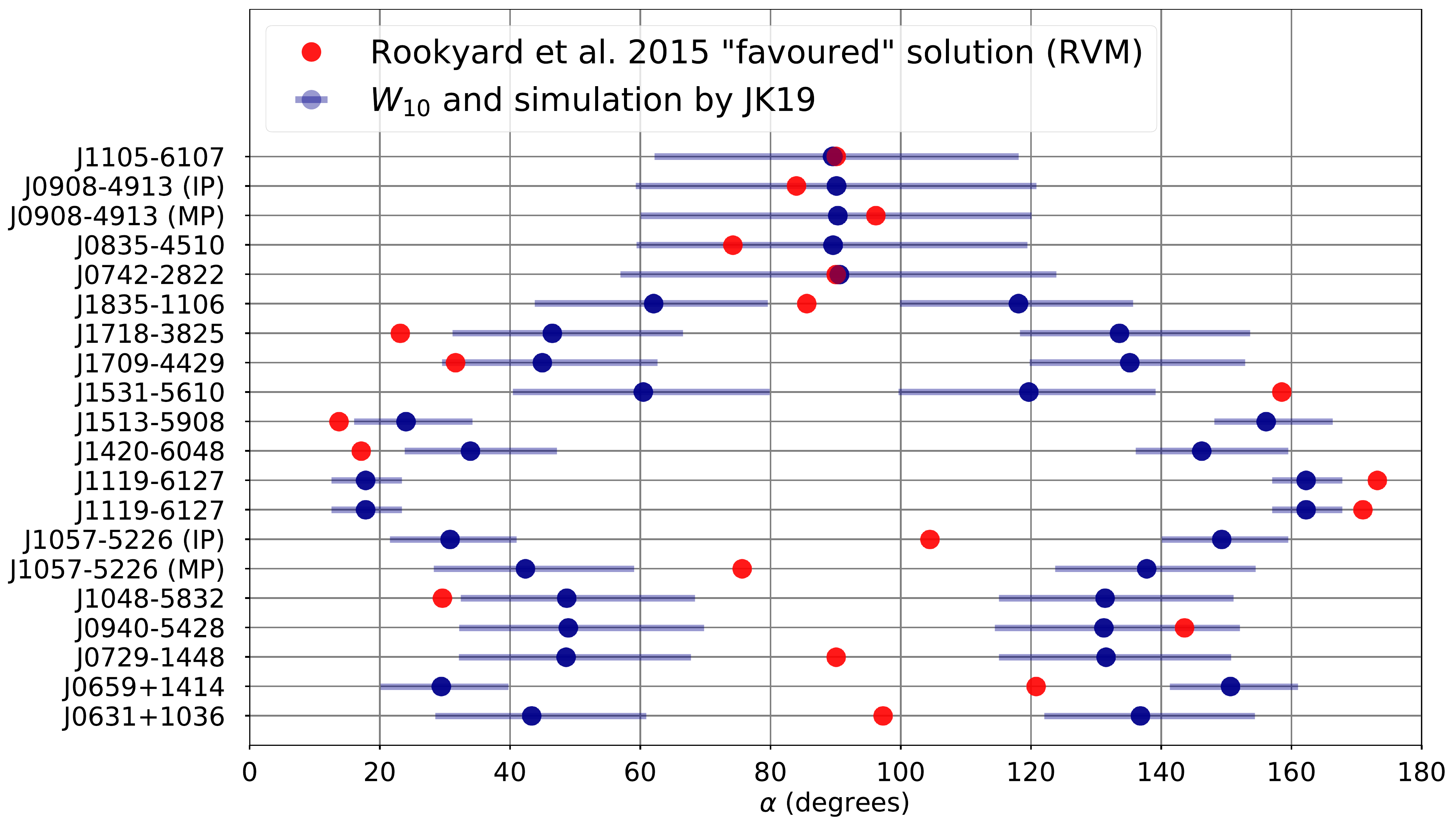}
\vspace{-0.5cm}
\caption{The obliquity parameters (50\% quantile of $\alpha$ with uncertainties corresponding to the 16\% and 84\% quantiles) for those TPA pulsars that were also investigated by \citet{Rookyard2015}. The blue points were obtained from our $W_{10}$ measurements of the frequency-averaged data, combined with the simulation input of \citetalias{Johnston2019} (their figure\,8). For the four upper pulsars, we obtained  $\alpha \approx 90\deg$, while for the lower pulsars we plot $\alpha$ and $180\deg-\alpha$. The red points indicate the ``favoured $\alpha$ solutions'' from Table~2 of \citet{Rookyard2015}. Note that the ``allowed solutions'' usually encompass a much larger range, up to the full $180 \deg$. \citet{Rookyard2015} used the Rotating Vector Model to derive their geometry constraints.
\label{fig:JK19alpha_rooky}}
\end{figure}

\begin{figure}
\includegraphics[width=\columnwidth]{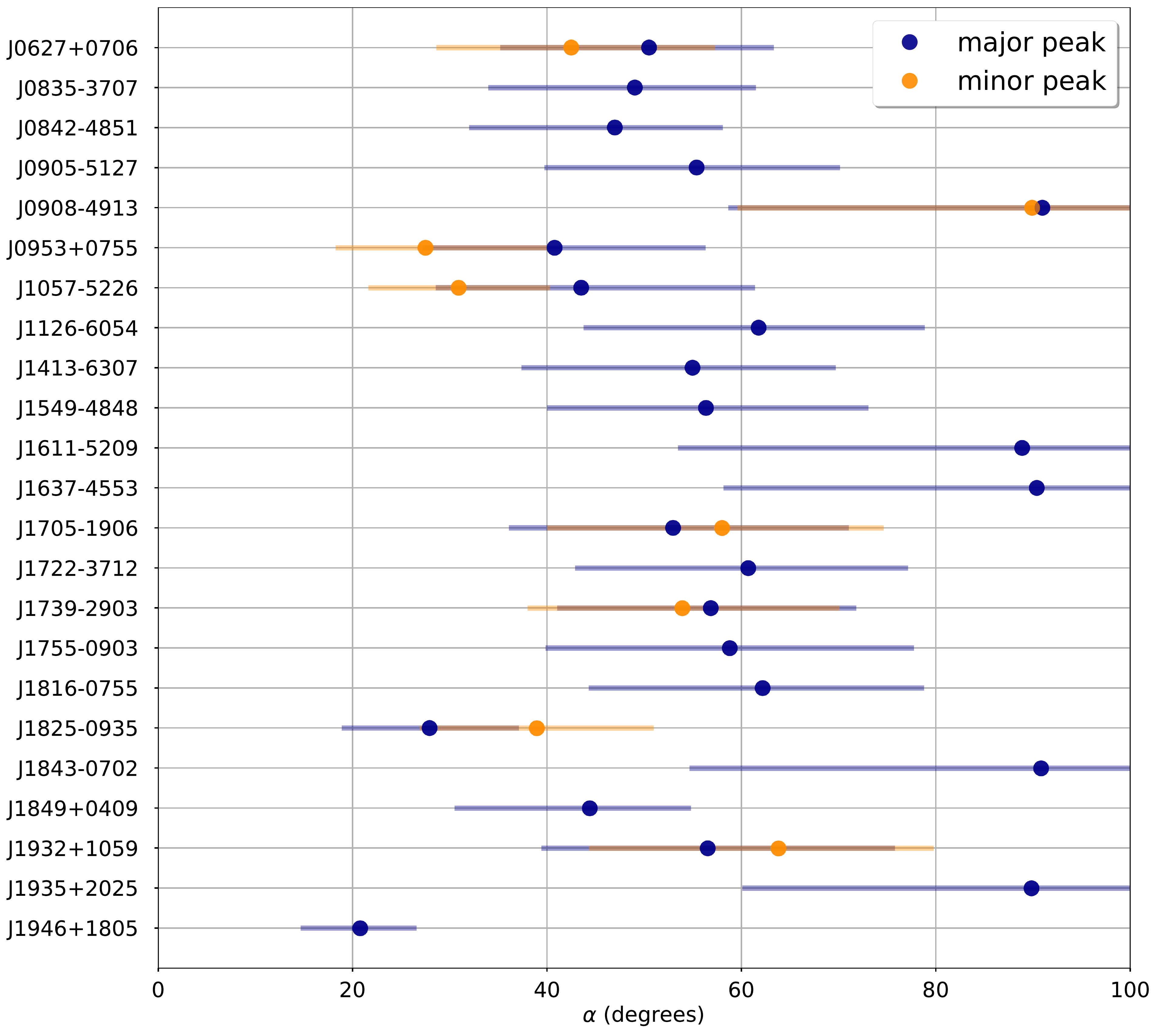}
\vspace{-0.5cm}
\caption{The obliquity (50\% quantile of $\alpha$ with uncertainties corresponding to the 16\% and 84\% quantiles) for those TPA pulsars that are identified as tentiative interpulse pulsars. 
The obliquity values for the main (blue) and suspected interpulse (orange, not all have $W_{10}$ measurement) pulse components were obtained from the respective $W_{10}$ measurements of the frequency-averaged data, combined with the simulation input of \citetalias{Johnston2019} (their figure\,8).
\label{fig:JK19alpha_interpulse}}
\end{figure}

Geometry estimates in the literature often rely on relations with the pulse widths as well. Compiling or deriving a comprehensive list of truely independent obliquity estimates is beyond the scope of this paper. 
Here, we restrict to two checks of our $\alpha$ estimates in Table~\ref{tab:alphas} with small samples. In Figure~\ref{fig:JK19alpha_rooky}, we plot a comparison of our population-based estimates with the results from \citet{Rookyard2015} who fit the observed polarization position angle swings with the Rotating Vector Model \citep{Radhakrishnan1969}. About 50\% of the ``favoured solutions'' from \citep{Rookyard2015} agree within (16\% to 84\% quantile) unertainties with our population-based estimates.
In Figure~\ref{fig:JK19alpha_interpulse} we plot those TPA-pulsars that are thought to have interpulses. Hence, for these pulsars one would expect $\alpha \approx 90\deg$.
Our population-based estimates show such $\alpha$ only for five of the 24 pulsars. Interpulse pulsars provide a good illustration that pulse widths and periods alone are not a sufficient set of parameters to constrain the viewing geometry.

\section{Colour slopes and Contrast spread}
\label{sec:colscat}
\begin{figure}
\includegraphics[width=\columnwidth]{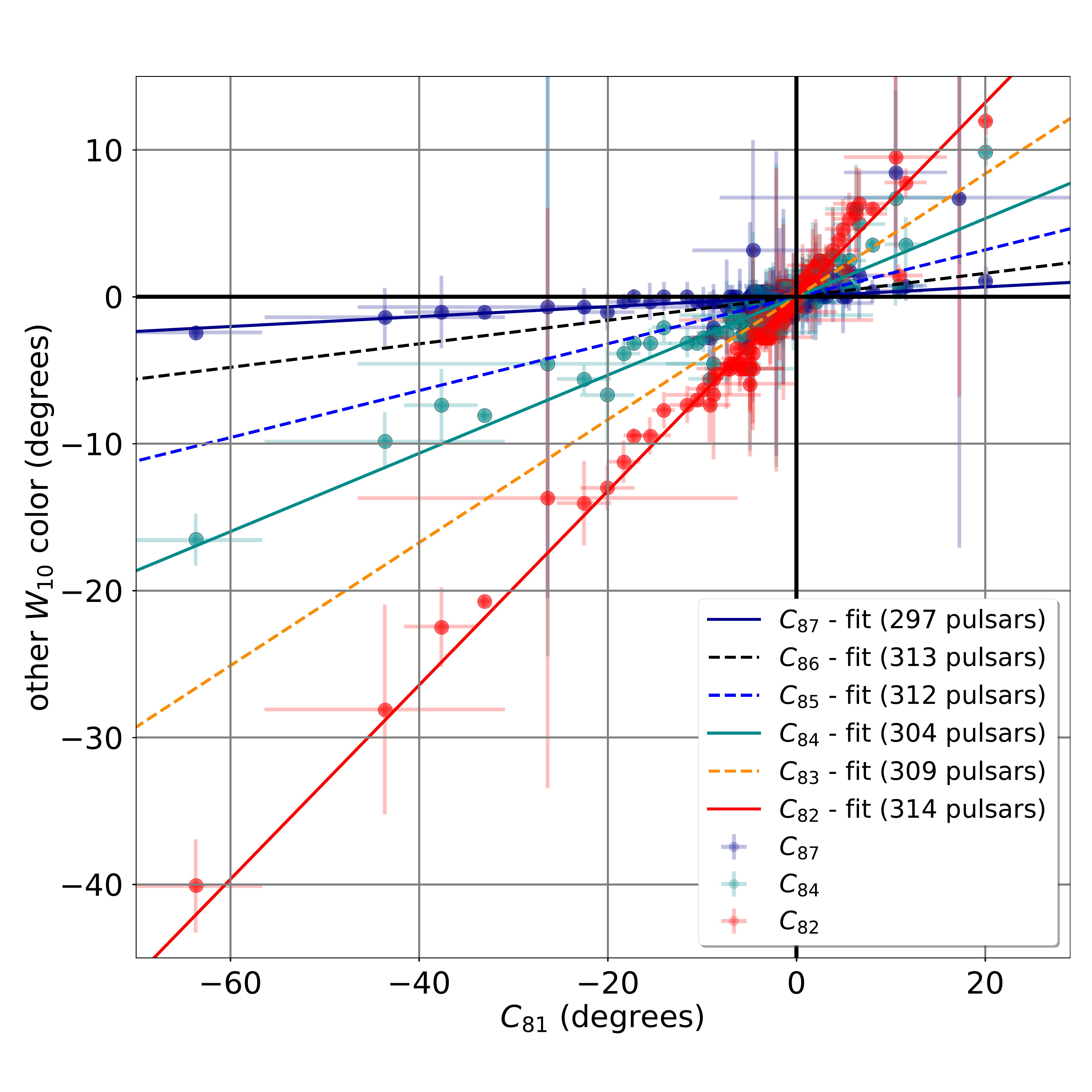}
\vspace{-0.5cm}
\caption{The $W_{10}$-colour distributions for the 8-channel data. The $x$-axis represents the largest possible colour $C_{81}$ (between frequency bands 1 and 8). Linear relations (ODR-fits, Table~\ref{tab:colourfits}) with an exemplary set of 6 other colours are shown on the $y$ axis. For clarity, we only show the measured values and their uncertainties for three of the six relations. Only TPA pulsars with valid $W_{10}$ measurements in all the frequency channels of the respective colours are plotted.
\label{fig:colours8ch}}
\end{figure}

For completeness, we show the $W_{10}$-colour distributions for the 8-channel data in Figure~\ref{fig:colours8ch}.\\

\begin{figure}
\includegraphics[width=\columnwidth]{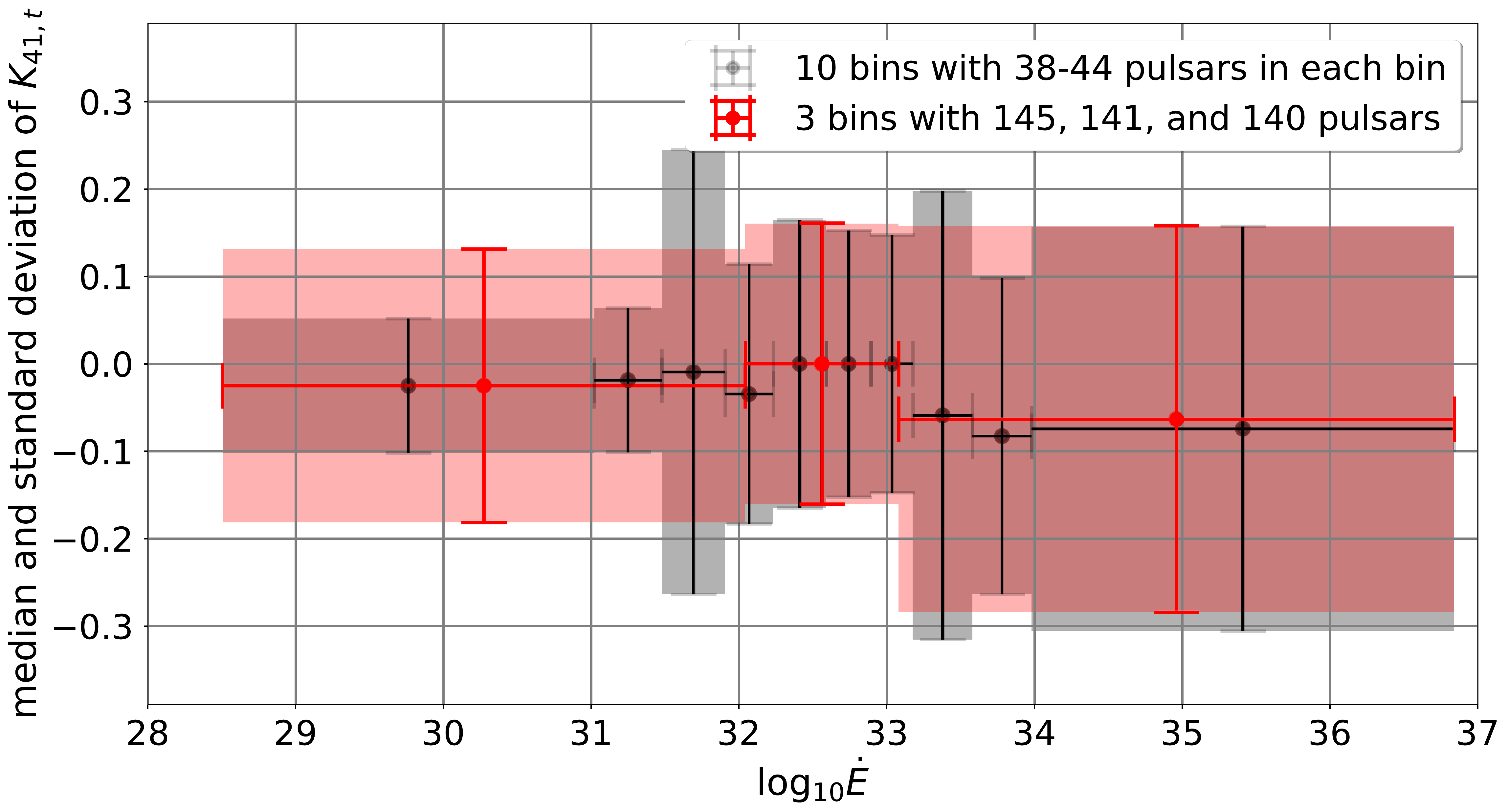}
\vspace{-0.5cm}
\caption{The median and spread of the contrast $K_{41,t}$ for different choices of $\dot{E}$-subsamples based on the 4-channel data. 
The three spin-down energy subsamples (light red boxes and crosses) were used for the KS and AD tests of Section~\ref{conEdotKStest}. In addition, we show 10 $\dot{E}$-bins (grey boxes and black points) that have roughly similar sample size as example of finer $\dot{E}$-binning. 
\label{fig:con41spreadEdot}}
\end{figure}

\label{contrastEdot}
In Section~\ref{conEdotKStest}, we found for the 4-channel data that the distribution of $K_{41,t}$ values for three $\dot{E}$ groups shows differences according to the KS and AD-tests. We checked for, and found no visually obvious $P$ and $\dot{P}$ dependencies of the contrast $K_{41,t}$. However, we noticed that the high-$\dot{E}$ pulsars may have a larger spread in $K_{41,t}$ than the other $\dot{E}$-groups.
In Figure~\ref{fig:con41spreadEdot}, we test this hypothesis and plot the median and standard deviation of $K_{41,t}$ for the three $\dot{E}$-groups, and in addition for another binning with ten $\dot{E}$-groups. 
The three $\dot{E}$-groups show a slightly larger spread in $K_{41,t}$ for the high-$\dot{E}$ pulsars, but the smaller binning reveals that this is mainly due to pulsars with $\dot{E}$ between $10^{33}$\,erg/s and $10^{34}$\,erg/s. Overall, the hint of larger contrast spread turned out to be not statistically significant. 

\section{Excluding an observational setup bias for the width behaviour}
\label{sec:exccolbias}
\begin{figure}
\includegraphics[width=\columnwidth]{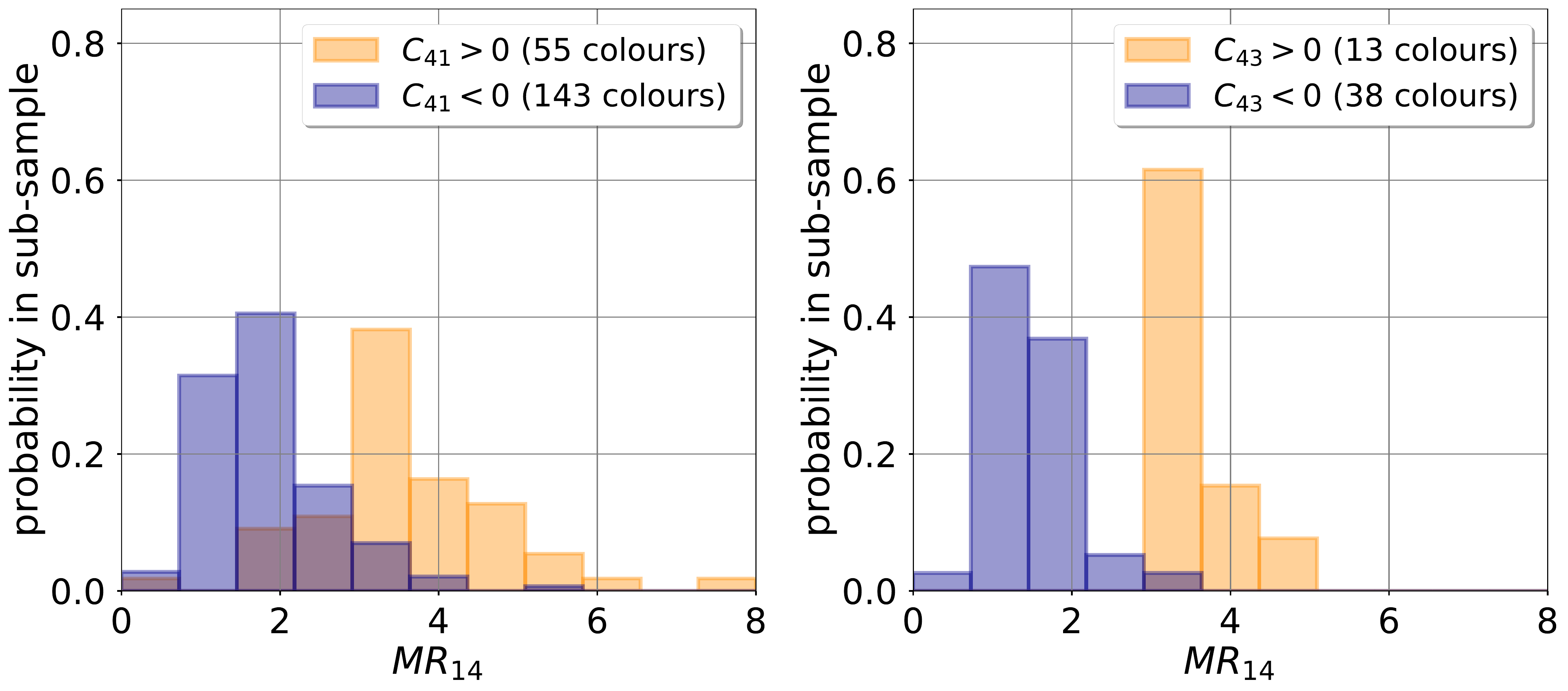}
\vspace{-0.5cm}
\caption{The positive (orange) and negative (blue) $W_{10}$-colour distributions over maximum ratio for the 4-channel data. 
The maximum ratio $MR_{14}$ is the maximum of the noiseless pulse profile in channel 1 divided by the respective maximum value in channel 4. The left and right panels show the sub-samples for colours with the largest and smallest frequency differences, $C_{41}$ and $C_{43}$, respectively.
\label{fig:colours4chmaxratio}}
\end{figure}

We test whether different sensitivities in the various frequency bands affect our results on the width difference, in particular the finding that for some pulsars there is an broadening of the pulse width with increasing frequency, while for most other pulsars the pulse widths broaden with decreasing frequency. 
We compared the width distribution over S/N for the lowest (channel 1) and highest frequency band (channel 4). While the peak of the S/Ns in the high-frequency band is lower than the S/N-peak in the low-frequency band, the width distribution does not show any notable differences. Thus, this comparison does not indicate a frequency-dependent S/N-effect.  
Our S/N, however, was estimated using the maximum of the noiseless profile and the GP-determined noise of the profile. If the noise is different in each channel (it is), the maximum  may be different too. The maximum, however, is employed to determine the 10\% level to measure the $W_{10}$. Imagining the same intrinsic Gaussian profile peaking out of different noise levels, the 10\% width may be measured at different heights of the intrinsically same Gaussian profile. To test for this sensitivity effect we compare the pulse maximum ratio distributions of the positive-colour and negative-colour sample on Figure~\ref{fig:colours4chmaxratio}. We use the maximum ratio between channel 1 to 4, $MR_{14}=M_1/M_4$. The maximum of the noiseless profile at the lowest frequency is typically smaller than the one at the highest frequency\footnote{Together with the lower S/N of the high frequency band this means that the noise level is larger for the high frequencies.}. The positive-colour sample ($W_{10}$ at the higher frequency is larger than the $W_{10}$ at the lower frequency) and the negative-colour sample show distinct distributions in Figure~\ref{fig:colours4chmaxratio}. The positive-colour sample has typically larger $MR_{14}$ than the negative-colour sample.
One would expect the opposite if the "peaking-out-of-different-noise-levels"-effect was the explanation for the observed monotonicity of the colours.
Therefore, we conclude that observed profile width changes with frequency are not a bias from the observational setup, i.e., they are not due to the sensitivity or noise behaviour of individual frequency channels.\\

\bsp	
\label{lastpage}
\end{document}